\documentclass[a4paper,11pt]{article}
\pdfoutput=1 

\usepackage{silence}
\usepackage{jheppub}

\usepackage{float}
\usepackage{relsize}
\usepackage{fontenc}
\usepackage{amsmath,amsfonts,amssymb,bm}
\usepackage{tabularx}
\usepackage{tabularray}
\usepackage{nicematrix}
\usepackage{adjustbox}
\usepackage{tikz}
\usepackage{slashed}
\usepackage{pgfplots}
\usepackage{mathtools}
\usepackage{siunitx}
\usepackage{physics}
\usepackage[compat=1.1.0]{tikz-feynman}
\usepackage[T1]{fontenc}
\usepackage{multirow}
\usepackage{booktabs}
\usepackage{cleveref}

\def\({\left(} 
\def\){\right)}

\newcommand\rmd { {\rm d} }

\newcommand{\nlo}{\text{NLO}}
\newcommand{\lo}{\text{LO}}
\newcommand{\Amp}{\mathcal{M}}

\tikzset{graviton/.style={decorate, decoration={snake, amplitude=.6mm, segment length=1.5mm, pre length=.3mm, post length=.3mm}, double}}

\usepackage{xcolor}
\xdefinecolor{redviolet}{RGB}{228,38,207}
\xdefinecolor{brickred}{RGB}{220,60,66}
\xdefinecolor{bluscuro}{RGB}{51,51,179}

\usepackage{hyperref}

\title{\boldmath Pion pair production in $e^+ e^-$ annihilation at next-to-leading order matched to Parton Shower}

\author[a,b]{Ettore Budassi,}
\author[b]{Carlo M. Carloni Calame,}
\author[a,b]{Marco Ghilardi,}
\author[a,b]{Andrea Gurgone,}
\author[a,b]{Guido Montagna,}
\author[c,d]{Mauro Moretti,}
\author[b]{Oreste Nicrosini,}
\author[b]{Fulvio Piccinini,}
\author[a,b]{and Francesco P. Ucci}

\affiliation[a]{Dipartimento di Fisica, Universit\`a di Pavia, Via A. Bassi 6, 27100 Pavia, Italy}
\affiliation[b]{INFN, Sezione di Pavia, Via A. Bassi 6, 27100 Pavia, Italy}
\affiliation[c]{Dipartimento di Fisica e Scienze della Terra, Universit\`a di Ferrara, Via Saragat 1, 44122 Ferrara, Italy}
\affiliation[d]{INFN, Sezione di Ferrara, Via Saragat 1, 44122 Ferrara, Italy}

\emailAdd{ettore.budassi01@universitadipavia.it}
\emailAdd{carlo.carloni.calame@pv.infn.it}
\emailAdd{marco.ghilardi01@universitadipavia.it}
\emailAdd{andrea.gurgone01@universitadipavia.it}
\emailAdd{guido.montagna@unipv.it}
\emailAdd{mauro.moretti@unife.it}
\emailAdd{oreste.nicrosini@pv.infn.it}
\emailAdd{fulvio.piccinini@pv.infn.it}
\emailAdd{francesco.ucci@pv.infn.it}

\abstract{The pion pair production in $e^+ e^-$ annihilation at flavour factories plays a crucial role in the determination of the hadronic contribution to the muon anomalous magnetic moment. The recent CMD-3 measurement of the pion form factor via energy scan displays a significant difference with the previous experimental determinations. In order to contribute to an improved theoretical description and simulation of energy scan experiments, we present a calculation of the $e^+ e^- \to \pi^+ \pi^- (\gamma)$ hadronic channel at next-to-leading order matched to a Parton Shower algorithm in QED and sQED. According to the recent advances in the literature, particular attention is paid to the treatment of the pion composite structure in loop diagrams beyond the commonly used factorised sQED approach, as well as to the modelling of multiple photon radiation through the Parton Shower algorithm. In particular, we carry out a detailed discussion on the inclusion of the pion form factor in the virtual sQED corrections according to two independent methods, inspired by the generalised vector meson dominance model and the dispersive approach, respectively. We find the two methods to be in remarkable agreement. We show phenomenological results for inclusive and differential observables which are relevant for precision energy scan measurements, focusing on the impact of the radiative corrections and the effect of the various approaches for the treatment of the pion form factor. Our calculation is implemented in an updated version of the Monte Carlo event generator \textsc{BabaYaga@NLO}, that can be used for fully exclusive simulations in data analysis.
}

\keywords{$e^+ e^-$ Experiments, Precision QED, NLO Corrections, Parton Shower}

\begin{document} 
\maketitle
\flushbottom

\section{Introduction}
\label{sec:intro}
The anomalous magnetic moment of the muon $a_\mu = (g-2)_\mu / 2$ is a quantity of crucial importance in contemporary particle physics~\cite{Jegerlehner:2009ry,Jegerlehner:2017gek,Aoyama:2020ynm}. The present experimental value, which is known with the striking accuracy of 0.19~ppm, stems from the average of the final E821 result at the Brookhaven National Laboratory~\cite{Bennett:2006fi} with the Run-1 and Run-2/3 measurements by the Fermilab muon $g-2$ experiment~\cite{Abi:2021,Aguillard:2023}. The Standard Model prediction of the muon magnetic anomaly, as compiled by the Muon $g - 2$ Theory Initiative in 2020~\cite{Aoyama:2020ynm}, yields a discrepancy of $5\sigma$ with the experimental world average quoted in~\cite{Aguillard:2023}.

The comparison between the experimental measurement and the theoretical prediction of the muon anomaly is complicated by the presence of strong interaction effects, which have to be calculated with non-perturbative methods. More specifically, the majority of the theoretical uncertainty of $a_\mu$ comes from the hadronic vacuum polarisation (HVP) effects. The $5\sigma$ deviation refers to the prediction that is obtained via a dispersive calculation, using time-like data. This prediction is based on multiple measurements of the cross section of the $e^+e^-~\to~\textit{hadrons}$ processes, as performed by various experiments. Among all the hadronic channels, the process of two pion production, i.e. $e^+ e^- \to \pi^+ \pi^-$, is responsible for more than $70~\%$ of the total value of HVP and gives the dominant contribution to the uncertainty of the time-like calculation~\cite{Jegerlehner:2017gek,Keshavarzi:2019abf,Davier:2019can,Benayoun:2019zwh,Keshavarzi:2024wow}.

Moreover, the lattice QCD determination of the HVP contribution by the BMW collaboration~\cite{Borsanyi:2021} shows significant tension with the time-like prediction. Its central value is in closer agreement with the experimental measurement rather than the dispersive theoretical average of~\cite{Aoyama:2020ynm}. Also, the recent high-precision lattice QCD calculation of the HVP contribution performed in~\cite{Boccaletti:2024guq}, which includes also a small long-distance contribution using input from experiments, leads to a discrepancy from the measurement of only 0.9 standard deviations. There are further efforts in the lattice QCD community aiming at consolidating the first-principle prediction of the hadronic contributions to $a_\mu$~\cite{Lehner:2020crt,Wang:2022lkq,Aubin:2022hgm,Colangelo:2022vok,Ce:2022kxy,ExtendedTwistedMass:2022jpw,FermilabLatticeHPQCD:2023jof,RBC:2023pvn,Benton:2023dci,Davier:2023cyp,Kuberski:2024bcj,Lahert:2024vvu,Spiegel:2024dec,RBC:2024fic,Davies:2024pvv,Itatani:2024fpr,Djukanovic:2024cmq,ExtendedTwistedMassCollaborationETMC:2024xdf,FermilabLattice:2024yho}. The complete prediction for $a_\mu$ of~\cite{Djukanovic:2024cmq} is in agreement with the results of~\cite{Borsanyi:2021,Boccaletti:2024guq}. 

On top of the discrepant scenario concerning the HVP contribution, one should also consider the measurement of the $e^+ e^- \to \pi^+ \pi^-$ cross section performed by the CMD-3 collaboration~\cite{CMD-3:2023alj,Ignatov:2023b} via energy scan at the VEPP-2000 collider. This experiment yields a pion form factor that significantly disagrees with the most precise measurements obtained with the radiative return method by BaBar~\cite{BaBar:2012bdw}, BESIII~\cite{BESIII:2015equ}, and KLOE~\cite{KLOE:2004lnj,KLOE:2008fmq,KLOE:2010qei,KLOE:2012anl,KLOE-2:2017fda}, as well as with previous energy scan determinations by CMD-2~\cite{CMD-2:2001ski,CMD-2:2005mvb,Aulchenko:2006dxz,CMD-2:2006gxt} and SND~\cite{Achasov:2006vp,SND:2020nwa}. By replacing the $\pi^+ \pi^-$ contribution measured by CMD-3 into the complete calculation of HVP, the resulting time-like prediction for $a_\mu$ is in good agreement -- at $1\sigma$ level -- with the experimental world average~\cite{Ignatov:2023b}. There are ongoing efforts~\cite{Colangelo:2022} and novel proposals to clarify the current theoretical situation of the HVP contribution, including the complementary approach via a space-like measurement of HVP using muon-electron scattering at the MUonE experiment~\cite{Calame:2015fva,Abbiendi:2016xup,MUonE:LoI,Banerjee:2020tdt,Gurgone:2024xdt,MUonE:Proposal}.
 
In spite of the continuous progress of first-principle lattice calculations and other avenues to estimate the hadronic contribution to $a_\mu$, the data-driven evaluation of HVP based on $e^+ e^-$ data will continue to play a pivotal role in the data-theory comparison of the muon magnetic anomaly. It is therefore crucial to grasp the origin of the discrepancy between the time-like dispersive determination of HVP and the lattice QCD calculations. This in turn implies a better understanding of the sources of inconsistency between the $e^+ e^- \to \textit{hadrons}$ cross section data measured by different experiments~\cite{Davier:2023fpl} .
 
In this context, the two pion production in $e^+ e^-$ annihilation requires particular attention, in light of the recent CMD-3 result for the pion form factor, future radiative return measurements or reanalyses by BaBar, BESIII, Belle II and KLOE and possible new energy scan results by CMD-3 and SND collaborations at VEPP-2000.
 
In the recent CMD-3 measurement, two Monte Carlo~(MC) generators have been used for the evaluation of the radiative corrections. \textsc{MCGPJ}~\cite{Arbuzov:2005pt} has been used for $e^+ e^- \to \pi^+ \pi^- / \mu^+ \mu^-$, while \textsc{BabaYaga@NLO}~\cite{Balossini:2006wc} has been used for $e^+ e^- \to e^+ e^- / \mu^+ \mu^-$. The QED processes enter the CMD-3 analysis as normalisation processes for the extraction of the pion form factor $|F_\pi (q^2) |^2$, since $|F_\pi (q^2) |^2 \propto N_{\pi^+ \pi^-} / N_{e^+ e^-}$. 
They are also crucial for a checking measurement with the QED-predicted ratio $R_{\rm QED} = N_{\mu^+ \mu^-} / N_{e^+ e^-}$. As a whole, the contribution of radiative corrections induces a systematic error of $0.3\%$ to the measurement of the pion form factor out of the $0.7\%$ which represents the total systematic error~\cite{Ignatov:2023b}. The $0.3\%$ systematic uncertainty due to radiative corrections stems from some difference between the predictions coming from \textsc{MCGPJ} and \textsc{BabaYaga@NLO}. These apparent discrepancies originate from the different implementation of algorithms to model multiple photon emission on top of exact $O(\alpha)$ corrections. More specifically, in the public version of \textsc{MCGPJ}, QED collinear Structure Functions~\cite{Kuraev:1985hb,AltarelliMartinelli,Nicrosini:1986sm,Skrzypek:1990qs,Skrzypek:1992vk,Cacciari:1992pz,Arbuzov:2010} are used. Thus, photon jets are emitted exactly along parent particles. On the other hand, \textsc{BabaYaga@NLO} makes use of a QED Parton Shower (PS) algorithm~\cite{Balossini:2006wc,CarloniCalame:2000pz,CarloniCalame:2001ny} according to which photons are generated exclusively, \textit{i.e.} by including their transverse degrees of freedom. This is the reason why the original version of \textsc{MCGPJ} was modified in the context of the CMD-3 measurement to account for the angular distribution of photons and improve the agreement with data, which are better captured by \textsc{BabaYaga@NLO}~\cite{Ignatov:2023b}. 

Furthermore, and more importantly, recent studies~\cite{Ignatov:2022iou,Colangelo:2022lzg} clearly showed the importance of a careful treatment of the internal structure of the pion in one-loop contributions. They stress the importance of going beyond the traditionally used approach given by the naive multiplication of the scalar QED (sQED) amplitudes for point-like pions with the pion form factor~\cite{Hoefer:2001mx,Bystritskiy:2005ib,Arbuzov:2020foj}. In particular, the standard prescription works well if the emitted photon is soft, as it does not resolve the structure of the pion. However, for other (hard) kinematic configurations and virtual box diagrams significant corrections beyond sQED come into play, as shown in~\cite{Ignatov:2022iou} by using the generalised vector meson dominance~(GVMD) model and in~\cite{Colangelo:2022lzg} according to a dispersive-inspired approach. The results of these calculations show a remarkable agreement with the results of the charge asymmetry of two pion production, which is dominated by corrections coming from initial-final state interference and box diagrams. In the measurement of the pion form factor by the CMD-3 Collaboration, the model of~\cite{Ignatov:2022iou} was used in the analysis to compare the predicted and measured $\pi^+ \pi^-$ charge asymmetry. This was done in order to overcome the limitations of the standard sQED approach inherent to the publicly available version of \textsc{MCGPJ}.

All the above considerations clearly underline the need for a careful theoretical description of photon radiation in the $e^+ e^- \to \pi^+ \pi^-$ process and for the development of next-generation MC generators in view of sub-percent precision measurements of the pion form factor in energy scan experiments\footnote{In~\cite{CMD-3:2023alj}, it is emphasised that {\it ``for the further reduction of theoretical systematics it is advisable to develop another precise $e^+ e^- \to \pi^+ \pi^-$ generator based on the theoretical framework beyond the scalar QED approach, as the point-like pion approximation is already not sufficient"}. Here, ``another'' indicates a MC code independent of \textsc{MCGPJ}.}.

With this motivation in mind, we present in this paper a new calculation of two pion production in $e^+ e^-$ annihilation at full next-to-leading-order (NLO) accuracy matched to PS and we discuss its phenomenological implications. This process is relevant for energy scan measurements at $e^+e^-$ colliders. The radiative return process, i.e.  $e^+ e^- \to \pi^+ \pi^- \gamma~(\gamma)$, will be investigated separately. Previous calculations of the radiative corrections to pion pair production in $e^+ e^-$ collisions can be found in~\cite{Brown:1974aq,Arbuzov:1997je,Hoefer:2001mx,Bystritskiy:2005ib,Arbuzov:2020foj}. None of them takes into account the findings of~\cite{Ignatov:2022iou,Colangelo:2022lzg} and contains other theoretical ingredients, such exclusive PS resummation, which are necessary for current and anticipated precision studies of the process. On the other hand, our results
 can be used for fully fledged simulations of $e^+ e^- \to \pi^+ \pi^- (\gamma)$ and are implemented in an improved version of the MC event generator \textsc{BabaYaga@NLO}. The latter is a well-known code for physics at flavour factories, where it is largely used for precision simulation of 
 QED processes for normalisation and measurement 
 cross-checks~\cite{CMD-3:2023alj,Ignatov:2023b}, as 
 well as for luminosity monitoring~\cite{SND:2024kbi,BESIII:2024lbn,Belle-II:2024vuc,KLOE-2:2016mgi}, with an estimated 
 theoretical accuracy of 0.1\%~\cite{Balossini:2006wc,Balossini:2008xr,WorkingGrouponRadiativeCorrections:2010bjp}.

The paper is organised as follows. In Sec.~\ref{sec:QED_sQED} we present the details of the calculation of NLO corrections, by using the standard sQED approach to pion contributions and including the pion form factor multiplicatively. In the same section, we describe the construction of the PS algorithm, with particular attention to the modelling of multiple photon emission from scalars. This represents an original extension of the \textsc{BabaYaga@NLO} formulation. The matching of NLO corrections to the PS is also addressed. In Sec.~\ref{sec:intffpi}  we focus on the treatment of the composite structure of the pion in the computation of the final state and initial-final state corrections. The calculation is performed both in the GVMD model and in a dispersive-inspired approach. In Sec.~\ref{sec:res} we show the numerical results of our study, as obtained with the upgraded \textsc{BabaYaga@NLO}, for both integrated and differential observables and with particular attention to the predictions for the charge asymmetry. The main conclusions and prospects of our work are drawn in Sec.~\ref{sec:conc}.

\section{Calculation in QED and factorised sQED}
\label{sec:QED_sQED}
In this section, the differential cross section for $e^+e^-\to\pi^+\pi^- (\gamma)$ is computed at NLOPS in QED and sQED. The pion vector form factor $F_\pi(q^2)$ is introduced to account for the internal structure of pions, which is the cause of the significant deviation between the measured cross section and the point-like approximation. It is defined as
\begin{equation}\label{eq:defff}
    \langle \pi^\pm(p') | j_\mathrm{em}^\mu(0) | \pi^\pm(p) \rangle =\pm (p'+p)^\mu F_\pi\left((p'-p)^2\right) ,
\end{equation}
where $j_\text{em}^\mu=(2\bar u\gamma^\mu u-\bar d\gamma^\mu d-\bar s \gamma^\mu s)/3$ is the electromagnetic current of light quarks~\cite{Colangelo:2022lzg}. Since on-shell photons do not resolve the internal structure of pions, the form factor must satisfy the condition ${F_\pi(0)=1}$. From a formal point of view, this is a consequence of the Ward identity. The issue of the introduction of the form factor in theoretical predictions for pion pair production in the presence of QED radiative corrections has been discussed in the literature~\cite{WorkingGrouponRadiativeCorrections:2010bjp}. The simplest approach, named factorised since it amounts to multiplying each matrix element by a global form factor, aims at satisfying the cancellation of infrared divergences between NLO virtual and real corrections, as will be detailed in the following. More recently, other approaches have been studied, aimed at keeping under control the infrared divergences and, at the same time, improving the description of pion production at the differential level. These approaches will be discussed in Sect.~\ref{sec:intffpi}, while in the present section, where the focus is on the implementation of NLO (s)QED corrections and their matching to higher order photonic corrections, we will adopt the factorised prescription.

\subsection{Born approximation}

\begin{figure}[t]
\centering
    \begin{tikzpicture}
  \begin{feynman}[small]
    \vertex (a);
    \vertex[right=1.25cm of a,style=blob] (b) {};
    \vertex[above left=0.75cm and 0.75cm of a] (c) {$e^+$};
    \vertex[below left=0.75cm and 0.75cm of a] (d) {$e^-$};
    \vertex[above right=1cm and 1cm of b] (e) {$\pi^+$};
    \vertex[below right=1cm and 1cm of b] (f) {$\pi^-$};
    \diagram* {
      (a) -- [photon,edge label=$\gamma$] (b),
      (d) -- [fermion,momentum'=$p_1$] (a),
      (a) -- [fermion,reversed momentum'=$p_2$] (c),
      (e) -- [scalar,reversed momentum'=$p_4$] (b) --[scalar, momentum'=$p_3$] (f),
    };
  \end{feynman}
\end{tikzpicture}
\caption{Tree level diagram for the process \eqref{eq:proc}, where the blob represents the form factor. This diagram is independent of the prescription of the form factor and is equivalent to the point-like sQED diagram multiplied by $F_\pi(s)$.}
\label{fig:eepp_ff}
\end{figure}
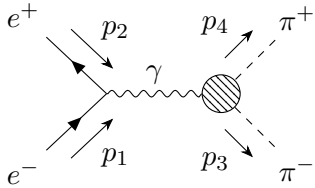

At LO, shown in Fig.~\ref{fig:eepp_ff}, the kinematics of the process is 
\begin{equation}\label{eq:proc}
    e^-(p_1)\,e^+(p_2) \to \gamma^* \to \pi^-(p_3)\,\pi^+(p_4) \,.
\end{equation}
The Mandelstam variables are computed as 
\begin{equation}
\begin{aligned}
    s &= \left(p_1+p_2\right)^2=4E^2 \,,\\
     t &= \left(p_1-p_3\right)^2=m_e^2+m_\pi^2-2E^2(1-\beta_e\beta_\pi \cos\vartheta) \,,\\
      u &= \left(p_1-p_4\right)^2=m_e^2+m_\pi^2-2E^2(1+\beta_e\beta_\pi \cos\vartheta) \,,
\end{aligned}
\end{equation}
where 
\begin{equation}
    \beta_e=\sqrt{1-\frac{4m_e^2}{s}}\, , \qquad \beta_\pi=\sqrt{1-\frac{4m_\pi^2}{s}}
\end{equation}
are the velocities of the electron and the pion, respectively, $E$ is the beam energy and $\vartheta$ is the $\pi^-$ angle w.r.t. the incoming electron with momentum $p_1$. 
By considering the massive kinematics, the differential cross section in Born approximation is given by
\begin{equation}\label{eq:born}
      \dv{\sigma_{\rm{LO}}}{\cos\vartheta}=\frac{\alpha^2\pi}{2 \mathcal{F}}\beta_\pi^3(1-\beta_e^2\cos^2\vartheta) |F_\pi(s)|^2 \,,
\end{equation}
where the pion form factor is evaluated at $q^2=s$ and $\mathcal{F}=2s\beta_e$ is the incoming flux factor\footnote{Notice that in the \textsc{BabaYaga} implementation the electron mass is neglected in the flux factor leading to a negligible difference of $\order{m_e^2/s}$.}. By inspection of Eq.~\eqref{eq:born}, the two pions are more likely emitted perpendicularly to the beam axis, in comparison with muon pair production in QED where muons travel preferably along the beam axis.

From Eq.~\eqref{eq:born} one can see that the Born cross section w.r.t. the pion angle $\vartheta$ is an even function around $\vartheta= \pi/2$ rad. By defining an angle-sensitive observable, one can study the effects of higher-order corrections and of the form factor parameterisation for this process. An observable of great experimental and theoretical interest for this process is the \textit{forward-backward asymmetry}, also called \textit{charge asymmetry}. It is defined as
\begin{equation}\label{eq:FBasym}
    A_{\rm FB}\left(\sqrt{s}\right)=\frac{\sigma_{\rm F}-\sigma_{\rm B}}{\sigma_{\rm F}+\sigma_{\rm B}} \,,
\end{equation}
where $\sigma_{\rm F}$ and $\sigma_{\rm B}$ are the integrated cross section in the forward ($\vartheta<\pi/2$) and backward ($\vartheta>\pi/2 $) angular regions. This quantity is equal to zero at tree level as the LO differential cross section ${\rm d}\sigma / {\rm d}\cos\vartheta$ is even under the transformation $\cos\vartheta\to-\cos\vartheta$. However, as discussed in the literature ~\cite{Ignatov:2022iou,Colangelo:2022lzg,Arbuzov:2020foj}, radiative corrections give rise to odd terms in $\cos\vartheta$ and induce NLO effects which can reach the one percent level.

\subsection{NLO photonic corrections}\label{sec:nlo}

At $\order{\alpha^3}$ the cross section receives contributions both from diagrams where virtual photons are exchanged and from real radiation contributions~\footnote{In the present discussion we do not consider the leptonic and hadronic VP contributions, as they are usually subtracted from the measured hadronic cross section~\cite{WorkingGrouponRadiativeCorrections:2010bjp,Aoyama:2020ynm}. However, these corrections can be  optionally switched on in \textsc{BabaYaga@NLO}; in this case they are included with the same approach adopted for $e^+ e^- \to \mu^+ \mu^-$~\cite{Balossini:2006wc}.}. Hence, one can write the NLO cross section as 
\begin{equation}
    \sigma_\nlo =\sigma_{2\to 2} + \sigma_{2\to 3}= \sigma_{e^+ e^- \to \pi^+ \pi^-} 
    +\sigma_{e^+ e^- \to \pi^+ \pi^-\gamma}\,,
\label{eq:NLOgeneral}
\end{equation}
In Eq.~(\ref{eq:NLOgeneral}), the two-body cross section reads
\begin{equation}
    \sigma_{2\to 2}= \frac{1}{\mathcal{F}}\left\{\int \dd \Phi_2 |\Amp_\lo|^2+\int \dd \Phi_2 \, 2 \Re \left( \Amp_\lo^\dagger \Amp_V(\lambda)\right)\right\} \equiv \sigma_\lo \left( 1 + \delta_V(\lambda) \right) \,,
\end{equation}
where $\dd \Phi_2$ is the $2 \to 2$ phase space. $\Amp_\lo$ and $\Amp_V(\lambda)$ represent the LO amplitude and NLO virtual amplitudes in which $\lambda$ is a fictitious photon mass that regularises the infrared divergences. The virtual NLO correction $\delta_V(\lambda)$, which depends on the infrared cutoff $\lambda$, is defined as
\begin{equation}\label{Eq:deltaVdef}
\delta_V(\lambda) = \frac{\displaystyle\int \dd \Phi_2 \, 2 \Re{\Amp_\lo^\dagger \Amp_V(\lambda)}}{\displaystyle\int \dd \Phi_2 \, |\Amp_\lo|^2}\, .
\end{equation}

The integration of the $2\to 3$ cross section is performed by means of the slicing scheme, introducing an arbitrarily small cutoff on the photon energy $\omega$, splitting the phase space into two regions as
\begin{align}\label{eq:2to3}
    \sigma_{2\to 3} &= \frac{1}{\mathcal{F}}\left\{  \int_{\lambda \leq \omega \leq \Delta E}\dd \Phi_3 \, |\Amp_{2\to 3}|^2+\int_{\omega > \Delta E} \dd \Phi_3  \, |\Amp_{2\to 3}|^2\right\} \\
    &\equiv\sigma_{S}(\lambda,\Delta E)+\sigma_{H}(\Delta E) \,,
\end{align}
where $\Delta E = \varepsilon \sqrt{s}/2$ is the soft-hard slicing separator, satisfying the condition $\lambda \ll \Delta E \ll \sqrt{s}$, from which the sum of the two contributions is independent. The two terms $\sigma_S(\lambda, \Delta E)$ and $\sigma_H(\Delta E)$ represent the soft and hard contributions to the cross section, respectively. The first one can be integrated analytically using the soft photon approximation, obtaining
\begin{equation}
\sigma_S(\lambda,\Delta E) = \delta_S(\lambda,\Delta E)\,\sigma_\lo \,,
\end{equation}
where $\delta_S(\lambda,\Delta E)$ is proportional to the integral over the photon phase space of the eikonal current~\cite{tHooft:1978jhc,Denner:1991kt}
\begin{equation}\label{Eq:deltaSdef}
    \delta_S(\lambda,\Delta E) = -\frac{\alpha}{2\pi^2}\left(\frac{1}{2}\sum_{i=1}^4 \sum_{j=1}^4\mathcal{I}_{ij}(k)\right)\, .
\end{equation}
The eikonal integral is defined as 
\begin{equation}\label{eq:eik_int}
    \mathcal{I}_{ij}(k)=\eta_i\eta_j\int_{k_0 \leq \Delta E}\frac{\dd^3k}{ k_0}\frac{p_i \cdot p_j}{(p_i\cdot k)(p_j\cdot k)}\, ,
\end{equation}
where $p_{i,j}$ are the massive momenta of the external charged particles, $k$ is the photon momentum and $k_0$ the photon energy. The symbol $\eta_i$ represents the physical charge (in positron charge units) flowing into the process from the $i$-th particle. 

The virtual and soft cross sections are separately IR divergent. However, the dependence on $\lambda $ exactly vanishes when the sum of the two contributions is considered. Therefore, one can write the soft-virtual contribution to the cross section as 
\begin{equation}
\sigma_{ SV}=\delta_{SV}\, \sigma_\lo \equiv (\delta_S + \delta_V)\, \sigma_\lo \, , 
\end{equation} 
without any dependence on the fictitious photon mass $\lambda$. Since the annihilation channel shown in Eq.~\eqref{eq:proc} is a neutral-current process, the corrections due to initial-state radiation~(ISR), final-state radiation~(FSR) and initial-final-state interference~(IFI) form gauge invariant subsets of the complete NLO calculation, allowing for a separate calculation of each part. The diagrams contributing to each subset are depicted in Tab.~\ref{tab:diagrams}. In particular, the cancellation of the infrared divergences between real and virtual diagrams separately occurs within each set. In this view, the differential cross section in the scattering angle can be written as
\begin{equation}\label{EQ:crossecsplitting}
    \dv{\sigma_\text{NLO}}{\cos\theta}= \dv{\sigma_\text{LO}}{\cos\theta} \left(1 + \delta_{SV}^\text{ISR} + \delta_{SV}^\text{FSR} +\delta_{SV}^\text{IFI}\right) + \dv{\sigma_H}{\cos\theta}\, .
\end{equation}
In the following, we discuss in detail the QED and sQED contribution of each gauge-invariant subset in the factorised form factor approach. To this end, we define the following shorthand notation for the Passarino--Veltman functions~\cite{Passarino:1978jh,tHooft:1978jhc}:
\begin{subequations}
\begin{align}\label{Eq:shorthand}
\mathcal{C}_0^e(x,y)&\equiv\text{C}_0(m_e^2,m_e^2,s,x,m_e^2,y) \,, \\
\mathcal{C}_0^\pi(x,y)&\equiv\text{C}_0(m_\pi^2,m_\pi^2,s,x,m_\pi^2,y) \,, \\
\mathcal{C}_0^{e,e}(x)&\equiv\text{C}_0(m_e^2,m_e^2,s,m_e^2,x,m_e^2) \,, \\
\mathcal{C}_0^{\pi,\pi}(x)&\equiv\text{C}_0(m_\pi^2,m_\pi^2,s,m_\pi^2,x,m_\pi^2) \,, \\
\mathcal{C}_0^{e,\pi}(z,x)&\equiv\text{C}_0(m_e^2,m_\pi^2,z,m_e^2,x,m_\pi^2) \,, \\
\mathcal{D}_0^{e,\pi}(z,x,y)&\equiv\text{D}_0(m_e^2,m_e^2,m_\pi^2,m_\pi^2,s,z,x,m_e^2,y,m_\pi^2) \,.
\end{align}
\end{subequations}
where $z=t,u$. 
This notation will be particularly useful in Sec.~\ref{sec:intffpi} for the description of the GVMD and FsQED approaches.

In this work, the matrix elements have been computed with both \textsc{Form}~\cite{Vermaseren:2000nd,Kuipers:2012rf,Ruijl:2017dtg} and \textsc{Mathematica}. Together with the latter, the \textsc{FeynArts}~\cite{Hahn:2000kx} package has been used in association with \textsc{FeynRules}~\cite{Alloul_2014}, for which a QED+sQED model has been developed. The reduction to scalar integrals has been performed with \textsc{FeynCalc}~\cite{Mertig:1990an,Shtabovenko:2016sxi,Shtabovenko:2020gxv,Shtabovenko:2023idz} and ${\textsc{Package-X}}$~\cite{Patel:2015tea,Patel:2016fam}. The one-loop scalar functions $\text{A}_0,\text{B}_0,\text{C}_0,\text{D}_0$ follow the conventions of~\cite{Denner:1991kt} and, in the code implementation, are alternatively evaluated with \textsc{Collier}~\cite{Denner:2016kdg,Denner:2002ii,Denner:2005nn,Denner:2010tr} or \textsc{LoopTools}~\cite{Hahn:1998yk}.

\begin{table}[t]
\renewcommand{\arraystretch}{3}%
    \centering
    \begin{NiceTabular}{cccccl}
\multicolumn{2}{c}{\textbf{Subset}}     & \multicolumn{3}{c}{\textbf{Diagrams}} &$F_\pi(q^2)$\\
\addlinespace[-5pt]\toprule\addlinespace[2pt]
\multirow{2}{*}{\adjustbox{valign=c}{\centering{\rotatebox[origin=l]{90}{\textbf{ISR}\hspace{25pt}}}}}
& \adjustbox{valign=c}{{\rotatebox[origin=c]{90}{{real}}}}  & 
 \adjustbox{valign=c}{
                 \begin{tikzpicture}
  \begin{feynman}[small]
    \vertex (a) ;
    \vertex[right=1cm of a] (b);
    \vertex[above left=0.75cm and 0.75cm of a] (c);
    \vertex[below left=0.75cm and 0.75cm of a] (d);
    \vertex[above right=0.75cm and 0.75cm of b] (e);
    \vertex[below right=0.75cm and 0.75cm of b] (f);
    \vertex[above left =0.5 and 0.5cm of a] (g);
    \vertex[below left =0.5 and 0.5cm of a] (h);
    \vertex[above left=0.35cm and 0.35cm of a](i);
    \vertex[above right=0.4cm and 0.6cm of i](j);
    \diagram* {
      (a) -- [photon] (b),
      (d) -- [fermion] (a),
      (a) -- [fermion] (c),
      (e) -- [scalar] (b) --[scalar] (f),
      (i) -- [photon] (j),
    };
  \end{feynman}
\end{tikzpicture}
       } &
       \adjustbox{valign=c}{
           \begin{tikzpicture}
  \begin{feynman}[small]
    \vertex (a);
    \vertex[right=1cm of a] (b);
    \vertex[above left=0.75cm and 0.75cm of a] (c);
    \vertex[below left=0.75cm and 0.75cm of a] (d);
    \vertex[above right=0.75cm and 0.75cm of b] (e);
    \vertex[below right=0.75cm and 0.75cm of b] (f);
    \vertex[above left =0.5 and 0.5cm of a] (g);
    \vertex[below left =0.5 and 0.5cm of a] (h);
      \vertex[below left=0.35cm and 0.35cm of a](i);
    \vertex[below right=0.4cm and 0.6cm of i](j);
    \diagram* {
      (a) -- [photon] (b),
      (d) -- [fermion] (a),
      (a) -- [fermion] (c),
      (e) -- [scalar] (b) --[scalar] (f),
      (i) --[photon] (j),
    };
  \end{feynman}
\end{tikzpicture}}
        & &\adjustbox{valign=c}{$F_\pi(m_{\pi\pi}^2)$}\\[10pt]
         &\adjustbox{valign=c}{{\rotatebox[origin=c]{90}{{virtual}}}} &  \adjustbox{valign=c}{
              \begin{tikzpicture}
    \hspace{-2pt}
  \begin{feynman}[small]
    \vertex (a);
    \vertex[right=1cm of a] (b);
    \vertex[above left=0.75cm and 0.75cm of a] (c);
    \vertex[below left=0.75cm and 0.75cm of a] (d);
    \vertex[above right=0.75cm and 0.75cm of b] (e);
    \vertex[below right=0.75cm and 0.75cm of b] (f);
    \vertex[above left =0.5 and 0.5cm of a] (g);
    \vertex[below left =0.5 and 0.5cm of a] (h);
    \diagram* {
      (a) -- [photon] (b),
      (d) -- [fermion] (a),
      (a) -- [fermion] (c),
      (e) -- [scalar] (b) --[scalar] (f),
      (g) --[photon,half right, looseness=1,] (h),
    };
  \end{feynman}
\end{tikzpicture}
        } & \multicolumn{2}{c}{}& \adjustbox{valign=c}{$F_\pi(s)$} \\
        \addlinespace[5pt]
        \midrule
        
 {\adjustbox{valign=c}{  \multirow{2}{*}{\rotatebox[origin=c]{90}{\textbf{FSR}\hspace{20pt}}}}} &\adjustbox{valign=c}{{\rotatebox[origin=c]{90}{{real}}}}& 
    \adjustbox{valign=c}{
          \begin{tikzpicture}
          \hspace{2.5pt}
  \begin{feynman}[small]
    \vertex (a) ;
    \vertex[right=1cm of a] (b);
    \vertex[above left=0.75cm and 0.75cm of a] (c);
    \vertex[below left=0.75cm and 0.75cm of a] (d);
    \vertex[above right=0.75cm and 0.75cm of b] (e);
    \vertex[below right=0.75cm and 0.75cm of b] (f);
    \vertex[above left =0.5 and 0.5cm of a] (g);
    \vertex[below left =0.5 and 0.5cm of a] (h);
    \vertex[below right=0.35cm and 0.35cm of b](i);
    \vertex[above right=0.4cm and 0.6cm of i](j);
    \diagram* {
      (a) -- [photon] (b),
      (d) -- [fermion] (a),
      (a) -- [fermion] (c),
      (e) -- [scalar] (b) --[scalar] (f),
      (i) -- [photon] (j),
    };
  \end{feynman}
\end{tikzpicture}  
    } & \adjustbox{valign=c}{
    \begin{tikzpicture}\hspace{2.6pt}
  \begin{feynman}[small]
    \vertex (a);
    \vertex[right=1cm of a] (b);
    \vertex[above left=0.75cm and 0.75cm of a] (c);
    \vertex[below left=0.75cm and 0.75cm of a] (d);
    \vertex[above right=0.75cm and 0.75cm of b] (e);
    \vertex[below right=0.75cm and 0.75cm of b] (f);
    \vertex[above left =0.5 and 0.5cm of a] (g);
    \vertex[below left =0.5 and 0.5cm of a] (h);
      \vertex[above right=0.35cm and 0.35cm of b](i);
    \vertex[below right=0.4cm and 0.6cm of i](j);
    \diagram* {
      (a) -- [photon] (b),
      (d) -- [fermion] (a),
      (a) -- [fermion] (c),
      (e) -- [scalar] (b) --[scalar] (f),
      (i) --[photon] (j),
    };
  \end{feynman}
\end{tikzpicture}
    }
&
  \adjustbox{valign=c}{
  \begin{tikzpicture}
  \begin{feynman}[small]
    \vertex (a);
    \vertex[right=1cm of a] (b);
    \vertex[above left=0.75cm and 0.75cm of a] (c);
    \vertex[below left=0.75cm and 0.75cm of a] (d);
    \vertex[above right=0.75cm and 0.75cm of b] (e);
    \vertex[below right=0.75cm and 0.75cm of b] (f);
    \vertex[above left =0.5 and 0.5cm of a] (g);
    \vertex[below left =0.5 and 0.5cm of a] (h);
\vertex[right=0.75 cm of b](i);
    \diagram* {
      (a) -- [photon] (b),
      (d) -- [fermion] (a),
      (a) -- [fermion] (c),
      (e) -- [scalar] (b) --[scalar] (f),
      (b) --[photon] (i),
    };
  \end{feynman}
\end{tikzpicture}
    }
    & 
  \adjustbox{valign=c}{ $F_\pi(s)$}\\[10pt]
  &\adjustbox{valign=c}{{\rotatebox[origin=c]{90}{{virtual}}}} & 
   \adjustbox{valign=c}{
    \begin{tikzpicture}
  \begin{feynman}[small]
    \vertex (a);
    \vertex[right=1cm of a] (b);
    \vertex[above left=0.75cm and 0.75cm of a] (c);
    \vertex[below left=0.75cm and 0.75cm of a] (d);
    \vertex[above right=0.75cm and 0.75cm of b] (e);
    \vertex[below right=0.75cm and 0.75cm of b] (f);
    \vertex[above right =0.5 and 0.5cm of b] (g);
    \vertex[below right =0.5 and 0.5cm of b] (h);
    \diagram* {
      (a) -- [photon] (b),
      (d) -- [fermion] (a),
      (a) -- [fermion] (c),
      (e) -- [scalar] (b) --[scalar] (f),
      (g) --[photon,half left, looseness=1,] (h),
    };
  \end{feynman}
\end{tikzpicture}}
    & 
    \adjustbox{valign=c}{
   \begin{tikzpicture}
  \begin{feynman}[small]
    \vertex (a);
    \vertex[right=1cm of a] (b);
    \vertex[above left=0.75cm and 0.75cm of a] (c);
    \vertex[below left=0.75cm and 0.75cm of a] (d);
    \vertex[above right=0.75cm and 0.75cm of b] (e);
    \vertex[below right=0.75cm and 0.75cm of b] (f);
    \vertex[above right =0.5 and 0.5cm of b] (g);
    \vertex[below right =0.5 and 0.5cm of b] (h);
    \diagram* {
      (a) -- [photon] (b),
      (d) -- [fermion] (a),
      (a) -- [fermion] (c),
      (e) -- [scalar] (b) --[scalar] (f),
      (g) --[photon,half left, looseness=1,] (b),
    };
  \end{feynman}
\end{tikzpicture}
    }
&
 \adjustbox{valign=c}{
    \begin{tikzpicture}
  \begin{feynman}[small]
    \vertex (a);
    \vertex[right=1cm of a] (b);
    \vertex[above left=0.75cm and 0.75cm of a] (c);
    \vertex[below left=0.75cm and 0.75cm of a] (d);
    \vertex[above right=0.75cm and 0.75cm of b] (e);
    \vertex[below right=0.75cm and 0.75cm of b] (f);
    \vertex[above right =0.5 and 0.5cm of b] (g);
    \vertex[below right =0.5 and 0.5cm of b] (h);
    \diagram* {
      (a) -- [photon] (b),
      (d) -- [fermion] (a),
      (a) -- [fermion] (c),
      (e) -- [scalar] (b) --[scalar] (f),
      (b) --[photon,half left, looseness=1,] (h),
    };
  \end{feynman}
\end{tikzpicture}
    }
    & 
   \adjustbox{valign=c}{  $F_\pi(s)$}\\
   \addlinespace[5pt]
    \midrule
    \addlinespace[7pt]
 \adjustbox{valign=c}{{\rotatebox[origin=c]{90}{\textbf{IFI}}}}&\adjustbox{valign=c}{{\rotatebox[origin=c]{90}{{virtual}}}} &
 \adjustbox{valign=c}{
    \begin{tikzpicture}
  \begin{feynman}[small]
    \vertex (a);
    \vertex[right=1cm of a] (b);
    \vertex[below=1cm of a] (i);
    \vertex[below=1cm of b] (j);
    \vertex[above left=0.75cm and 0.75cm of a] (c);
    \vertex[below left=0.75cm and 0.75cm of i] (d);
    \vertex[above right=0.75cm and 0.75cm of b] (e);
    \vertex[below right=0.75cm and 0.75cm of j] (f);
    \diagram* {
      (a) -- [photon] (b),
      (d) -- [fermion] (i) -- [fermion] (a) -- [fermion] (c),
      (e) -- [scalar] (b) --[scalar] (j) -- [scalar] (f),
      (i) -- [photon] (j),
    };
  \end{feynman}
\end{tikzpicture}      
 }
 &
 \adjustbox{valign=c}{
    \begin{tikzpicture}
  \begin{feynman}[small]
    \vertex (a);
    \vertex[right=1cm of a] (b);
    \vertex[below=1cm of a] (i);
    \vertex[below=1cm of b] (j);
    \vertex[above left=0.75cm and 0.75cm of a] (c);
    \vertex[below left=0.75cm and 0.75cm of i] (d);
    \vertex[above right=0.75cm and 0.75cm of b] (e);
    \vertex[below right=0.75cm and 0.75cm of j] (f);
    \diagram* {
      (a) -- [photon] (j),
      (d) -- [fermion] (i) -- [fermion] (a) -- [fermion] (c),
      (e) -- [scalar] (b) --[scalar] (j) -- [scalar] (f),
      (i) -- [photon] (b),
    };
  \end{feynman}
\end{tikzpicture}     
 }
 &
 \adjustbox{valign=c}{
          \begin{tikzpicture}
  \begin{feynman}[small]
    \vertex (a);
    \vertex[below=1cm of a] (i);
    \vertex[above left=0.75cm and 0.75cm of a] (c);
    \vertex[below left=0.75cm and 0.75cm of i] (d);
    \vertex[below right=0.5cm and 0.75cm of a] (e);
    \vertex[above right=1.25 cm and 1cm of e] (f);
    \vertex[below right=1.25 cm and 1cm of e] (g);
    \diagram* {
      (d) -- [fermion] (i) -- [fermion] (a) -- [fermion] (c),
      (a) -- [photon] (e) -- [photon] (i),
      (f) -- [scalar] (e) --[scalar] (g),
    };
  \end{feynman}
\end{tikzpicture}
 }
 &
 \adjustbox{valign=c}{ $F_\pi(s)$}\\
 \addlinespace[5pt]
 \bottomrule
    \end{NiceTabular}
    \smallskip
    \caption{Virtual and real NLO diagrams in the factorised approach with the form factor evaluated at the appropriate virtuality multiplying each subset of diagrams. Counterterm diagrams are understood.}
    \label{tab:diagrams}
\end{table}

\subsubsection*{Initial-State Radiation (ISR)}

The subset of initial-state QED corrections includes diagrams where the photon is emitted either from the electron or the positron. The IR divergence arising from the soft photon emission is cancelled by the NLO vertex correction. The real and virtual diagrams that constitute the ISR gauge-invariant subset are shown in the first two rows of Tab.~\ref{tab:diagrams}.  The expressions of the counterterms in the on-shell renormalisation scheme are understood. 

In the factorised approximation for the pion form factor, the squared momentum of the photon propagator is given by the invariant mass of the pion pair $m^2_{\pi\pi}$. Hence, the sQED vertex is multiplied by $F_\pi(m^2_{\pi\pi})$. 
Since in the soft regime $m_{\pi\pi}^2\simeq s$, the pion form factor in the virtual amplitude and in the real emission amplitude in the soft-photon limit is evaluated at virtuality $q^2=s$. In the interference of the virtual correction diagrams in the ISR row of Tab.~\ref{tab:diagrams} with the tree level, the infrared structure of the matrix element is modified only by a multiplicative factor $|F_\pi(s)|^2$, while the standard QED vertex correction is kept intact. Since the electron mass is much smaller than the pion mass and the centre-of-mass energies involved, the ISR contribution is expected to be the dominant NLO effect, as it is proportional to $\log(s/m_e^2)$. 
On top of that, for energies larger than the $\rho$ resonance mass $\sqrt{s} > m_\rho$, the ISR is also enhanced outside of the soft regime when $m_{\pi\pi}^2\simeq m_\rho^2$ by the larger value of $F_\pi(m_\rho^2)$.

After the on-shell renormalisation of the electron wave function, the explicit expression for the virtual ISR correction $\delta_{V}^\text{ISR}$ reads
\begin{equation}
    \begin{aligned}
\delta_{V}^\text{ISR}(\lambda)&=\frac{\alpha}{2\pi}
\Re\Biggl\{\kappa\left[\,\text{B}_0(s,m_e^2,m_e^2) - \,\text{B}_0\left(m_e^2,\lambda^2,m_e^2\right)\right]\\
&+ 2(2m_e^2-s)\,\mathcal{C}_0^{e,e}(\lambda^2) 
+ 4m_e^2 \, \frac{\partial\text{B}_0}{\partial p^2}(p^2,\lambda^2,m_e^2) \biggl|_{p^2=m_e^2}
\Biggr\} \,,
    \end{aligned}
\end{equation}
where
\begin{align}
\kappa = \frac{3s}{4m_e^2-s}-\frac{4m_e^2\,(12m_\pi^2s -3s^2 +2(t-u)^2 )}{(4 m_e^2-s) (4 m_\pi^2 s-s^2+(t-u)^2)} \,.
\end{align}

\subsubsection*{Final-State Radiation (FSR)} 

The final-state photonic corrections are pure sQED contributions. As shown in the second two rows of Tab.~\ref{tab:diagrams}, the sQED corrections receive contributions also from the four-point vertex interactions, unlike the QED ISR diagrams. 

The first two real emission diagrams have the same IR behaviour as the QED ones, whereas the bremsstrahlung diagram with the quartic $\gamma\gamma\pi\pi$ vertex is not divergent in the soft limit. Similarly, the emission of a virtual photon via the quartic vertex and its re-absorption from one of the two pion legs is IR-finite. For this reason, the IR divergent structure of the final-state radiation is similar to the one of ISR, with the eikonal current being identical to the spinor case. This is a general fact: a soft photon is not sensitive to the spin degree of freedom of the emitter, since the eikonal current is universal. All of the real emission diagrams in this subset are multiplied by the pion form factor evaluated at $q^2=s$, since the radiation from pion legs does not change the momentum that flows into the scalar vertex. The previous argument for the soft correction applies also for $\delta_S^\text{FSR}$.

In the virtual diagrams the evaluation of the form factor may give rise to ambiguities: all diagrams of the fourth row of Fig.~\ref{tab:diagrams} present two or more sQED vertices, to none of which a real photon is attached. Therefore it is non-trivial to decide at which virtuality $F_\pi(q^2)$ has to be evaluated. However, in the factorised approximation, the only possible choice to preserve IR cancellations is to assign $F_\pi(s)$ to virtual diagrams in analogy with the soft ones. 

The inclusion of the FSR corrections needs to be treated carefully when measuring the pion form factor, as it represents a $\mathcal{O}(\alpha)$ corrections to $F_\pi(q^2)$ itself. The cross section measured by energy scan experiments usually includes the FSR contribution, which is therefore needed to describe the experimental data. On the other hand, for applications based on dispersion relations and involving the total cross section of $e^+e^- \to \textit{hadrons}$, the latter is usually taken to be inclusive w.r.t.  FSR of additional photons~\cite{Aoyama:2020ynm,CMD-2:2001ski}. For this reason, the implementation of~$e^+e^- \to \pi^+\pi^-$ in \textsc{BabaYaga@NLO} allows to turn on and off the FSR contribution, depending on the experimental and theoretical needs.

After the on-shell renormalisation of the pion wave function, the explicit expression for the virtual FSR correction $\delta_{V}^\text{FSR}$ reads
\begin{equation}\label{eq:fsr_sqed}
\begin{aligned}
\delta_{V}^\text{FSR}(\lambda) &=
\frac{\alpha}{\pi} \frac{1}{4 m_\pi^2 - s} \Re\Bigg\{ \left(2 m_\pi^2 - s \right) \Big[\left(4 m_\pi^2 - s\right) \mathcal{C}_0^{\pi,\pi}(\lambda^2)
    -2\hspace{1pt} \text{B}_0(s,m_\pi^2, m_\pi^2)  \\
    &+ 2\hspace{1pt} \text{B}_0(m_\pi^2,\lambda^2,m_\pi^2)  \Big] 
    + 2 m_\pi^2 (4 m_\pi^2-s)\, \frac{\partial\text{B}_0}{\partial p^2}\left(p^2, \lambda^2, m_\pi^2\right)\biggl|_{p^2=m_\pi^2}
    \Bigg\}\, .
\end{aligned}
\end{equation}

\subsubsection*{Initial-Final Interference (IFI)} 

The last class of diagrams which are included in the fixed NLO calculation comprises box diagrams and the initial-final-state interference of real radiation. The real squared matrix element $\abs{\mathcal{M}_{2\to3}}^2$ in Eq.~\eqref{eq:2to3} contains the interference between photon radiation from both the electron and pion legs. These IR-divergent contributions are cancelled by the box diagrams represented in the last row of Tab.~\ref{tab:diagrams}, in which two photons connect the initial and final external legs. Only the first two diagrams are IR divergent, while the third one is finite and represents a very small contribution, being proportional to $m_e^2$.

Also for the first two diagrams in the IFI row of Tab.~\ref{tab:diagrams} it is not obvious how one can insert the form factor into the calculation in the factorised assumption. However, as discussed in~\cite{Ignatov:2022iou}, the IR-divergent integration regions come with one of the two photons with momentum $k\to 0 $, hence one of the two form factors always goes to $F_\pi(0)\to 1$, while the other one is evaluated at $q^2=s$. The hard bremsstrahlung cross section $\sigma_{H}$, on the other hand, is proportional to three different terms containing $|F_\pi(m^2_{\pi\pi})|^2$, $|F_\pi(s)|^2$ and $\Re{F_\pi(s)^*F_\pi(m^2_{\pi\pi})}$ for the hard emission from electron legs, pion legs and initial-final interference, respectively. 

The explicit expression for the virtual IFI correction $\delta_{V}^\text{IFI}$ reads
\begin{equation}\label{eq:deltaIFIs}
\begin{aligned}
\delta_{V}^\text{IFI}(\lambda) &=\frac{\alpha}{2\pi}\hspace{1pt}\frac{4s}{4 m_\pi^2 s-s^2+(t-u)^2} \, \Re\biggl\{\frac{4 m_e^2 (t-u)}{4 m_e^2-s}\hspace{1pt}\Bigl[\text{B}_0(m_e^2,\lambda^2,m_e^2)\\[3pt] 
&- \text{B}_0(s,\lambda^2,\lambda^2)\Bigr]
+ \frac{8 m_e^4-8 m_e^2 s +s^2}{4m_e^2-s} \left(t-u\right) \mathcal{C}_0^e(\lambda^2,\lambda^2)\\[2pt]
&+ \left(2 m_\pi^2-s\right) \left(t-u\right) \mathcal{C}_0^\pi(\lambda^2,\lambda^2)
- 2\left(m_e^2-t\right) \left(m_e^2+m_\pi^2-t\right) \mathcal{C}_0^{e,\pi}(t,\lambda^2)\\[5pt]
&+ 2\left(m_e^2-u\right) \left(m_e^2+m_\pi^2-u\right) \mathcal{C}_0^{e,\pi}(u,\lambda^2)\\[-1pt]
&+\kappa_t^0 \left(m_e^2+m_\pi^2-t\right) \mathcal{D}_0^{e,\pi}(t,\lambda^2,\lambda^2) -\kappa_u^0 \left(m_e^2+m_\pi^2-u\right) \mathcal{D}_0^{e,\pi}(u,\lambda^2,\lambda^2)\biggr\}\, ,
\end{aligned}
\end{equation}
where
\begin{align}
\begin{split}
\kappa_t^0 &\equiv (2m_\pi^2-t+u)\,m_e^2 + m_\pi^4 + (m_\pi^2-t)^2 - tu \,,\\
\kappa_u^0 &\equiv (2m_\pi^2-u+t)\,m_e^2 + m_\pi^4 + (m_\pi^2-u)^2 - tu \,.
\end{split}
\end{align}

Summarising the above discussion, in the factorised prescription the soft plus virtual point-like cross section, which is already finite, is multiplied by $|F_\pi(s)|^2$. In this view, the IR cancellations occur in the point-like theory whose cross section gets an overall form factor evaluated at the centre-of-mass energy of the collision. As can be seen by Eq.~\eqref{Eq:deltaVdef} and Eq.~\eqref{Eq:deltaSdef}, the factorised approach amounts to multiply both the numerator and the denominator by $|F_\pi(s)|^2$, hence $\delta_{SV}$ is the same as the point-like approximation, the form factor appearing only in the LO cross section. This simplified treatment of the pion form factor as a multiplicative function does not take into account the actual momentum transfer of each virtual particle especially when dealing with box diagrams. Moreover, in the calculation of the charge asymmetry $A_{\rm FB}(\sqrt{s})$, as defined in Eq.~\eqref{eq:FBasym}, the dependence on the form factor at NLO is given only by the real emission diagrams, as it vanishes in the ratio of the virtual cross section to the LO. Sec.~\ref{sec:intffpi} will provide a more careful theoretical treatment of the pion form factor in the computation of NLO corrections.

\subsection{Parton Shower algorithm}\label{sec:ps}

In the present section, the main features of the PS algorithm adopted in our calculation will be described. The formulation closely follows the approach implemented in the code \textsc{BabaYaga@NLO}, already applied to other processes in $e^+ e^-$ collisions, such as two fermion~\cite{Balossini:2006wc} and photon pair~\cite{Balossini:2008xr} production, as well as $W/Z$ production~\cite{CarloniCalame:2006zq,CarloniCalame:2007cd} and Higgs boson decay~\cite{Boselli:2015aha} at hadron colliders, and adapts it to pion pair production. Recent reviews about the main features of \textsc{BabaYaga@NLO} can be found in~\cite{CarloniCalame:2017ioy,Frixione:2022ofv}.

As in any PS description, the basic ingredient of the algorithm is the Sudakov form factor. It can be written as:
\begin{eqnarray}
    \Pi (\varepsilon, Q^2) \, = \, {\rm exp} \left\{ - \frac{\alpha}{2 \pi} \, \int_0^{1-\varepsilon} \rmd z \, P(z) \, \int 
    \rmd \Omega_{k} \, {\cal I}(k) \right\} \,.
    \label{eq:SFF}
\end{eqnarray}
The Sudakov form factor represents the probability of \textit{no resolved emission}. This means that, by considering an energy cutoff $\varepsilon 
\, \sqrt{s} / 2$, with  $\varepsilon \ll 1$, the probability that a  charged particle
with virtuality $Q^2$ emits either a real photon with energy below the cutoff or a virtual photon is given by $\Pi (\varepsilon, Q^2)$. The Sudakov form factor accounts for the exponentiation of these contributions. The $Q^2$ variable is a virtuality scale which stems from the integration of the function ${\cal I}(k)$, as explained in the following.

In Eq.~(\ref{eq:SFF}), $P(z)$ is the splitting function. It describes how energy is shared in the photon emission process and it depends on the spin of the emitter. Therefore, for $e^+ e^- \to \pi^+ \pi^- (\gamma)$ two different splitting functions are needed. The (unregularised) splitting function for photon emission off a fermion $f \to f + \gamma$ is given by~\cite{CZ:1975,Altarelli:1977zs,Leenaarts:2019}
\begin{eqnarray}
P_f (z) \, = \, \frac{1+z^2}{1-z} \,,
\end{eqnarray}
which is defined in the approximation $m_f=0$. It is infrared divergent in the limit $z \to 1$, $z$ being the energy fraction of the daughter fermion involved in the branching. 

For a point-like scalar particle, \textit{i.e.} for 
photon emission from a pion according to sQED, 
the splitting function in massless approximation reads as follows~\cite{Leenaarts:2019,Chen:2016wkt}
\begin{eqnarray}
P_s (z) \, = \, \frac{2 \, z}{1-z} \,.
\end{eqnarray}

If one takes the soft limit $z\to1$, the two splitting functions $P_f(z)$ and $P_s(z)$ coincide. This can be understood in terms of the universal (spin independent) nature of soft radiation. Indeed, the integration over $z$ of the two splitting functions provides
\begin{eqnarray}
  I_+^{\rm QED} (\varepsilon) &=& \int_0^{1-\varepsilon} \, 
  \rmd z \, P_f (z)
  = - 2 \ln \varepsilon - \frac{3}{2} + 2\varepsilon - \frac{1}{2}\varepsilon^2 \,, \\
  I_+^{\rm sQED} (\varepsilon) &=& \int_0^{1-\varepsilon} \,
    \rmd z \, P_s (z)
  = - 2 \ln \varepsilon - 2 + 2\varepsilon \,.
\end{eqnarray}
The two results only differ by non-singular contributions. Massive splitting functions~\cite{Dittmaier:1999mb,Catani:2000ef} are not used in our approach, as finite mass corrections to the emission dynamics are treated through the factor ${\cal I} (k)$, as described below.

The function ${\cal I} (k)$ in Eq.~(\ref{eq:SFF}) is a non-infrared, dipole radiation factor which is modelled from the structure of the Yennie-Frautschi-Suura (YFS) eikonal current~\cite{Yennie:1961ad}. Its explicit expression is given by
\begin{eqnarray}
{\cal I}(k) \, = \,
\sum_{i,j} \,
\eta_i \eta_j~
\frac{p_i\cdot p_j}{(p_i\cdot k)(p_j\cdot k)} \, k_0^2 \,,
\label{idik}
\end{eqnarray}
where $p_{i,j}$ are the massive momenta of the external charged particles, $k$ is the photon momentum, $k_0$ the photon energy and the symbol $\eta_i$ is the same as in Eq.~\eqref{eq:eik_int}. By construction, Eq.~(\ref{idik}) accounts for the interference of radiation coming from different legs and  also the diagonal contributions of the kind $m^2 / (p \cdot k)^2$, rendering the inclusion of finite mass corrections in the splitting functions unnecessary.

By integrating ${\cal I}(k)$ over the photon angular
variables, as in Eq.~(\ref{eq:SFF}), the non-diagonal contributions provide the collinear logarithms $\ln (s / m_a^2)$, where $m_a = m_e, m_\pi$, due to initial- and final-state radiation, respectively, as well as angular dependent logarithms coming from initial-final-state interference. The integration of the mass corrections gives constant terms which, after multiplication with the $I_{+}$ functions, improves the treatment of radiative corrections 
in the infrared region.

By means of the above theoretical ingredients, it is possible to obtain predictions in the leading logarithmic (LL) approximation, whereas the dominant contributions due to soft and collinear radiation are resummed to all orders of perturbation theory. As shown in~\cite{CarloniCalame:2000pz}, the Sudakov form factor can be used in association with a splitting function to obtain for each leg an iterative MC solution (known as PS algorithm) of the 
DGLAP equation~\cite{Altarelli:1977zs,Gribov:1972ri,Dokshitzer:1977sg} for the non-singlet structure function, which describes multiple photon emission in the collinear approximation. In the present implementation based on two different splitting functions, one gets from the PS simulation the contribution due to two electron structure functions for ISR and two scalar structure functions 
for FSR. However, by virtue of the coherence effects included in Eq.~(\ref{idik}), the PS implementation also exponentiates the dominant correction due to infrared cancellation between box diagrams and initial-final state interference. The PS resummation of initial-final contributions mimics the 
soft-photon exponentiation 
of interference contributions in the QED
collinear Structure Function approach~\cite{GN:1990}
and generalises the algorithm of 
\textsc{Babayaga@NLO} for the simulation of 
pure QED processes\footnote{ Strictly 
speaking, the present treatment in \textsc{Babayaga@NLO} 
of interference exponentiation holds for a non-resonant 
LO cross section. Since the Born cross section of 
$e^+ e^- \to \pi^+ \pi^-$ has a non-trivial resonant shape, 
this point deserves further investigation and is left to a future study.}.

An important feature of the PS algorithm is the possibility of going beyond the strictly collinear limit and including the transverse momentum of the emitting particles and photons at each branching. In \textsc{Babayaga@NLO}, the exclusive kinematics is simulated by generating the angular spectrum of the emitted photons according to Eq.~(\ref{idik}).

\subsection{Matching NLO corrections to Parton Shower}\label{sec:nlops}

According to the PS algorithm sketched above, the cross section is corrected to account for the emission of an arbitrary number of photons. It can be written in the following form:
\begin{eqnarray}
    \rmd\sigma_{\rm PS} \, = \, \Pi(\varepsilon, Q^2) \, \sum_{n=0}^\infty
  \frac{1}{n!}\;\left|{\cal M}_n^{\rm PS}\right|^2\;\rmd\Phi_n(\{p\},\{k\}) \,,
  \label{eq:PS}
\end{eqnarray}
where $\{p\},\{k\}$ displays the set of the final state momenta and $\rmd\Phi_n$ is the exact $(n+2)$-particle phase-space element of the underlying process accompanied by radiation of $n$ real photons. 

In Eq.~(\ref{eq:PS}), $|{\cal M}_n^{\rm PS}|^2$ is the PS approximation to the squared amplitude of the process describing the emission of $n$ hard photons, i.e. with energy fraction larger than $\varepsilon $, in the LL approximation. According to the factorisation theorems of soft and collinear singularities, it can be written in a factorised form in terms of splitting functions, to describe the photon energy spectrum and the function ${\cal I} (k)$, to characterise the photon angular distribution. Its explicit expression can be found in~\cite{Balossini:2006wc,Frixione:2022ofv}. The integral over the phase space has a lower limit for the photon energies set to $\varepsilon \frac{\sqrt{s}}{2}$, in order to ensure the cancellation of the infrared divergences.

As detailed in~\cite{Balossini:2006wc}, Eq.~(\ref{eq:PS}) can be improved to match exact NLO corrections with the PS resummation. The master formula implemented in \textsc{BabaYaga@NLO} reads as follows:
\begin{eqnarray}
    \rmd\sigma_{\rm NLOPS}
  = F_{SV}\;\Pi(\varepsilon, Q^2)
  \sum_{n=0}^\infty\frac{1}{n!}\;\left(\prod_{i=1}^n F_{{ H},i}\right)\;
  |{\cal M}_n^{\rm PS}(\{p\},\{k\})|^2\rmd\Phi_n(\{p\},\{k\}) \,,
  \label{eq:masterformula}
\end{eqnarray}
where $F_{ SV}$ and $F_{{ H},i}$ are soft+virtual and hard bremsstrahlung correction factors, respectively. They are given by
\begin{eqnarray}
F_{ SV} = 1+\frac{\rmd\sigma^{\rm NLO}_{ SV}-\rmd\sigma^{{\rm PS},\alpha}_{ SV}}{\rmd\sigma^{\rm LO}} \,,~~~~~~
F_{{ H},i} =
1+\frac{|{\cal M}_1^{\rm NLO}(k_i)|^2 - |{\cal M}_{1}^{{\rm PS},\alpha}(k_i)|^2}{|{\cal M}_{1}^{{\rm PS},\alpha}(k_i)|^2} \,.
\label{eq:fsh}
\end{eqnarray}
The quantities $F_{ SV}$ and $F_{{H},i}$ carry the information of the exact fixed-order calculation described in Sec.~\ref{sec:nlo} and in the next section, since they are defined in terms of the $2 \to 2$ NLO soft+virtual cross section $\rmd\sigma^{\rm NLO}_{\rm SV}$ and the $2 \to 3$ exact bremsstrahlung amplitude ${\cal M}_1^{\rm NLO}(k_i)$. In Eq.~(\ref{eq:fsh}), $\rmd\sigma^{{\rm PS},\alpha}_{\rm SV}$ is the $O(\alpha)$ expansion of Eq.~(\ref{eq:PS}) in the soft+virtual limit and $|{\cal M}_{1}^{{\rm PS},\alpha}(k_i)|^2$ is the PS approximation of the one-photon emission process.

It is worth noting that the coefficients $F_{ SV}$ and $F_{{H},i}$ are, by construction, free of collinear and infrared logarithms and that the ${\cal O}(\alpha)$ expansion of Eq.~(\ref{eq:masterformula}) exactly reproduces the NLO calculation, without double counting. Furthermore, Eq.~(\ref{eq:masterformula}) preserves the exponentiation of LL contributions as in a pure PS approach and can be conveniently implemented into a MC code, providing exclusive event generation at NLOPS accuracy.

It should also be noticed that the factorised matching 
of NLO corrections with the universal LL contributions 
of the PS algorithm allows us to 
effectively incorporate in our approach the bulk of the most 
important NNLO corrections and given by 
infrared-enhanced $\alpha^2 \, \log(s/m_e^2)$ sub-leading contributions~\cite{Montagna:1996gw}. As shown in~\cite{Balossini:2006wc,Balossini:2008xr,WorkingGrouponRadiativeCorrections:2010bjp}, the size of such corrections 
does not exceed the 0.1\% level for  $e^+ e^-$ and 
$\gamma \gamma$ production and can be seen as an estimate of the theoretical accuracy of 
\textsc{BabaYaga@NLO} in the modelling of QED processes.

In this work, the matching of NLO contributions to PS 
is performed in terms of three calculations of fixed-order corrections to pion pair production, that differ in the 
treatment of the pion internal structure in loops, as 
detailed in the following.

\section{Handling the pion composite structure in loops}
\label{sec:intffpi}
\newcommand{\ptlike}{0}
\newcommand{\FF}{\text{FF}}
\newcommand{\ifi}{\text{IFI}}
\newcommand{\fsr}{\text{FSR}}
\newcommand{\isr}{\text{ISR}}
\newcommand{\disp}{\text{FsQED}}

In Sec.~\ref{sec:QED_sQED}, the treatment of the pion form factor within the calculation of radiative corrections to the process $e^+ e^- \to \pi^+ \pi^-$ has been discussed adopting the factorised prescription with respect to the point-like sQED matrix elements. As already stressed, this approach has the virtue of respecting the cancellation of infrared divergences between virtual and real radiation matrix elements. However, in this scheme all kinematical configurations with different pion scattering angles are equally weighted, losing possible correlations with the angular asymmetries generated by (s)QED radiative corrections. In fact, the  factorised approach yields a good description of observables integrated over symmetric angular ranges, like $\pi^+ \pi^-$ invariant mass distributions, but it does not reproduce correctly more exclusive observables, such as the charge asymmetry~\cite{Arbuzov:2020foj,Colangelo:2022lzg,Ignatov:2022iou}, defined in Eq.~\eqref{eq:FBasym}. 
    
In order to improve the theoretical accuracy in the modelling of the data through Monte Carlo event generation, the form factor needs to be included in the loop amplitudes. This has been achieved in the first place by adopting an approach inspired by the generalised vector meson dominance~(GVMD) model, in which a pion form factor is inserted for each photon propagator attached to a sQED vertex~\cite{Ignatov:2022iou} (see also~\cite{Patil:1975ge} for first explorations). In this approach the pion form factor is approximated by means of a sum of Breit-Wigner~(BW) functions, corresponding to the included hadronic resonances, in order to perform the loop integration through standard techniques. A systematic investigation, less model dependent and fully consistent with quantum field theory, has been carried out by means of dispersion relations in~\cite{Colangelo:2014dfa,Colangelo:2015ama} for light-by-light scattering and in~\cite{Colangelo:2022lzg} for $e^+ e^- \to \pi^+ \pi^-$. An important result of such studies is the proof that the pion pole contribution to the transition amplitude can be  equivalently obtained by means of the point-like sQED amplitude dressed with pion form factors at every vertex of the diagrams involving pions. The latter approach is named FsQED, according to~\cite{Colangelo:2014dfa}. 

In the following sections we present a few details about the calculation and implementation in \textsc{BabaYaga@NLO} of the one-loop corrections to $e^+ e^- \to \pi^+ \pi^-$ with FsQED and GVMD approaches. In both methods, following the same argument of the factorised approximation, the virtual $(\delta_V^\text{ISR})$ and real soft $(\delta_S)$ corrections are not modified by the more complicated expressions of the corrections with the form factor in the loop diagrams w.r.t. the ones in point-like sQED. For this reason, we focus on the calculation of $\delta_V^\text{FSR}$ and $\delta_V^\text{IFI}$ inserting the form factor in the loops. A graphical representation of the contributions to be evaluated in both approaches for $\delta_V^\text{FSR}$ and $\delta_V^\text{IFI}$ is given in Fig.~\ref{FIG:FSR_FF} and Fig.~\ref{FIG:IFI_FF}, respectively. 
The calculation of the FsQED and GVMD contributions has common features, which we illustrate in the following discussion. 
\begin{figure}[t]
\begin{equation*}
\begin{gathered}
\begin{tikzpicture}
\begin{feynman}[small]
    \vertex (a);
    \vertex[right=1cm of a,style=blob] (b) {};
    \vertex[above left=1cm and 1cm of a] (c);
    \vertex[below left=1cm and 1cm of a] (d);
    \vertex[above right=1cm and 1cm of b] (e);
    \vertex[below right=1cm and 1cm of b] (f);
    \vertex[above right =0.6and 0.6 of b,style=blob] (g) {};
    \vertex[below right =0.6and 0.6cm of b,style=blob] (h) {};
    \diagram* {
      (a) -- [photon] (b),
      (d) -- [fermion] (a),
      (a) -- [fermion] (c),
      (e) -- [scalar] (g) --[scalar] (b),
      (b) -- [scalar] (h) --[scalar] (f),
      (g) --[photon,half left, looseness=1,] (h),
    };
\end{feynman}
\end{tikzpicture}
\end{gathered}
\qquad
\begin{gathered}
\begin{tikzpicture}
\begin{feynman}[small]
    \vertex (a);
    \vertex[right=1cm of a,style=blob] (b) {};
    \vertex[above left=1cm and 1cm of a] (c);
    \vertex[below left=1cm and 1cm of a] (d);
    \vertex[above right=1cm and 1cm of b] (e);
    \vertex[below right=1cm and 1cm of b] (f);
    \vertex[above right =0.6 and 0.6 of b,style=blob] (g) {};
    \vertex[below right =0.6 and 0.6cm of b] (h) ;
    \diagram* {
      (a) -- [photon] (b),
      (d) -- [fermion] (a),
      (a) -- [fermion] (c),
      (e) -- [scalar] (g) --[scalar] (b),
      (b) -- [scalar] (h) --[scalar] (f),
      (g) --[photon,half left, looseness=1,] (b),
    };
\end{feynman}
\end{tikzpicture}
\end{gathered}
\qquad 
\begin{gathered}
\begin{tikzpicture}
\begin{feynman}[small]
    \vertex (a);
    \vertex[right=1cm of a,style=blob] (b) {};
    \vertex[above left=1cm and 1cm of a] (c);
    \vertex[below left=1cm and 1cm of a] (d);
    \vertex[below right=1cm and 1cm of b] (e);
    \vertex[above right=1cm and 1cm of b] (f);
    \vertex[below right =0.6 and 0.6 of b,style=blob] (g) {};
    \vertex[above right =0.6 and 0.6cm of b] (h) ;
    \diagram* {
      (a) -- [photon] (b),
      (d) -- [fermion] (a),
      (a) -- [fermion] (c),
      (e) -- [scalar] (g) --[scalar] (b),
      (b) -- [scalar] (h) --[scalar] (f),
      (g) --[photon,half right, looseness=1,] (b),
    };
\end{feynman}
\end{tikzpicture}
\end{gathered}
\end{equation*}
\caption{Final state radiation diagrams in sQED with the form factor introduced into the loop integration, where the blobs represent point-like vertices dressed with form factors. A $F_\pi(q^2)$ contribution for each photon entering the blobs is understood.}
\label{FIG:FSR_FF}
\end{figure}
\begin{figure}[t]
\begin{equation*}
\begin{gathered}
\begin{tikzpicture}
\begin{feynman}[small]
   \vertex (a);
   \vertex[right=1cm of a,style=blob] (b) {};
   \vertex[below=1cm of a] (i);
   \vertex[below=1cm of b,style=blob] (j) {};
   \vertex[above left=0.75cm and 0.75cm of a] (c);
   \vertex[below left=0.75cm and 0.75cm of i] (d);
   \vertex[above right=0.75cm and 0.75cm of b] (e);
   \vertex[below right=0.75cm and 0.75cm of j] (f);
   \diagram* {
     (a) -- [photon] (b),
     (d) -- [fermion] (i) -- [fermion] (a) -- [fermion] (c),
     (e) -- [scalar] (b) --[scalar] (j) -- [scalar] (f),
     (i) -- [photon] (j),
   };
\end{feynman}
\end{tikzpicture}
\end{gathered}
\qquad
\begin{gathered}
\begin{tikzpicture}
\begin{feynman}[small]
   \vertex (a);
   \vertex[right=1cm of a,style=blob] (b) {};
   \vertex[below=1cm of a] (i);
   \vertex[below=1cm of b,style=blob] (j) {};
   \vertex[above left=0.75cm and 0.75cm of a] (c);
   \vertex[below left=0.75cm and 0.75cm of i] (d);
   \vertex[above right=0.75cm and 0.75cm of b] (e);
   \vertex[below right=0.75cm and 0.75cm of j] (f);
   \diagram* {
     (a) -- [photon] (j),
     (d) -- [fermion] (i) -- [fermion] (a) -- [fermion] (c),
     (e) -- [scalar] (b) --[scalar] (j) -- [scalar] (f),
     (i) -- [photon] (b),
   };
\end{feynman}
\end{tikzpicture}
\end{gathered}
\qquad 
\begin{gathered}
\begin{tikzpicture}
\begin{feynman}[small]
   \vertex (a);
   \vertex[below=1cm of a] (i);
   \vertex[above left=0.75cm and 0.75cm of a] (c);
   \vertex[below left=0.75cm and 0.75cm of i] (d);
   \vertex[below right=0.35cm and 0.8cm of a, style=blob] (e) {};
   \vertex[above right=1.25 cm and 1cm of e] (f);
   \vertex[below right=1.25 cm and 1cm of e] (g);
   \diagram* {
     (d) -- [fermion] (i) -- [fermion] (a) -- [fermion] (c),
     (a) -- [photon] (e) -- [photon] (i),
     (f) -- [scalar] (e) --[scalar] (g),
   };
\end{feynman}
\end{tikzpicture}
\end{gathered}
\end{equation*}
\caption{Initial-final state interference box diagrams in sQED with the form factor inside the loop integration, as in Fig.~\ref{FIG:FSR_FF}.}
\label{FIG:IFI_FF}
\end{figure}

A  careful treatment of the divergences arising from the loop integration is required. At first, to find the proper ultraviolet (UV) counterterms, the form factor has to be embedded also in the pion self energy diagram, as shown in Eq.~\eqref{LOOP:pionSE}
\begin{equation}\label{LOOP:pionSE}
\begin{gathered}
\Sigma_\pi(p^2)\quad=\quad
\begin{tikzpicture}[baseline={(0, -0.1)}]
\begin{feynman}[small]
    \vertex (a) at (-1.3, 0);
    \vertex (b) at (1.3, 0);
    \vertex[style=blob] (c) at (-0.6,0) {} ;
    \vertex (e) at (0,0.25) ;
    \vertex[style=blob] (d) at (0.6,0) {};
    \diagram* {
      (a) -- [scalar] (c) -- [scalar] (d) -- [scalar] (b),
      (c) -- [photon, half left, looseness=1.55] (d),
    };
\end{feynman}
\end{tikzpicture}\, ,
\end{gathered}
\end{equation}
where the diagram with the quartic vertex is not shown as it vanishes in the limit $\lambda \to 0$ and, moreover, being independent of the external momentum, does not contribute to the pion wavefunction counterterm. In the on-shell scheme, the pion wavefunction counterterm is shifted by a IR-finite amount, according to the relation
\begin{equation}\label{eq:pionwave}
    \delta Z_{\phi,\FF}(\lambda) = - \pdv{\Sigma_\pi(p^2)}{p^2}\bigg|_{p^2=m_\pi^2}= \delta Z_{\phi,\ptlike}(\lambda) + \text{IR-finite terms}\, , 
\end{equation} 
where the subscript $\FF$ denotes either the GVMD approach or the FsQED one, whereas the subscript $\ptlike$ denotes the point-like contribution, namely
\begin{align}\label{eq:ct_sqed}
    \delta Z_{\phi,\ptlike}(\lambda) = \frac{\alpha}{2\pi} \left\{\text{B}_0(m_\pi^2,\lambda^2,m_\pi^2)+2 \hspace{1pt}m_\pi^2\,\frac{\partial\text{B}_0}{\partial p^2}(p^2, \lambda^2, m_\pi^2)\biggl|_{p^2=m_\pi^2} \right\} \,.
\end{align}
The virtual corrections, fully differential in the phase space, are written as
\begin{equation}\label{def:Delta_diff}
    \delta_{V,\FF}^i(\lambda) = \frac{ 2\Re{F^*_\pi(s)\Amp_{\lo,\ptlike}^\dagger \, \Amp_{V,\FF}^i(\lambda)}}{\left|F_\pi(s)\right|^2|\Amp_{\lo,\ptlike}|^2}\, ,  \qquad 
    \genfrac{}{}{0pt}{}{i = \fsr,\,\ifi}{\FF = \text{GVMD},\,\text{FsQED}}
\end{equation}
where we explicitly factorise the tree-level form factors from the point-like matrix element to stress the presence of $F_\pi(q^2)$ inside $\Amp_V$. From this definition, the renormalised FSR virtual correction given by the diagrams of Fig.~\ref{FIG:FSR_FF}, is therefore  modified as
\begin{equation}\label{Eq:FSR_FF}
\delta^\text{FSR}_{V,\FF} = \delta^\text{FSR}_{V,0} + \text{IR-finite terms}\, ,
\end{equation}
where the result of the loop integration is an additional IR-finite piece, which is independent of the scattering angle. 

Unlike the FSR diagrams that always present a $s$-channel vertex and two additional ones, for a total of three form factors, box diagrams have only two $F_\pi(q_i^2)$, with $i=1,2$. This means that, unlike Eq.~\eqref{Eq:FSR_FF}, at a first glance it is not obvious that the IFI correction can be always written as the sum of a point-like contribution and a form-factor correction. However, since the soft correction $\delta_S^\text{IFI}(\lambda)$ is the same as the point-like one, the soft-virtual IFI correction can be written as
\begin{equation}\label{IFIfinite}
\delta_{SV,\FF}^\text{IFI}=\delta_{SV,\ptlike}^\text{IFI} \hspace{0.5pt}\big|_{\rm IR} + \text{IR-finite terms}\, .
\end{equation}
where $|_{\rm IR}$ denotes the IR-divergent part of the point-like contribution $\delta_{SV,\ptlike}^\text{IFI}$ whose dependence on $\lambda^2$ has to vanish. This holds only if the coefficient of the IR-divergent term in the virtual amplitude matches the $\log\lambda$ coefficient in Eq.~\eqref{Eq:deltaSdef}, namely
\begin{equation}\label{eq:ifi_ir}
\delta_{S,\text{IR}}^\text{IFI}=\mathcal{C}^\ifi_{\text{IR}}\log\frac{4 \Delta E^2}{\lambda^2}\, .
\end{equation}
The coefficient $\mathcal{C}_\text{IR}^\ifi$  for the IFI correction is given by the following expression
\begin{equation}\label{EQ:IR_SOFT}
\begin{alignedat}{2}
\mathcal{C}_\text{IR}^\ifi= \frac{2\alpha}{\pi} &\Biggl[&&
        \frac{ \left(m_e^2+m_\pi^2-t\right) }{f(t)}\log\frac{m_e^2+m_\pi^2-t + f(t)}{2 m_e m_\pi}\\
         &-&&\frac{ \left(m_e^2+m_\pi^2-u\right) }{f(u)}\log\frac{m_e^2+m_\pi^2-u + f(u)}{2 m_e m_\pi}\Biggr]\, ,
        \end{alignedat}
\end{equation}
where 
 \begin{equation}
      f(z) = \sqrt{z^2+(m_e^2-m_\pi^2)^2-2 (m_e^2+m_\pi^2)z}\,.
  \end{equation}
Finally notice that the virtual initial-final-state amplitude is the sole angular-odd correction that is modified by the loop insertion. Therefore, the charge asymmetry in the GVMD and FsQED approaches crucially depends on $\delta_V^\ifi$.

By inspection of the formulae, the insertion of the pion form factor in loop diagrams amounts to non-trivial combinations of virtual sQED-like contributions that involve the exchanges of photons having an effective mass. In the GVMD approach this effective mass represents a vector meson mass, while in the FsQED approach it is introduced by the dispersion relation as an integration variable. We first compute the pion wavefunction counterterm, as it is needed to renormalise the final state vertex corrections. Specifically we obtain 
\begin{align}\label{eq:ct_phmass}
    \delta \bar{Z}_{\phi}(s') = \frac{\alpha}{4\pi} \left\{2\hspace{1pt}\text{B}_0(m_\pi^2,s',m_\pi^2)+
    (4 m_\pi^2-s') \frac{\partial\text{B}_0}{\partial p^2}(p^2, s', m_\pi^2)\biggl|_{p^2=m_\pi^2} \right\} \, ,
\end{align}
where $s'$ represents the squared effective mass of the virtual photon. 

We define in the following the kernel functions for the renormalised FSR and IFI corrections computed with massive photons as
\begin{align}
\bar{\delta}_{V}^{\hspace{1pt}\fsr}(s') &= \frac{ 2\Amp_{\lo,\ptlike}^\dagger \, \Amp_{V,0}^\fsr(s')}{|\Amp_{\lo,\ptlike}|^2} \,, \\[2pt]
\bar{\delta}_{V}^{\hspace{1pt}\ifi}(s',s'') &= \frac{ 2\Amp_{\lo,\ptlike}^\dagger \,\Amp_{V,\ptlike}^\ifi(s',s'')}{|\Amp_{\lo,\ptlike}|^2} \,,\label{Eq:IFIKernel}
\end{align}
where $s'$ and $s''$ are the squared effective masses of the virtual photons. The $\ptlike$ in the subscript denotes that the virtual matrix elements have the analytic structure of the point-like ones but with effectively massive virtual photons, one with  $m_\gamma^2=s'$ and one with $m_\gamma^2=s''$. The explicit expression of the renormalised FSR in terms of scalar one-loop functions reads
\begin{equation}\label{eq:fsr_phmass}
\begin{aligned}
\Bar{\delta}_{V}^{\hspace{1pt}\fsr}(s')=    \frac{\alpha}{2\pi} \Bigg\{ &
    \frac{4 m_\pi^2 - 2 s - s'}
    {4 m_\pi^2 - s} \Big[ \left(4 m_\pi^2 - s - 2 s'\right) \mathcal{C}_0^{\pi,\pi}(s')\\
    &
    - 2 \text{B}_0\left(s, m_\pi^2, m_\pi^2\right)\Big] - \frac{2 \left(s + s'\right)}{4m_\pi^2-s} \text{B}_0\left(m_\pi^2, m_\pi^2, s'\right)  \\
    &+ 2 \text{B}_0\left(m_\pi^2, m_\pi^2, s'\right)+ \left(4 m_\pi^2 - s'\right) \frac{\partial\text{B}_0}{\partial p^2}\left(p^2, m_\pi^2, s'\right)\biggl|_{p^2=m_\pi^2}
    \Bigg\}\, .
\end{aligned}
\end{equation}
Assuming photons with different squared effective masses $s'$ and $s''$, the IFI kernel, as defined in Eq.~\eqref{Eq:IFIKernel}, has the following expression
\begin{equation}
\label{Eq:IFI_phmass}
\begin{aligned}
\Bar{\delta}_{V}^{\hspace{1pt}\ifi}(s',s'')=&\frac{\alpha}{2\pi}\frac{4s}{(4 m_\pi^2 s-s^2+(t-u)^2)}\\ \biggl\{&\frac{2 m_e^2 (t-u)}{4 m_e^2-s} \left[\text{B}_0(m_e^2,m_e^2,s')+\text{B}_0(m_e^2,m_e^2,s'')-2 \text{B}_0(s,s',s'')\right]\\
&+\mathcal{C}_0^e(s',s'')  (t-u)\frac{8 m_e^4+2 m_e^2 \left(s'+s''-4s\right)+s^2}{4m_e^2-s}\\
&+\mathcal{C}_0^\pi(s',s'') \left(2 m_\pi^2-s\right) (t-u)\\
&-\Bigl[\mathcal{C}_0^{e,\pi}(t,s')+\mathcal{C}_0^{e,\pi}(t,s'')\Bigr] \left(m_e^2-t\right) \left(m_e^2+m_\pi^2-t\right)\\
&+\Bigl[\mathcal{C}_0^{e,\pi}(u,s')+\mathcal{C}_0^{e,\pi}(u,s'')\Bigr]\left(m_e^2-u\right) \left(m_e^2+m_\pi^2-u\right)\\
&+\mathcal{D}_0^{e,\pi}(t,s',s'')\left(m_e^2+m_\pi^2-t\right)\kappa_t(s'+s'')\\
&-\mathcal{D}_0^{e,\pi}(u,s',s'')\left(m_e^2+m_\pi^2-u\right)\kappa_u(s'+s'')\biggr\}\, ,
\end{aligned}
\end{equation}
where we have introduced the auxiliary functions
\begin{align}\label{Eq:auxiliary}
\begin{split}
\kappa_t(z) &\equiv m_e^2(2m_\pi^2-t+u-z) + m_\pi^4 + (m_\pi^2-t)^2 + t (z-u) \,,\\
\kappa_u(z) &\equiv m_e^2(2m_\pi^2-u+t-z) + m_\pi^4 + (m_\pi^2-u)^2 + u (z-t) \,.
\end{split}
\end{align}
We remind that the shorthand notation for the Passarino-Veltman scalar functions is given in Eq.~\eqref{Eq:shorthand}. In the following, we explain in detail the calculation of the FSR and IFI corrections in both approaches, without neglecting any external mass.

\subsection{NLO calculation in the GVMD model}\label{sec:gvmd}

A method to improve the factorised approximation, inspired by the GVMD model~\cite{Sakurai:1972wk}, has been proposed in~\cite{Ignatov:2022iou}. The key idea is to approximate the pion form factor as a sum of a finite number $n_r$ of BW functions. Accordingly, we write the pion form factor as 
\begin{align}
F^\text{BW}_\pi(q^2) = \sum_{v=1}^{n_r} F^\text{BW}_{\pi,v}(q^2)  = \frac{1}{c_t} \sum_{v=1}^{n_r} c_v \frac{\Lambda_v^2}{\Lambda_v^2 - q^2} \:,
\label{eq:bwsum}
\end{align}
where $\Lambda_v^2 = m_v^2 - i m_v \Gamma_v$ and $c_v = |c_v|e^{i\phi_v}$. The division by $c_t = \sum_v c_v$ ensures the normalisation condition $F_\pi(0)=1$. In the GVMD model, each term $F^\text{BW}_{\pi,v}(q^2)$ corresponds to the propagator of a vector meson $v$ with mass $m_v$ and width $\Gamma_v$, multiplied by a complex coupling $c_v$. As proposed in~\cite{Ignatov:2022iou}, the form factor $F^\text{BW}_\pi(q^2)$ can be inserted in each sQED vertex according to the rules
\begin{subequations}
\begin{alignat}{2}
\begin{gathered}
\begin{tikzpicture}[baseline=(a)]
\begin{feynman}[inline=(a)]
    \vertex (a);
    \vertex[right=1.0cm of a,style=blob] (b) {};
    \vertex[above right=0.75cm and 0.75cm of b] (c);
    \vertex[below right=0.75cm and 0.75cm of b] (d);
    \diagram* {
      (a) -- [photon] (b), 
      (c) -- [scalar] (b),
      (b) -- [scalar] (d),
    };
\end{feynman}
\end{tikzpicture}
\end{gathered}
\quad &= \quad 
\begin{gathered}
\begin{tikzpicture}[baseline=(a)]
\begin{feynman}[inline=(a)]
    \vertex (a);
    \vertex[right=0.6cm of a,style=dot] (b) {};
    \vertex[right = 0.6cm of b, style=dot] (e) {};
    \vertex[above right=0.75cm and 0.75cm of e] (c);
    \vertex[below right=0.75cm and 0.75cm of e] (d);
    \diagram* {
      (a) -- [photon] (b),
      (b) --[graviton] (e),
      (c) -- [scalar] (e),
      (e) -- [scalar] (d),
      (a) -- [fermion,opacity=0.0] (e)
    };
\end{feynman}
\end{tikzpicture}
\end{gathered}
\quad &&= \quad\;
\begin{gathered}
\begin{tikzpicture}[baseline=(a)]
\begin{feynman}[inline=(a)]
    \vertex (a);
    \vertex[right=1.0cm of a] (b);
    \vertex[above right=0.75cm and 0.75cm of b] (c);
    \vertex[below right=0.75cm and 0.75cm of b] (d);
    \diagram* {
      (a) -- [photon, momentum'=$q$] (b),
      (c) -- [scalar] (b),
      (b) -- [scalar] (d),
    };
\end{feynman}
\end{tikzpicture}
\end{gathered}
\times F^\text{BW}_\pi(q^2) \,,
\\[2pt]
\begin{gathered}
\begin{tikzpicture}[baseline=(a)]
\begin{feynman}[inline=(a)]
     \vertex[style=blob] (a) {};
      \vertex[above left=0.75cm and 0.75cm of a] (b);
     \vertex[below left=0.75cm and 0.75cm of a] (e);
     \vertex[above right=0.75cm and 0.75cm of a] (c);
     \vertex[below right=0.75cm and 0.75cm of a] (d);
     \diagram* {
       (e) -- [photon] (a), 
       (b) -- [photon] (a), 
       (c) -- [scalar] (a),
       (a) -- [scalar] (d),
     };
\end{feynman}
\end{tikzpicture}
\end{gathered}
\quad\;\;\; &= \quad\;\;
\begin{gathered}
\begin{tikzpicture}[baseline=(a)]
\begin{feynman}[inline=(a)]
     \vertex[style=dot] (a) {};
      \vertex[above left=0.75cm and 0.75cm of a] (b);
     \vertex[below left=0.75cm and 0.75cm of a] (e);
       \vertex[above left=0.35cm and 0.35  cm of a, style=dot] (f) {};
     \vertex[below left=0.35cm and 0.35  cm of a, style=dot] (g){};
     \vertex[above right=0.75cm and 0.75cm of a] (c);
     \vertex[below right=0.75cm and 0.75cm of a] (d);
     \diagram* {
       (e) -- [photon] (g),
       (b) -- [photon] (f),
       (e) -- [photon] (g),
       (b) -- [photon] (f),
       (c) -- [scalar] (a),
       (a) -- [scalar] (d),
       (g) -- [graviton] (a),
       (f) -- [graviton] (a),
       (e) -- [photon, opacity=0.0] (a), 
       (b) -- [photon, opacity=0.0] (a) 
     };
\end{feynman}
\end{tikzpicture}
\end{gathered}
\quad &&= \;\;
\begin{gathered}
\begin{tikzpicture}[baseline=(a)]
\begin{feynman}[inline=(a)]
        \tikzfeynmanset{
    momentum/arrow shorten=0.25,
    }
     \vertex (a);
      \vertex[above left=0.75cm and 0.75cm of a] (b);
     \vertex[below left=0.75cm and 0.75cm of a] (e);
     \vertex[above right=0.75cm and 0.75cm of a] (c);
     \vertex[below right=0.75cm and 0.75cm of a] (d);
     \diagram* {
       (e) -- [photon, momentum'=$q_2$] (a),
       (b) -- [photon, momentum'=$q_1$] (a),
       (c) -- [scalar] (a),
       (a) -- [scalar] (d),
     };
\end{feynman}
\end{tikzpicture}
\end{gathered}
\times F^\text{BW}_\pi(q_1^2)\,F^\text{BW}_\pi(q_2^2) \,.
\end{alignat}
\end{subequations}
This is equivalent to multiplying the point-like amplitude by a certain number of form factors before evaluating the loop integral. As pointed out in~\cite{Ignatov:2022iou}, such a multiplication preserves gauge invariance. Namely, the FSR and IFI virtual corrections can be written as
\begin{align}
\Amp_{V,\text{GVMD}}^\text{FSR} &= \int \dd^D q \, \Amp_{V,0}^\text{FSR} \: F_\pi^\text{BW}(s) \sum^{n_r}_{v,w=1} F_{\pi,v}^\text{BW}(q^2)\, F_{\pi,w}^\text{BW}(q^2) \,, \\[2pt]
\Amp_{V,\text{GVMD}}^\text{IFI} &= \int \dd^D q \, \Amp_{V,0}^\text{IFI} \,\sum^{n_r}_{v,w=1} \, F_ {\pi,v}^\text{BW}(q^2)\, F_{\pi,w}^\text{BW}((q-p_3-p_4)^2) \,, 
\end{align}
where $\Amp_{V,0}^i$ denotes the one-loop point-like amplitude for $i=\text{FSR},\text{IFI}$ and $q$ is the momentum flowing in the loop. Each form factor takes as input the momenta flowing in the virtual photon attached to the corresponding vertex. Note that the FSR correction is multiplied by a further form factor $F_{\pi,v}^\text{BW}(s)$, which factorises over the whole calculation. Since each term $F^\text{BW}_{\pi,v}(q^2)$ has a simple propagator-like structure, the one-loop amplitudes can be computed through standard techniques. In this regard, we remark that the insertion of two pion form factors in one diagram does not increase the number of points of the scalar one-loop functions with respect to the point-like calculation. For instance, the virtual IFI correction $\Amp_{V}^\text{IFI}$ does not contain any 6-point function, because each term $F^\text{BW}_{\pi,v}(q^2)$ is multiplied by a photon propagator with the same momentum. Hence, we can simplify each additional propagator by using the identity
\begin{align}\label{eq:prop_ifi}
\frac{1}{q_i^2-\lambda^2}\frac{1}{q^2_i-\Lambda_i^2} = \frac{1}{\Lambda_i^2-\lambda^2}\bigg[\frac{1}{q_i^2-\Lambda_i^2} -\frac{1}{q_i^2-\lambda^2}\bigg] \,,
\end{align}
with $q_i=\{q,\hspace{1pt}q-p_3-p_4\}$ and $\Lambda_i=\{\Lambda_v,\hspace{1pt}\Lambda_w\}$. The same reasoning can be applied to the virtual FSR correction. The only difference is that in this case there are two form factors evaluated at $q^2$, where $q$ also corresponds to the momentum of the virtual photon. Specifically, assuming~$\Lambda_v \neq \Lambda_w$, we have
\begin{align}\label{eq:prop_fsr_neq}
\begin{split}
\frac{1}{q^2-\lambda^2}\frac{1}{q^2 -\Lambda_v^2} \frac{1}{q^2 -\Lambda_w^2} = 
\frac{1}{\Lambda_v^2 -\Lambda_w^2} \bigg[
&\frac{1}{\Lambda_v^2-\lambda^2}
\bigg(\frac{1}{q^2-\Lambda_v^2} -\frac{1}{q^2-\lambda^2}\bigg)\\[2pt]
-&\frac{1}{\Lambda_w^2-\lambda^2}
\bigg(\frac{1}{q^2-\Lambda_w^2}-\frac{1}{q^2-\lambda^2}\bigg) \bigg] \,,
\end{split}
\end{align}
which is singular for $\Lambda_v = \Lambda_w$. In this case, one can use the identity
\begin{align}\label{eq:prop_fsr_eq}
\begin{split}
\frac{1}{q^2-\lambda^2}\bigg(\frac{1}{q^2 -\Lambda_v^2}\bigg)^{\!2} &= 
\frac{1}{\Lambda_v^2 -\lambda^2} \bigg[ \bigg(\frac{1}{q^2 -\Lambda_v^2}\bigg)^{\!2}
- \frac{1}{\Lambda_v^2 -\lambda^2} \bigg(\frac{1}{q^2 -\Lambda_v^2} - \frac{1}{q^2 -\lambda^2} \bigg)
 \bigg] \\
 &= \frac{1}{\Lambda_v^2 -\lambda^2} \bigg[ \frac{\partial}{\partial \Lambda_v^2}\frac{1}{q^2 -\Lambda_v^2}
- \frac{1}{\Lambda_v^2 -\lambda^2} \bigg(\frac{1}{q^2 -\Lambda_v^2} - \frac{1}{q^2 -\lambda^2} \bigg)
 \bigg] \,, 
\end{split}
\end{align}
where the derivative on the r.h.s. can be taken outside the regularised loop integral.

We extend the calculation presented in~\cite{Ignatov:2022iou} by including the FSR corrections and the electron mass effects in all contributions. Furthermore, the NLO calculation is matched to the PS algorithm described in Sec.~\ref{sec:ps} by using the same procedure introduced in Sec.~\ref{sec:nlops} for the factorised  approach. 

The calculation of the FSR correction follows the procedure described between Eq.~\eqref{LOOP:pionSE} and Eq.~\eqref{Eq:FSR_FF}. In order to apply Eq.~\eqref{eq:prop_fsr_neq} and Eq.~\eqref{eq:prop_fsr_eq} to simplify the calculation, we need to distinguish between the case $\Lambda_v \neq \Lambda_w$ and $\Lambda_v = \Lambda_w$. According to Eq.~\eqref{def:Delta_diff}, the virtual FSR correction in the GVMD approach can be written as
\begin{align}\label{eq:gvmd_fsr}
\begin{split}
    \delta_{V,\text{GVMD}}^\text{FSR}(\lambda) &= \frac{ 2\Re{F_\pi(s)^*\Amp_{\lo,\ptlike}^\dagger \, \Amp_{V,\text{GVMD}}^\text{FSR}(\lambda)}}{\left|F_\pi(s)\right|^2|\Amp_{\lo,\ptlike}|^2}  \\[2pt]
    &= \sum_{v=1}^{n_r} \sum_{w=1}^{n_r} \Re{\frac{c_v \hspace{1pt} c_w}{c_t^2}\,\Delta_{V,\text{GVMD}}^\text{FSR}(\Lambda_v^2,\Lambda_w^2)}
    \,.
\end{split}
\end{align}
In the case $\Lambda_v \neq \Lambda_w$, we obtain 
\begin{align}\label{eq:fsr_neqmass}
\Delta_{V,\text{GVMD}}^\text{FSR}(\Lambda_v^2,\Lambda_w^2) = \Bar\delta_V^{\hspace{1pt}\text{FSR}}(\lambda^2) +
\frac{1}{\Lambda^2_v - \Lambda^2_w} \bigg[
\Lambda_w^2 \, \Bar\delta_V^{\text{FSR}}(\Lambda_v^2)
- \Lambda_v^2 \, \Bar\delta_V^{\hspace{1pt}\text{FSR}}(\Lambda_w^2)
\bigg] \,,
\end{align}
while for $\Lambda_v =\Lambda_w$ we have
\begin{align}\label{eq:fsr_eqmass}
\Delta_{V,\text{GVMD}}^\text{FSR}(\Lambda_v^2,\Lambda_v^2) = \Bar\delta_V^{\hspace{1pt}\text{FSR}}(\lambda^2)
- \Bar\delta_V^{\hspace{1pt}\text{FSR}}(\Lambda_v^2)
+ \Lambda_v^2 \, \frac{\partial}{\partial \Lambda_v^2} \Bar\delta_V^{\text{FSR}}(\Lambda_v^2) \,,
\end{align}
where $\Bar\delta_V^{\hspace{1pt}\text{FSR}}(s')$ is given by Eq.~\eqref{eq:fsr_phmass}. We remark that Eq.~\eqref{eq:fsr_eqmass} is consistent with Eq.~\eqref{eq:fsr_neqmass}, as the former can be obtained by expanding the latter in the limit $\Lambda_w^2 \to \Lambda_v^2$. As expected, all IR singularities are contained in the point-like contribution $\delta_{V,\ptlike}^{\hspace{1pt}\text{FSR}}(\lambda)$, which can be recovered from Eq.~\eqref{eq:gvmd_fsr} by noting that
\begin{align}
\sum_{v=1}^{n_r} \sum_{w=1}^{n_r} \Re{\frac{c_v \hspace{1pt} c_w}{c_t^2}\,\Bar\delta_V^{\hspace{1pt}\text{FSR}}(\lambda^2)}
= \delta_{V,\ptlike}^{\hspace{1pt}\text{FSR}}(\lambda) \,,
\end{align}
for the normalisation condition $\sum_v c_v = \sum_w c_w = c_t$.

As discussed in~\cite{Ignatov:2022iou}, the computation of the two-photon exchange diagrams in the GVMD approach provides an improved theoretical description of the charge asymmetry, which is consistent with the CMD-3 measurement~\cite{CMD-3:2023alj}. According to Eq.~\eqref{def:Delta_diff}, the virtual IFI correction in the GVMD approach can be written as 
\begin{align}\label{eq:ifi_gvmd}
\begin{split}
    \delta_{V,\text{GVMD}}^\text{IFI}(\lambda) &= \frac{ 2\Re{F_\pi(s)^*\Amp_{\lo,\ptlike}^\dagger \, \Amp_{V,\text{GVMD}}^\text{IFI}(\lambda)}}{\left|F_\pi(s)\right|^2|\Amp_{\lo,\ptlike}|^2} \\[2pt]
    &= \sum_{v=1}^{n_r} \sum_{w=1}^{n_r} \Re{\frac{c_v \hspace{1pt} c_w}{c_t^2 F_\pi(s)}\,\Delta_{V,\rm{GVMD}}^\text{IFI}(\Lambda_v^2,\Lambda_w^2)} \,.
\end{split}
\end{align}
As a direct consequence of Eq.~\eqref{eq:prop_ifi}, we obtain the simple relation
\begin{align}\label{eq:deltagrossa_gvmd}
\Delta_{V,\text{GVMD}}^\text{IFI}(\Lambda_v^2,\Lambda_w^2) =
\Bar\delta_{V}^\text{\hspace{1pt}IFI}(\lambda^2,\lambda^2)
-\Bar\delta_{V}^\text{\hspace{1pt}IFI}(\Lambda_v^2,\lambda^2)
-\Bar\delta_{V}^\text{\hspace{1pt}IFI}(\lambda^2,\Lambda_w^2)
+\Bar\delta_{V}^\text{\hspace{1pt}IFI}(\Lambda_v^2,\Lambda_w^2) \,,
\end{align}
where $\Bar\delta_{V}^\text{\hspace{1pt}IFI}(s',s'')$ is given by Eq.~\eqref{Eq:IFI_phmass}. All terms are IR divergent apart from the last one. In the soft limit, the combination of the IR divergent terms, that we indicate with $|_\text{IR}$, gives 
\begin{equation}
\delta_{V,\text{GVMD}}^\ifi\biggr|_{\text{IR}}=\frac{1}{F_\pi(s)}\left\{\Bar{\delta}_V^\text{\hspace{1pt}ISR}(\lambda^2,\lambda^2)\biggr|_\text{IR}(F_\pi(s)+F_\pi(0)-1)\right\} \,,
\end{equation}
in which the $\lambda^2$ dependence exactly cancels with the real soft correction, using the condition $F_\pi(0)=1$.

The approximation $F_\pi(q^2) \simeq F_\pi^\text{BW}(q^2)$ is not only a limitation in the phenomenological description of the pion form factor. Since $\Im F_\pi^\text{BW}(q^2< 4m_\pi^2) \neq 0$, the GVMD approach does not naturally implement the unitarity of the scattering matrix. Although the sub-threshold imaginary part can be numerically reduced by choosing suitable input parameters ($c_v$, $m_v$, $\Gamma_v$), this is undeniably a theoretical limitation of the GVMD approach. A more formal treatment of the pion form factor is given by the FsQED approach, which will be discussed in the next section.

\subsection{NLO calculation in the FsQED approach}\label{sec:disp}

The pion vector form factor can be parameterised by means of a dispersion relation, as proposed in~\cite{Colangelo:2022}. Restricting to the elastic final state regime and assuming the normalisation $F_\pi(0)=1$, the pion form factor can be written as a once-subtracted dispersion relation
\begin{equation}
F_\pi(q^2) = 1 + \frac{q^2}{\pi} \int_{4m_\pi^2}^\infty \frac{\dd s'}{s'}\frac{\Im F_\pi(s')}{s'-q^2-i\varepsilon'}\, ,
\end{equation}
where $q$ is the momentum flowing in the photon propagator entering in a sQED vertex. The dispersion relation comes with the following sum rule
\begin{align}
\frac{1}{\pi} \int_{4m_\pi^2}^\infty \frac{\dd s'}{s'} \Im F_\pi(s')= 1\, ,
\label{eq:sumrule}
\end{align} 
which enforces $F_\pi(0)=1$ and implies that $F_\pi(s)$ vanishes for $s\to\infty$. In practice, the dispersion integral is computed numerically up to a certain high-energy cutoff $\Lambda^2$. Since the pion form factor parameterisation relies on experimental data, we take such a cutoff equal to the maximum centre-of-mass energy at which $F_\pi(q^2)$ is measured. 

In tree-level amplitudes, since the photon virtuality is fixed by the four-momentum conservation, the form factor is always factorised and evaluated at a given virtuality, \textit{i.e.} there is no gain in introducing its dispersive representation. When dealing with loops, the factor $F_\pi(q^2)/q^2$ has to be regularised to avoid IR divergences, so a small photon mass $\lambda$ is introduced in the following way
\begin{equation}
   \frac{ F_\pi(q^2)}{q^2}\,\to\,  \frac{1}{q^2-\lambda^2+i \varepsilon'} - \frac{1}{\pi} \int_{4m_\pi^2 - \lambda^2}^\infty \frac{\dd s'}{s'}\frac{\Im F_\pi(s'+\lambda^2)}{q^2-s' -\lambda^2+i\varepsilon'}\, ,
   \label{Fregularised}
\end{equation}
where the above relation holds under the standard dispersive assumptions:  $F_\pi(q^2)$ is analytic on the whole complex plane with the exception of a branch cut on the real axis for $q^2>4 m_\pi^2$ and $F_\pi(q^2)/q^2$ has null residue at $q^2\to\infty$. The latter condition holds as $F(\lambda^2)=1$. We emphasise that Eq.~\eqref{Fregularised} differs from Eq.~(2.4) of \cite{Colangelo:2022lzg} by  terms of ${\cal O}(\lambda^2)$ in the dispersive integral, which, however, are of no practical relevance in the limit $\lambda \to 0$. In particular, in the same limit, $F(\lambda^2) \to F(0) = 1$.

At first, we briefly discuss the computation for the final state radiation. The diagrams of Fig.~\ref{FIG:FSR_FF}, in addition to the tree level structure, present a photon propagator and two form factors. We recall that the dispersive representation of the pion form factor is introduced to enable the calculation of loop integrals, thus it is not used for any form factor independent of loop momenta. The FSR correction due to the diagrams represented in Fig.~\ref{FIG:FSR_FF} can be written as 
\begin{equation}\Tilde{\delta}_{V,\disp}^\fsr(\lambda) =\frac{(2\pi \mu)^{4-D}}{i\pi^2} 2\Re\int \dd^D q \frac{ \Amp_{\lo,\ptlike}^\dagger \, \overline{\Amp}_{V,\disp}^\fsr(q,\lambda)}{|\Amp_{\lo,\ptlike}|^2}\frac{F^2_\pi(q^2)}{q^2-\lambda^2+i\varepsilon}\, ,
\end{equation}
where $\overline{\Amp}_{V}^\fsr(q,\lambda)$ is the sum of the FSR amplitudes from which we have factored out the photon propagator and the form factors. The UV divergence is regulated by $D=4-2\epsilon_\text{UV}$ space-time dimensions. The $s$-channel form factor in the virtual amplitude in interference with the Born diagram cancels against $1/|F_\pi(s)|^2$ from the definition of the correction. The double insertion of the dispersion relation yields, in addition to the point-like contribution, two pieces in which the photon propagator $(q^2-\lambda^2)^{-1}$ is replaced by the dispersive one, effectively representing a massive photon 
\begin{equation}\label{Eq:FSRdecomposition}
\begin{aligned}
\hspace{-1mm}\Tilde{\delta}_{V,\disp}^\fsr =&2\hspace{1pt}\mathcal{C}_D \Re\!\int\!\dd^D q \frac{ \Amp_{\lo,\ptlike}^\dagger \, \overline{\Amp}_{V,\disp}^\fsr(q,\lambda)}{|\Amp_{\lo,\ptlike}|^2}\biggl[
     \frac{1}{q^2-\lambda^2}-\frac{2}{\pi}\int_{4m_\pi^2}^\infty\frac{\dd s' }{s' }\frac{\Im F_\pi(s')}{q^2-\lambda^2-s'+i\varepsilon'}\\
   &  +\frac{1}{\pi^2}\int_{\Omega_\infty}\frac{\dd s' }{s' }\frac{\dd s'' }{s'' }\frac{\Im F_\pi(s')\Im F_\pi(s'')}{s''-s'-i\varepsilon''+i\varepsilon' }\left(\frac{s''}{q^2-\lambda^2-s''+i\varepsilon''}-\frac{s'}{q^2-\lambda^2-s'+i\varepsilon'}\right)
     \biggr]\, ,
\end{aligned}
\end{equation}
where $\Omega_\infty=[4m_\pi^2,\infty)\times [4m_\pi^2,\infty)$. Here we can neglect $\lambda^2$ in the argument of $\Im F_\pi$ and poles are dealt with the $+i\varepsilon$ prescription. The factor $\mathcal{C}_D$ is defined as $\mathcal{C}_D=\frac{(2\pi \mu)^{4-D}}{i\pi^2}$. The loop and dispersive integrals can be exchanged, yielding a dispersive integral of the massive photon amplitudes. In order to get a UV-finite result, the above correction has to be summed to the pion wavefunction counterterm, for which the decomposition of Eq.~\eqref{Eq:FSRdecomposition} is still valid. We obtain, after adding the proper UV counterterm $2\delta Z_\phi(s')$ in each term,
\begin{equation}
\begin{aligned}
{\delta}^\fsr_{V,\disp} =&    \biggl\{
    \Bar{\delta}_V^\fsr(0)-\frac{2}{\pi}\int_{4m_\pi^2}^\infty
\frac{\dd s' }{s' }\Im F_\pi(s')\Bar{\delta}_V^\fsr(s')\\
   &  +\frac{1}{\pi^2}\int_{\Omega_\infty}\dd s'\frac{\dd s'' }{s'' }\frac{\Im F_\pi(s')\Im F_\pi(s'')}{s''-s'-i\varepsilon''+i\varepsilon' }\left( \Bar{\delta}_V^\fsr(s'') s'' - \Bar{\delta}_V^\fsr(s')s'\right)
     \biggr\}\, ,
\end{aligned}
\end{equation}
where the expression for the massive FSR kernel $\Bar{\delta}_V^\fsr(s')$ is given in Eq.~\eqref{eq:fsr_phmass}. 

An analogous discussion can be done for the box diagrams of Fig.~\ref{FIG:IFI_FF}. Those diagrams present two photon propagators with as many form factors evaluated at different virtualities: for this reason, it is convenient to think in terms of $F_\pi(q^2)/q^2$ acting as a modification of the photon propagator. Therefore, IFI  diagrams are the sum of three different contributions, given by the presence of zero, one or two dispersive integrations representing the same number of massive photons. Thus, one can extract the form factor coming from the Born amplitude from the rest of the correction, recasting each contribution as
\begin{equation}\label{def:deltabar}
 \delta_{V,\disp}^\ifi = \frac{ 2\Re{F_\pi^*(s)\Amp_{\lo,\ptlike}^\dagger \, \Amp_{V,\disp}^\ifi}}{\left|F_\pi(s)\right|^2|\Amp_{\lo,\ptlike}|^2} \equiv  \frac{\Re{F^*_\pi(s)\Delta_{V,\disp}^\ifi}}{|F_\pi(s)|^2}\, .
\end{equation}
Therefore, the expression for $\Delta_{V,\disp}^\ifi$ as a sum of the (dispersively integrated) kernels $\Bar{\delta}^\ifi_V(s',s'')$ of Eq.~\eqref{Eq:IFI_phmass}, reads
\begin{equation}
    \begin{aligned}
\Delta_{V,\disp}^\ifi=\Bar{\delta}^\ifi_V(\lambda^2,\lambda^2) 
&-\frac{1}{\pi} \int_{4 m_\pi^2}^{\infty} \frac{\dd s'}{s'} \Im F_\pi(s') \left[\Bar{\delta}^\ifi_V(s',\lambda^2)+\Bar{\delta}^\ifi_V(\lambda^2,s')\right]\\
     &+  \frac{1}{\pi^2}\int_{\Omega_\infty}
\frac{\dd s'}{s'}  \frac{\dd s''}{s''}\Im F(s') \Im F(s'')\Bar{\delta}^\ifi_V(s', s'')\, .
    \end{aligned} \label{eq:realdisp}
\end{equation}
Many subtleties are hidden in this decomposition: at first, conversely to the case of FSR, the pole-pole contribution does not reproduce the sQED correction. This is due to the absence of a third form factor in box diagrams, so the factor $F_\pi^*(s)$ associated with the tree-level diagram is not matched by the virtual amplitude. The IR structure of the IFI in the FsQED approach needs to be treated with care. By explicitly writing the real part of the correction in Eq.~\eqref{def:deltabar} 
\begin{equation}\label{Eq:FF_decomposition} 
    \delta_{V,\disp}^\ifi =\frac{1}{|F_\pi(s)|^2}\left[\Re F_\pi(s)\Re\Delta_{V,\disp}^\ifi+ \Im F_\pi(s)\Im\Delta_{V,\disp}^\ifi\right]
\end{equation}
we see that the IR divergence is spread  between the real and imaginary parts of the correction, which we indicate with the capital delta $\Delta$ notation. 

In order to single out the IR singularities in each contribution, we start with the analysis of the IR structure of the real part of the correction, $\Re\Delta_{V,\disp}^\ifi$, as the pole-pole IR contribution can be directly obtained by the substitution $s' \to \lambda^2$. Moreover, we note that the correction kernel is symmetric, \textit{i.e.} $\Bar{\delta}_V^\ifi(\lambda^2,s')=\Bar{\delta}_V^\ifi(s',\lambda^2)$. The divergence arises in two regions of the loop integration, namely for $q\to 0$ and $q \to p_3+p_4$:
\begin{equation}
\begin{aligned}
  \Bar{\delta}^\ifi_{V,\text{IR}}(\lambda^2,s') =& \frac{\alpha}{\pi}\frac{ s}{s-s'-\lambda^2+i\varepsilon}\times\\
  &\bigg[\left(m_e^2+m_\pi^2-t\right) \mathcal{C}_0^{e,\pi}(t,\lambda^2)-\left(m_e^2+m_\pi^2-u\right) \mathcal{C}_0^{e,\pi}(u,\lambda^2)\bigg]_\text{IR}\\
    =&\frac{s}{2(s-s'+i\varepsilon')}\mathcal{C}_\text{IR}\log\frac{\lambda^2}{s}\,. 
\end{aligned}
\end{equation}
Since the {\em logarithmic} IR divergence is contained in the $\text{C}_0$ functions, we can safely neglect the terms of ${\cal O}(\lambda^2)$ in the dispersive denominator. 

Moreover, the IR-divergent piece of the pole-pole contribution is given by 
\begin{equation}
\Bar{\delta}_{V,\text{IR}}^\ifi(\lambda^2,\lambda^2) =\mathcal{C}_\text{IR}\log\frac{\lambda^2}{s}\,.
\end{equation}
We can add and subtract the IR-divergent part of each correction kernel, obtaining the sum of the pure IR divergence and a finite contribution
\begin{equation}
\begin{aligned}
    \Re\Delta_{V,\disp}^\ifi &= \Re\Bar{\delta}_V^\ifi(\lambda^2,\lambda^2)-\Re\Bar{\delta}^\ifi_{V,\text{IR}}(\lambda^2,\lambda^2)+\Re\Bar{\delta}^\ifi_{V,\text{IR}}(\lambda^2,\lambda^2)  \\
    &- \frac{2}{\pi}\Re\int_{4m_\pi^2}^\infty \frac{\dd s'}{s'}\Im F_\pi(s') \left[ \Bar{\delta}^\ifi_V(\lambda^2,s') -  \Bar{\delta}^\ifi_{V,\text{IR}}(\lambda^2,s')+ \Bar{\delta}^\ifi_{V,\text{IR}}(\lambda^2,s')\right]\\
    & +  \frac{1}{\pi^2}\int_{\Omega_\infty}\frac{\dd s'}{s'}  \frac{\dd s''}{s''}\Im F(s') \Im F(s'')\Re\Bar{\delta}_V^\ifi(s', s'')\,.
    \end{aligned}
    \label{eq:deltaVinterm}
\end{equation}
The integral in the second line exhibits a pole around $s'=s$ for the pole-dispersive part, while the double dispersive is free from poles. The integration around the pole is dealt with the $i\varepsilon$ prescription, inherited from the dispersive integral\footnote{In the preliminary version of this work, we omitted the term proportional to $\Im f(s_\pm)$ in Eq.~\eqref{eq:deltaVinterm2}. We thank Fedor Ignatov for noticing this error.}, given by 
\begin{equation}\label{Eq:Poles}
  \lim_{\varepsilon'\to 0_+}\Re\int {\dd s'} \frac{f(s')}{s-s'+i\varepsilon'} = \text{P.V.}\int \left(\frac{\Re f(s')}{s-s'}\right) +\frac{\pi}{2} \Im f(s_+) +\frac{\pi}{2} \Im f(s_-)\, ,
\end{equation}
where care is taken to deal with a function which is discontinuous exactly on the pole location. We use Eq.~\eqref{Eq:Poles} with the integrand 
\begin{equation}
    f(s')=-\frac{2}{\pi}\frac{ \Im F_\pi(s')}{s'}\,  [\Bar{\delta}_V^\ifi(\lambda^2,s')-\Bar{\delta}_{V,\text{IR}}^\ifi(\lambda^2,s')](s-s'+i\varepsilon') 
\end{equation} whose imaginary part vanishes at $s_+$ and $\Im f(s_\pm)=\lim_{s'\to s_\pm}\Im f(s')$.
The real part of the IR-divergent pole-dispersive contribution, i.e. the third term of the second line in Eq~\eqref{eq:deltaVinterm} is extracted as  
\begin{equation}
\begin{aligned}
    -\frac{2}{\pi}\Re\int_{4m_\pi^2}^\infty \frac{\dd s'}{s'}\Im F_\pi(s') \Bar{\delta}_{V,\text{IR}}^\ifi(\lambda^2,s')=&    -\frac{1}{\pi}\Re\int_{4m_\pi^2}^\infty \frac{\dd s'}{s'}\frac{\Im F_\pi(s')}{s-s'+i\varepsilon'}\mathcal{C}_\text{IR}\log\frac{\lambda^2}{s}\\
    =&\left(\Re F_\pi(s)-1\right) \mathcal{C}_\text{IR}\log\frac{\lambda^2}{s} \,,
\end{aligned}
    \end{equation}
where we recognised the real part of the dispersion relation and the principal value is taken at $s'=s$. The IR term of the pole-dispersive contribution which is linear in the real part of the form factor exactly cancels the divergence of the pole-pole term, namely the third term in the first row of Eq.~\eqref{eq:deltaVinterm}. Therefore, we have extracted the divergence of the real part of the IFI correction in a quite immediate way
\begin{equation}
\begin{aligned}
\Re\Delta_{V,\disp}^\ifi=    & \Re F_\pi(s)\mathcal{C}_\text{IR}\log\frac{\lambda^2}{s}\\ &+\left[\Re\Bar{\delta}_V^\ifi(\lambda^2,\lambda^2)-\Re\Bar{\delta}^\ifi_{V,\text{IR}}(\lambda^2,\lambda^2)\right]_\text{fin.} \\
    &- \frac{2}{\pi}\,\textrm{P.V.}\int_{4m_\pi^2}^\infty \frac{\dd s'}{s'}\Im F_\pi(s') \left[\Re \Bar{\delta}^\ifi_V(\lambda^2,s') -  \Re\delta^\ifi_{V,\text{IR}}(\lambda^2,s')\right]_\text{fin.}\\
    &- \frac{\Im F_\pi(s)}{s}\left[\lim_{s'\to s^-}\Im\bar{\delta}_V^\text{IFI}(\lambda^2,s')(s-s')\right]\\
    & +  \frac{1}{\pi^2}\int_{\Omega_\infty} \frac{\dd s'}{s'}  \frac{\dd s''}{s''}\Im F(s') \Im F(s'')\Re\Bar{\delta}^\ifi_V(s', s'')\,,
    \end{aligned}
    \label{eq:deltaVinterm2}
\end{equation}
where the parentheses $[...]_\text{fin.}$ in Eq.~\eqref{eq:deltaVinterm2} contain IR-finite combinations \footnote{The pole-dispersive contribution of Eq.~\eqref{eq:deltaVinterm2} differs  from Eq.~(3.10) of~\cite{Colangelo:2022lzg} because of the term $- \Im F_\pi(s)/s\left[\lim_{s'\to s^-}\Im\bar{\delta}_V^\text{IFI}(\lambda^2,s')\right]$. The two expressions would coincide, up to terms that vanish in the $m_e\to 0$ limit, 
if the pole prescription~\eqref{Eq:Poles} had also been adopted in~\cite{Colangelo:2022lzg}.}.

We emphasise that the above discussion assumes that the limit $\lambda \to 0  $ (or equivalently $\epsilon_\text{IR}\to 0$ in dimensional regularisation) is taken \textit{before} the limit $\varepsilon'\to 0$  in the dispersive integrals. We have verified numerically that, in the case of photon mass regularisation $\lambda \neq 0$, reversing the order of the limits we obtain the same result within the numerical accuracy of the MC integration. Notice that, keeping a {\em finite} photon mass, in Eq.~\eqref{eq:realdisp} we need to replace $\Bar{\delta}^\ifi_V(\lambda^2,s')$ with $ \Bar{\delta}^\ifi_V(\lambda^2,s'+\lambda^2)$ and
$\Bar{\delta}^\ifi_V(s',s'')$ with $ \Bar{\delta}^\ifi_V(s'+\lambda^2,s''+\lambda^2)$. We have also verified that a shift of order $\lambda^2$ in the argument of $\Im F(s')$, namely replacing $\Im F(s')$ with $ \Im F(s'+\lambda^2)$, does not affect the relevant limit. 

Extracting the divergence of the imaginary part of the box diagrams in the FsQED approach is way more challenging. This is because the IR divergence is not contained in the imaginary part of the amplitude itself but emerges from the dispersive integration, due to the \textit{"end-point singularity"}, as pointed out in~\cite{Colangelo:2022lzg}. Similarly to the real part, one can add and subtract $\Im F_\pi(s) \Im \Bar{\delta}^\ifi_V(\lambda^2,s')$ in the pole-dispersive integrand 
\begin{equation}\label{Eq:IFI_disp}
    \begin{aligned}
\Im\Delta_{V,\disp}^\ifi=\Im\Bar{\delta}^\ifi_V(\lambda^2,\lambda^2) &-\frac{2}{\pi} \Im\int_{4 m_\pi^2}^{s} \frac{\dd s'}{s'} \left(\Im F_\pi(s') - \Im F_\pi(s) \right)\Bar{\delta}_V^\ifi(\lambda^2,s')\\
& -\frac{2}{\pi}\Im F_\pi(s) \Im \int_{4 m_\pi^2}^{s} \frac{\dd s'}{s'}  \Bar{\delta}^\ifi_V(\lambda^2,s')\\
  &   +  \frac{1}{\pi^2}\int_{\Omega_s}\frac{\dd s'}{s'}  \frac{\dd s''}{s''}\Im F(s') \Im F(s'')\Im\Bar{\delta}_V^\ifi(s', s'')\, .
    \end{aligned}
\end{equation}
where $\Omega_s=[4m_\pi^2,s)\times [4m_\pi^2,s)$. Conversely to the real part, the imaginary part is dispersively integrated up to $s$, since it vanishes for $s'>s$ due to the optical theorem.

In the above equation, the double pole and double dispersive contribution are IR-finite. The subtracted term is free of any divergence because of the vanishing difference of the two form factors in the limit $s'\to s$. Therefore, we can isolate the term containing the IR divergence as 
\begin{equation}\label{Eq:IFI_disp_IR}
 \Im\Delta_{V,\disp}^\ifi \Big|^{\text{pole-disp}}_{\text{IR}}= -\frac{2}{\pi}\Im F_\pi(s)  \Im \int_{4 m_\pi^2}^{s} \frac{\dd s'}{s'} \Bar{\delta}^\ifi_V(\lambda^2,s')\, ,
\end{equation}
where $\Bar{\delta}^\ifi_V(\lambda^2,s')$ has the same functional form as the expression given in Eq.~\eqref{Eq:IFI_phmass} with the substitution $(s',s'')~\to~(\lambda^2,s')$. 
By inspection of Eq.~\eqref{Eq:IFI_phmass}, we can isolate the contributions of the pole-dispersive part which have a non-zero imaginary part.

The imaginary part of the two- and three-point functions depending on $s$ can be easily computed. We find 
\begin{subequations}
    \begin{align}
          \Im \text{B}_0(s,s',\lambda^2) &= \pi\left(1-\frac{s'}{s}\right)\, , \\
    \Im \mathcal{C}_0^{p,p}(s')&  =\frac{1}{s\beta_p}\log\frac{1-\beta_p}{1+\beta_p}\, , \qquad p=e,\pi\, .
\end{align}
\end{subequations}
Since the dependence on $s'$ in these expressions is at most linear, no particular attention is required for their dispersive integration that yields finite pieces. 

The IR divergence is entirely contained in the dispersive integral of the $\text{D}_0$ function, due to the end-point singularity. From the second row of Eq.~\eqref{Eq:IFI_disp} and the expression of $\Bar{\delta}_V^\ifi(\lambda^2,s')$ that can be read off Eq.~\eqref{Eq:IFI_phmass}, we see that the terms proportional to $s'$ in the coefficient of the four-point functions cancel the dispersive mass at the denominator. To this end, we identify two integrals whose explicit computation can be found in Appendix~\ref{app:dimreg} 
\begin{subequations}\label{EQ:disp_ints_sec3}
    \begin{align}
 \mathcal{I}_{1/s'}(x) &=  \Im\int_{4m_\pi^2}^s \frac{\dd s'}{s' } \mathcal{D}_0^{e,\pi}(x,s',s'')= \frac{1}{4 s}\mathcal{L}(x) + \mathcal{I}_2(x) \, , \\
   \mathcal{I}_{s'/s'}(x)& =  \Im\int_{4m_\pi^2}^s \dd s'\mathcal{D}_0^{e,\pi}(x,s',s'')= \frac{1}{4}\mathcal{L}(x) \,,
    \end{align}
\end{subequations}
with $x=t,u$. The symbol $\mathcal{L}(x)$ stands for the eikonal integral, as defined in~\cite{tHooft:1978jhc}, using the conventions of~\cite{Denner:1991kt}. When computing the above integrals, one has to be very careful in handling the finite terms arising from the $\lambda\to 0$ expansion, as discussed in detail in Appendix~\ref{app:dimreg}. We also remark that, as an internal check of the independence of the result from the regularisation scheme used for IR singularities, we obtained exactly the same result with dimensional regularisation, translating the $1/\epsilon_\text{IR}$ poles into $\log\lambda$ \cite{Dittmaier:1999mb}, as in Eq.~\eqref{eq:translation_rule}. We additionally verified that the complete result for the IFI correction is free from collinear singularities if we take the limit $m_e\to0$.
The finite terms found in the present work differ from the result of Eq.~(3.8) in~\cite{Colangelo:2022lzg} 
originally obtained with photon mass regularisation. However, our computation in the limit $m_e\to 0$ agrees with the latest result obtained by the same authors in dimensional regularisation~\cite{Colangelo:2022lzg}.

The full expression of the IR-divergent pole-dispersive imaginary part, namely the second row of Eq.~\eqref{Eq:IFI_disp}, is given by 
\begin{equation}
\Im \Delta_{V,\disp}^\ifi \Big|^{\text{pole-disp}}_{\text{IR}}=-\frac{2}{\pi}\Im F_\pi(s)\biggl\{C_{1/s'}(t)\,\mathcal{I}_{1/s'}(t)+C_{s'/s'}(t)\,\mathcal{I}_{s'/s'}(t) - (t\to u) \biggr\}\, ,
\end{equation}
where the coefficients are given by
\begin{subequations}
\begin{align}
    C_{1/s'}(t)&=\frac{4 s \left(m_e^2+m_\pi^2-t\right) \left(m_e^2 \left(2 m_\pi^2-t+u\right)+2 m_\pi^4-2 m_\pi^2 t+t \left(t-u\right)\right)}{4 m_\pi^2 s-s^2+(t-u)^2}\, , \\
    C_{s'/s'}(t)&=-\frac{4 s \left(m_e^2-t\right) \left(m_e^2+m_\pi^2-t\right)}{4 m_\pi^2 s-s^2+(t-u)^2}
\end{align}
\end{subequations}
and the integrals have been calculated above. 
Summing all contributions, we find that the IR coefficient for the imaginary part is 
\begin{equation}
  \Im F_\pi(s)\, \mathcal{C}_\text{IR}\log\frac{\lambda^2}{s}\, .
\end{equation}
Gathering the IR-divergent parts coming from $\Re\Delta_{V,\disp}^\ifi$ and $\Im\Delta_{V,\disp}^\ifi$ and inserting them in Eq.~\eqref{Eq:FF_decomposition}, we find
\begin{equation}
\delta^\ifi_{V, \disp}\Big|_{\text{IR}} =\left( \frac{|\Re F_\pi(s)|^2}{|F_\pi(s)|^2}\, \mathcal{C}_\text{IR}+\frac{|\Im F_\pi(s)|^2}{|F_\pi(s)|^2}\, \mathcal{C}_\text{IR}\right) \log\frac{\lambda^2}{s}= \mathcal{C}_\text{IR} \log\frac{\lambda^2}{s}\, ,
\end{equation}
that cancels against the soft IR coefficient of Eq.~\eqref{EQ:IR_SOFT}.

All the IR-finite integrals needed in this section are computed via MC integration, generating one value for the $s'$ and $s''$ variables for each point of the phase space. As for the other approaches, the NLO calculation in the FsQED approach is matched to the PS.

\section{Numerical results}
\label{sec:res}
This section shows illustrative numerical results for the $e^+e^-\to \pi^+\pi^-$ scattering, obtained with the updated version of the \textsc{BabaYaga@NLO} event generator. 
After defining the input parameters and the event selection criteria in Sec.~\ref{pionff_eventselection}, the results obtained for the integrated cross section are exhibited in Sec.~\ref{resFsQED}. 
The results for the differential cross section of experimentally relevant observables are reported in Sec.~\ref{res_scalar_differential}. 
Lastly, in Sec.~\ref{res_asymmetry}, we show the impact on the charge asymmetry of the different approaches to the pion form factor.

\subsection{Pion form factor and event selection}\label{pionff_eventselection}

The choice of a suitable parameterisation for the pion form factor $F_\pi(q^2)$ is needed to obtain realistic numerical results for the $e^+e^- \to \pi^+\pi^-$ scattering. 
To this end, several form factors have been implemented in \textsc{BabaYaga@NLO}. The list includes parameterisation inspired by the experimental determinations of BaBar~\cite{BaBar:2012bdw}, BESIII~\cite{BESIII:2015equ}, CMD-2~\cite{CMD-2:2001ski,CMD-2:2005mvb,Aulchenko:2006dxz,CMD-2:2006gxt}, CMD-3~\cite{CMD-3:2023alj,Ignatov:2023b}, KLOE~\cite{KLOE:2004lnj,KLOE:2008fmq,KLOE:2010qei,KLOE:2012anl,KLOE-2:2017fda}, and SND~\cite{Achasov:2006vp,SND:2020nwa}, as well as the combinations published in~\cite{Czyz:2010hj,Colangelo:2018mtw}. 
Any other form factor can be easily implemented in the code, either as an analytical function or as a numerical table.

Since this article focuses on energy scan experiments, we choose a parameterisation based on the CMD-3 measurement to illustrate the numerical results of our implementation of the $e^+e^- \to \pi^+\pi^-$ process. 
In addition, the CMD-2 measurement is used to study possible systematic effects on the radiative corrections due to different values of the input form factor. 
Nevertheless, our discussion is valid for any other form factor.

As illustrated in Sec.~\ref{sec:intffpi}, each approach to the pion form factor requires some constraints on the parameterisation of the $F_\pi(q^2)$ function. 
Any form factor that respects the condition $F_\pi(0)=1$ can be used in the factorised approximation. 
The GVMD is certainly the more demanding approach because it needs a sum of BW functions as form factor parameterisation, as specified in Eq.~\eqref{eq:bwsum}. 
On top of the $F_\pi(0)=1$ condition, the FsQED approach also requires the validity of the dispersive sum rule, as defined in Eq.~\eqref{eq:sumrule}.

For the factorised and FsQED approaches, we write $F_\pi(s)$ as a sum of Gounaris--Sakurai~(GS) functions~\cite{Gounaris:1968mw}, as usually done in experimental fits. 
Since we are interested in energies smaller than $1.2$~GeV, we consider the resonance of the $\rho$, $\rho'$ and $\rho''$ mesons, as well as the $\rho-\omega$ and $\rho-\phi$ interferences. 
The explicit parameterisation reads
\begin{equation}
\begin{aligned}
\label{eq:fgs}
F_\pi^\textrm{GS}(q^2) =\frac{1}{1+c_{\rho '}+c_{\rho ''}} \Bigg[& \Biggl(1+ \sum_{v=\omega,\phi}c_v\,\frac{q^2}{m_v^2}\textrm{BW}_v\left(q^2\right)\Biggr) \textrm{BW}_\rho^\textrm{GS}\left(q^2\right)\\
&+ c_{\rho'}\, \textrm{BW}_{\rho'}^\textrm{GS}\left(q^2\right)+ c_{\rho''}\, \textrm{BW}_{\rho''}^\textrm{GS}\left(q^2\right) \Bigg] \,,
\end{aligned}
\end{equation}
where $\textrm{BW}_v(q^2)$ denotes a Breit-Wigner (BW) function with mass $m_v$ and width $\Gamma_v$, namely
\begin{align}
\textrm{BW}_v(q^2) = \frac{m_v^2}{m_v^2 - i m_v \Gamma_v - q^2 } \,,
\end{align}
while $\textrm{BW}^\textrm{GS}_v(q^2)$ indicates a GS function with the same mass and width, namely
\begin{align}
\textrm{BW}_v^\textrm{GS}(q^2) = \frac{m_v^2 + d(m_v)\,m_v\,\Gamma_v}{m_v^2 - q^2 + f(q^2,m_v,\Gamma_v) - i\, m_v\, \Gamma(q^2,m_v,\Gamma_v)} \,,
\end{align}
where the definition of the auxiliary functions $d(m_v)$, $f(q^2,m_v,\Gamma_v)$, and $\Gamma(q^2,m_v,\Gamma_v)$ can be found in~\cite{Gounaris:1968mw}. 
The amplitude of each resonance is a complex number, \textit{i.e.} ${c_v = |c_v|\,e^{i\varphi_v}}$. 
For the GVMD approach, we approximate $F_\pi(q^2)$ as a sum of BW functions, namely
\begin{align}
\begin{aligned}
\label{eq:fbw}
F_\pi^\textrm{BW}(q^2) = \frac{\textrm{BW}_\rho (q^2) + c_\omega \textrm{BW}_\omega (q^2) + c_\phi \textrm{BW}_\phi (q^2) + c_{\rho'} \textrm{BW}_{\rho'} (q^2)+ c_{\rho''} \textrm{BW}_{\rho''} (q^2)}{1+c_\omega+c_\phi+c_{\rho'}+c_{\rho''}} \,.
\end{aligned}
\end{align}

We choose the input parameters according to the CMD-3 measurement or alternatively to the former CMD-2 determination. 
Following~\cite{CMD-3:2023alj}, the parameterisation for CMD-3 also takes into account the CMD-2 measurement for $1.1~\rm{GeV}<\sqrt{s}<1.35 $~GeV~\cite{CMD-2:2005mvb} and the DM-2 data for $\sqrt{s}\geq1.35$~GeV~\cite{1989321}.
On the other hand, the CMD-2 parameterisation considers the full dataset of the CMD-2 experiment for the $e^+e^- \to \pi^+\pi^-$ channel~\cite{CMD-2:2001ski,CMD-2:2005mvb,Aulchenko:2006dxz,CMD-2:2006gxt}.
The $\rho-\phi$ interference and the $\rho''$ resonance are considered only for CMD-3, as they are not visible in the CMD-2 data. 
Although the parameters are determined by fitting the experimental data, we do not claim that $F_\pi^\textrm{GS}(q^2)$ or $F_\pi^\textrm{BW}(q^2)$ are a proper determination of the pion form factor. 
They have to be understood as fixed parameterisations inspired by real data, used to study the impact of the radiative corrections and the effect of the different form factor treatments in a realistic scenario. 
Hence, we do not analyse the discrepancy with experimental data or give an error to the input parameters. 
We also remark that the specific parameterisations shown in this article do not affect the adopted solutions for the simulation of $e^+e^-\to\pi^+\pi^-$, as the pion form factor is treated as a user input in \textsc{BabaYaga@NLO}.

The chosen input parameters are reported in Tab.~\ref{tab:ffpi-param}, while Fig.~\ref{fig:ffpi-inputs} shows a qualitative comparison between the various parameterisations and the experimental data for both ${\text{CMD-2}}$ and ${\text{CMD-3}}$. 
The parameters were chosen prioritising the formal requirements of each form factor approach rather than the data description.\footnote{This is also the reason why we do not use the fit published by the CMD-3 collaboration in~\cite{CMD-3:2023alj}. 
Although their fit function is very similar to~\eqref{eq:fgs}, it has a constant additive term, which is not allowed by the dispersive sum rule~\eqref{eq:sumrule}.} 
In this regard, both versions of $F_\pi^\textrm{GS}(q^2)$ satisfy the dispersive sum rule~\eqref{eq:sumrule} and the sub-threshold condition $\Im F_\pi(q^2<4m_\pi^2) = 0$ with a per mille accuracy, which is sufficient for our purpose. 
On the other hand, both versions of $F_\pi^\textrm{BW}(q^2)$ respect the latter condition only at the per cent level due to the analytical structure of the BW function.
The dispersive sum rule is also enforced on $F_\pi^\textrm{BW}(q^2)$ at the 1\% level for CMD-2 and the at the 0.1\% for CMD-3.
Although this is not explicitly required by the GVMD approach, such a condition ensures to obtain a consistent separation between the real and imaginary part of $F_\pi(q^2)$.

As can be seen in Fig.~\ref{fig:ffpi-inputs}, the relative difference between $F_\pi^\textrm{GS}(q^2)$ and $F_\pi^\textrm{BW}(q^2)$ for CMD-2 is about 1\% at the centre of the $\rho$ resonance and in the range  5-10\% in the tails. 
The difference is further reduced by a factor of 2\hspace{1pt}-\hspace{1pt}5, depending on the energy region, for CMD-3. 
These discrepancies can be relevant when computing the integrated cross section, but lose importance when calculating ratios such as the forward-backward asymmetry.

The numerical values of the physical constants are set to
\begin{gather}
\alpha = 1/137.03599908, \quad
m_e = 0.51099895~\text{MeV}, \quad
m_\pi = 139.57039~\text{MeV}.
\end{gather}

\begin{table}
\begin{center}
\begin{tabular}{cc@{\hspace{0.4cm}}ccc>{\hspace{0.4cm}}ccccc}
 \multicolumn{2}{c}{} & 
 \multicolumn{3}{c}{\textbf{CMD-2}} &
 \multicolumn{4}{c}{\textbf{CMD-3}}\\
 \multicolumn{2}{c}{} & 
 $\rho$ &
 $\omega$ &
 $\rho'$ &
{$\rho$} &
{$\omega$} &
{$\phi$} &
{$\rho'$} &
{$\rho''$}\\[5pt]
 \toprule
\parbox[t]{2mm}{\multirow{4}{*}{\rotatebox[origin=c]{90}{\textbf{GS}}}} 

& $m_v$ & 775.49 & 782.66 & 1369.8 
        & 773.98 & 782.22 & 1019.5 & 1456.7 & 1870.74\\

&$\Gamma_v$  & 145.70 & 8.560   &  385.21
             &  147.86 & 8.174   & 5.275 & 524.05 & 170.49 \\

 &$|c_v|$ & - &  0.0016 & 0.0887
          & - & 0.0016 & 0.00059 & 0.097 & 0.037\\
 
 &$\varphi_v$ & - & 0.179 & 3.159
              & - & 0.057 & 2.836 & 3.541 & 2.277 \\ 

\addlinespace[5pt]\cmidrule{1-10}\addlinespace[4pt]
\parbox[t]{2mm}{\multirow{4}{*}{\rotatebox[origin=c]{90}{\textbf{BW}}}} 

& $m_v$ & 758.08 & 782.80 & 1253.8 
        & 755.71 & 782.07 & 1019.5 & 1338.64 & 1745.02\\

&$\Gamma_v$  & 136.81 & 8.004   &  530.86
             &  142.86 & 7.997   & 6.251 & 982.29  & 397.85\\

 &$|c_v|$ & - &  0.0079 & 0.144
          & - & 0.0085 & 0.00089 & 0.259 & 0.098 \\
 
 &$\varphi_v$ & - & 2.014 & 3.021
              & - & 1.782 & 5.561 & 3.340 & 0.817\\

 \bottomrule
\end{tabular}
\end{center}
\caption{Input parameters for the pion vector form factor $F_\pi(q^2)$ written as a sum of GS or BW functions using CMD-2 or CMD-3 scenarios as described in the text. Masses and widths are reported in MeV.}
\label{tab:ffpi-param}
\end{table}

\begin{figure}[t]
    \centering
    \includegraphics[width =\textwidth]{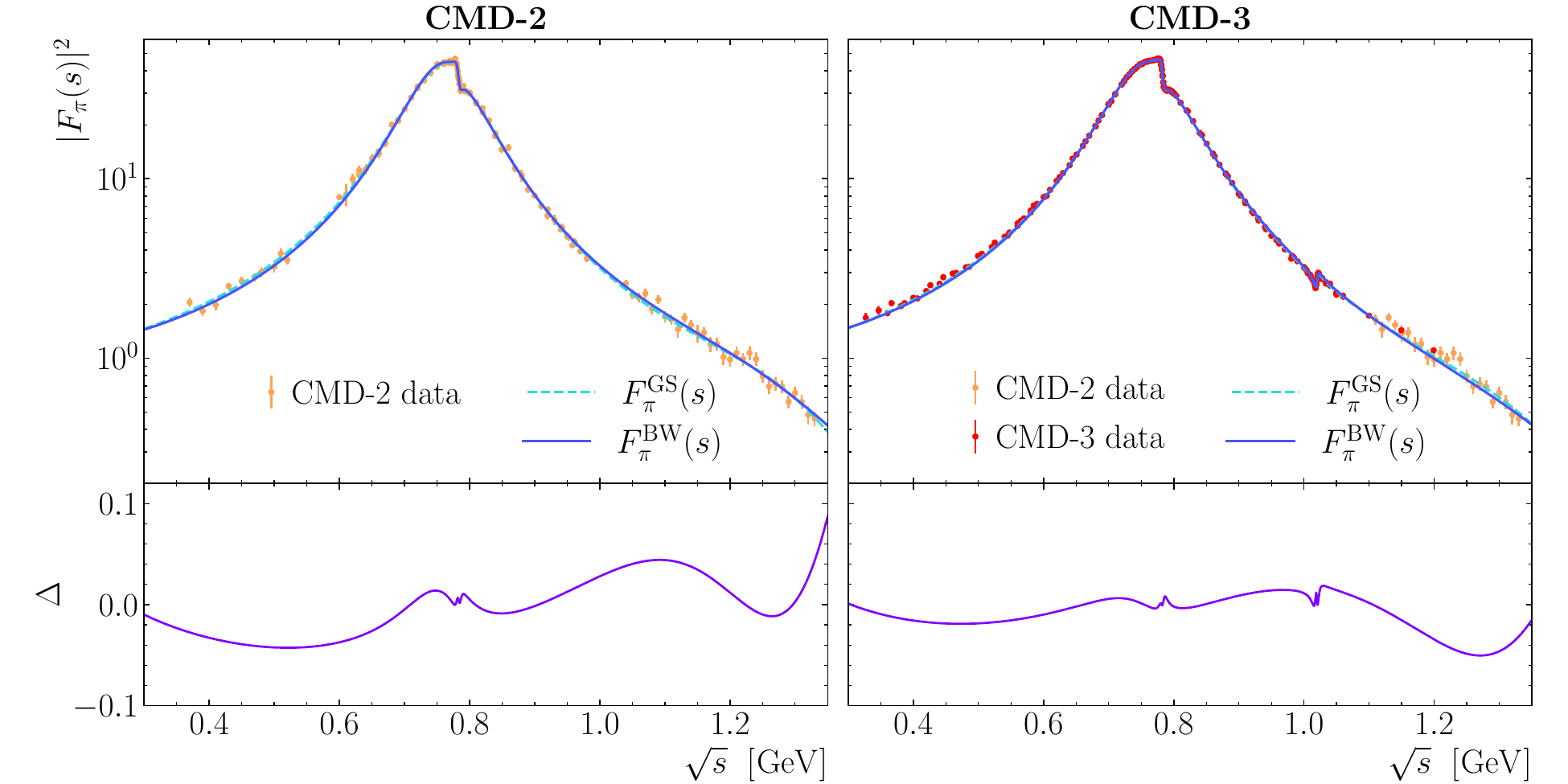}
    \caption{The $F_\pi(q^2)$ parameterisations used as inputs in the numerical simulations, both for CMD-2 (left) and CMD-3 (right), compared to the experimental data. The lower panels show the relative difference between $F_\pi^\textrm{GS}(q^2)$ and $F_\pi^\textrm{BW}(q^2)$ for both experiments, namely $\Delta \equiv \left|F_\pi^{\rm BW}\right|^2 / \left|F_\pi^{\rm GS}\right|^2 - 1$.}
     \label{fig:ffpi-inputs}
\end{figure}

In order to obtain realistic numerical results, we define a set of kinematical cuts, inspired by the CMD-3 event selection criteria~\cite{CMD-3:2023alj}, namely
\begin{subequations}
\begin{align}
   p^{\pm} & \equiv |\bm{p}^{\pm}| > 0.45 E \, ,\\[5pt] 
   \vartheta_\text{avg} &\equiv \frac12( \pi -\vartheta^++\vartheta^-) \in [1,\pi-1] \, ,\\[5pt] 
    \delta \vartheta &\equiv |\vartheta^+ + \vartheta^- - \pi|< 0.25 \, ,\\[5pt]
    \delta \phi &\equiv \left||\phi^+-\phi^-|-\pi\right|<0.15 \, ,
    \label{eq:selcrit}
\end{align}
\end{subequations}
where $E=\sqrt{s}/2$ is the beam energy, $\bm{p}_\pm$ are the $\pi^\pm$ three-momenta, $\vartheta^\pm$ are the $\pi^\pm$ polar scattering angles, and $\phi^\pm$ are the $\pi^\pm$ azimuthal scattering angles. 
We also have defined the event average angle $\vartheta_\text{avg}$, the polar acollinearity $\delta\vartheta$ and the azimuthal acollinearity $\delta\phi$, also known as acoplanarity.
In the following, we study the integrated cross section $\sigma$, the differential distributions  ${\rm d}\sigma/{\rm d}\vartheta_\text{avg}$ and ${\rm d}\sigma/{\rm d}m_{\pi\pi}$, where $m_{\pi \pi}$ is the pion invariant mass, and the charge asymmetry $A_{\rm FB}$. 

\subsection{Results for the integrated cross section}\label{resFsQED}

In this section, results for the cross section of the $e^+ e^- \to \pi^+ \pi^- (\gamma)$ process, in the centre-of-mass energy range $0.35$~GeV $\leq$ $\sqrt{s}$ $\leq$ $1.2$~GeV, are discussed. The MC simulations are performed with the \textsc{BabaYaga@NLO} generator, according to the event selection defined in Sec.~\ref{pionff_eventselection}. In particular we focus on the effects of (s)QED radiative corrections at different levels of precision. To this aim, the simulations are performed with only the factorised approach of the pion form factor. For illustrative purposes we compare the results obtained with the two parameterisations shown in  Sec.~\ref{pionff_eventselection}, in order to investigate the level of sensitivity of the radiative effects to different form factor parameterisations. 

The upper panels of Fig.~\ref{fig:FxsQED_scan} show the cross section for $e^+ e^- \to \pi^+ \pi^- (\gamma)$ as a function of $\sqrt{s}$,  with three different approximations: LO (purple), NLO (cyan) and NLOPS (light green). The line referring to NLO accuracy is overlapping with the one referring to NLOPS, within the resolution scale of the picture. As a general comment, the values of the scales $m_e$, $m_\pi$ and $\sqrt{s}$ induce a natural hierarchy between ISR and FSR. In fact, they are ruled by the collinear logarithms $\ln(s/m_e^2)$ and $\ln (s/m_\pi^2)$, which, at the $\rho$ peak, have the values of about $14.7$ and $3.4$, respectively. Moreover, the presence of the $\rho$ resonance introduces the typical effects of line-shape distortion due to ISR, extensively studied at the $\phi$ resonance at DA$\Phi$NE (see for instance Chapter~10 of~\cite{Maiani:1995ve}) and at the $Z$ pole at LEP (see for example~\cite{Montagna:1998sp}): referring to the QED corrected cross section as a double convolution of the tree-level kernel cross section with the leading logarithmic electron and positron structure functions, the QED corrections are expected to be negative below and positive above the $\rho$ peak, respectively, because of the weight given by the tree-level kernel cross section in the convolution, with a shift towards higher values of the peak position of the radiatively corrected line-shape w.r.t. the tree-level one. 

\begin{figure}[t]
    \centering
    \includegraphics[width =\textwidth]{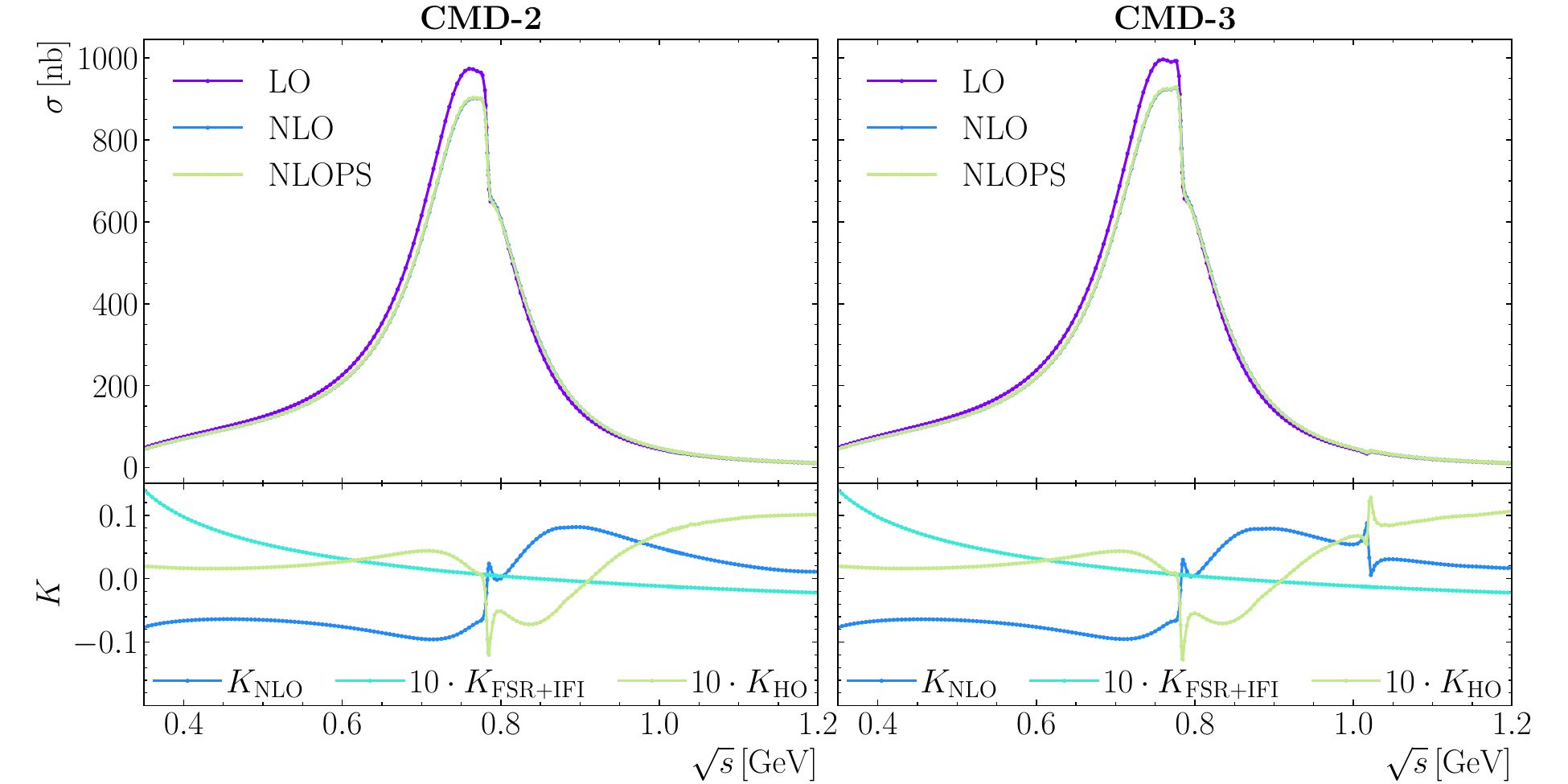}
    \caption{In the top plots, the total cross section for the process $e^+e^-\to\pi^+\pi^-$ as a function of the centre-of-mass energy $\sqrt{s}$ at the LO (purple), NLO (cyan) and NLOPS (light green) is shown. In the bottom plots, the $K$-factors at different orders are shown. 
    The left panel refers to the CMD-2 form factor, while the right panel to the CMD-3 form factor.}
    \label{fig:FxsQED_scan}
\end{figure}

In order to quantitatively estimate the impact of the different classes of radiative corrections, the following  $K$-factors are introduced: 
\begin{subequations}\label{eq:Kratios}
\begin{align}
    \label{eq:k-nlo}
    K_{\rm NLO} &= \frac{\sigma_{\rm NLO} - \sigma_{\rm LO}}{\sigma_{\rm LO}}\,,\\[5pt]
    \label{eq:k-fsrifi}
    K_{\rm FSR+IFI} &= \frac{\sigma_{\rm NLO} - \sigma_{\rm ISR}}{\sigma_{\rm LO}}\,,\\[5pt]
    \label{eq:k-nlops}
    K_{\rm HO} &= \frac{\sigma_{\rm NLOPS} - \sigma_{\rm NLO}}{\sigma_{\rm LO}}\,,
\end{align}
\end{subequations}
where $\sigma_{\rm NLO}$ is the complete NLO cross section,  $\sigma_{\rm ISR}$ is the NLO cross section including only the ISR contribution, and $\sigma_{\rm NLOPS}$ is the NLO cross section matched to the Parton Shower. For observables that are even under the transformation $\cos\vartheta_{\rm avg} \to - \cos\vartheta_{\rm avg}$, as it is the case of the cross section integrated over a symmetric angular range, the IFI contribution vanishes and therefore $K_{\rm FSR+IFI}$ measures the size of the FSR corrections. As discussed in Sec.~\ref{sec:nlo}, the NLO calculation includes only photonic corrections, without photon vacuum polarisation contributions.

In the lower panels of Fig.~\ref{fig:FxsQED_scan}, the quantitative effect of the NLO corrections w.r.t. the LO predictions is displayed (cyan  lines). The effect of resonance shape distortion driven by the ISR contribution can be quantified at the level of $-8$\% below the $\rho$ peak and positive above, reaching a maximum of the order of $+8$\% at $\sqrt{s} \sim 0.9$~GeV and decreasing at larger values of $\sqrt{s}$. The FSR correction, green line, is small, as expected, at the $0.1\%$ level, increasing with decreasing $\sqrt{s}$, up to about $1\%$ at $\sqrt{s} \sim 0.4$~GeV. Towards the $\pi^+ \pi^-$ threshold the FSR correction goes as  $\alpha \, \pi / (2 \, \beta_\pi)$, as expected in the non-relativistic limit $\beta_\pi \to 0$. In this limit, the Sommerfeld enhancement should also be considered~\cite{sommerfeld1921atombau,Gamow:1928zz,Sakharov:1948plh,Arbuzov:2011ff}. However, while this effect is relevant for higher mass hadron production, it is completely negligible for $e^+ e^- \to \pi^+ \pi^- (\gamma)$, since the minimum pion energy detection threshold keeps the detected events far away from the production threshold $\sqrt{s}=2m_\pi\sim 0.278 \,\rm{GeV}$. The higher order effects introduced by the Parton Shower are contained at the few $0.1$~\% level, as expected. 

With the CMD-3 inspired parameterisation shown in the plot to the right, it is clearly visible the $\rho - \phi$ interference effect. Comparing the effects between the panels on the l.h.s. and r.h.s., we can safely conclude that the QED radiative effects are, with good approximation, independent of the pion form factor parameterisation. For this reason, the following results are shown for the CMD-3 parameterisation only.
 
\subsection{Results for differential cross sections}\label{res_scalar_differential}

\begin{figure}[t]
    \centering    \includegraphics[width=0.95\textwidth]{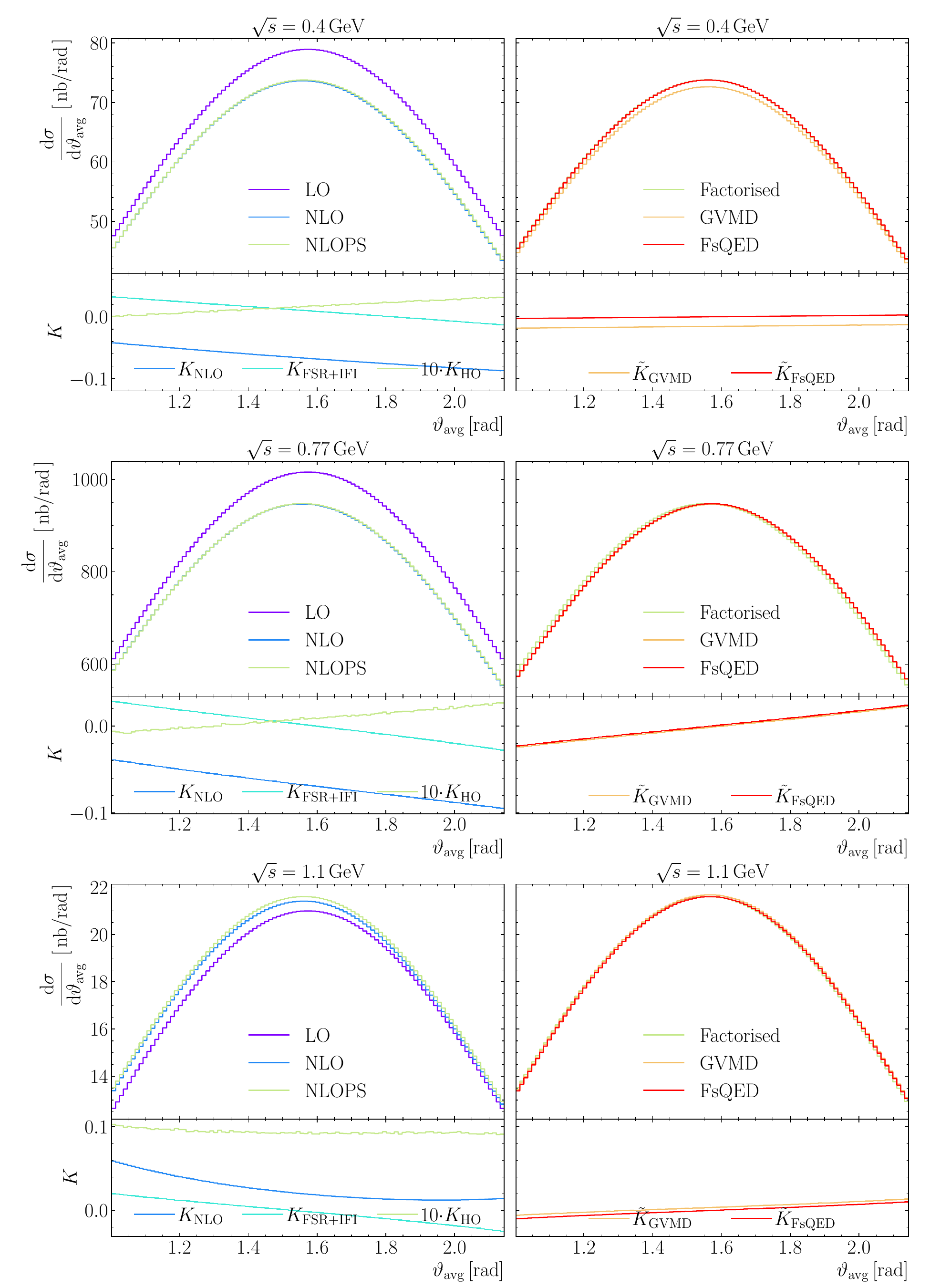}
       \caption{The differential cross section $\sigma(e^+e^-\to\pi^+\pi^-(\gamma))$ w.r.t. the average angle $\vartheta_{\rm avg}$ with different approximations for radiative corrections and with different form factor approaches, for three values of $\sqrt{s}$. Details on the panels are given in the text.}
    \label{fig:FsQED_thav}
\end{figure}

In order to quantitatively study the impact of the different contributions to the simulation of the process $e^+ e^- \to \pi^+ \pi^- (\gamma)$, both in terms of radiative corrections and of approaches to the pion form factor, in Fig.~\ref{fig:FsQED_thav} we analyse the features of the differential cross section w.r.t. the average final state angle $\vartheta_{\rm avg}$. Three different centre-of-mass energies are selected, $\sqrt{s}=0.4$ GeV, $\sqrt{s}=0.77$ GeV and $\sqrt{s}=1.1$ GeV (first, second and third panel line, respectively), to investigate the dependence of the results on the position of $\sqrt{s}$ w.r.t. the $\rho$ resonance peak. On the left panels, different contributions to the radiative corrections are shown using the factorised approach for the pion form factor at LO (purple), NLO (blue), NLO including only FSR and IFI (cyan) and NLOPS (green). The upper plots show the absolute predictions while the bottom plots show the $K$-factors at different orders, as defined in \eqref{eq:Kratios} where the integrated cross sections are replaced with the differential ones. For all the three values of $\sqrt{s}$ the NLO correction decreases as $\vartheta_{\rm avg}$ increases. It is negative for $\sqrt{s}$ below and at the $\rho$ peak while it is positive above the peak. The sign of the NLO contributions is strictly correlated with the energies at which they are evaluated. This is due to the already mentioned mechanism of radiation below or above the $\rho$ resonance. The relative corrections to the LO span the range between about $-10\%$ and $-4\%$ for $\sqrt{s} = 0.4$~GeV and $\sqrt{s} = 0.77$~GeV. At $\sqrt{s} = 1.1$~GeV the NLO correction is positive, in the range between $1\%$ and $6\%$. The FSR+IFI corrections have slopes which are similar to the full NLO ones, changing sign in passing from low to large values of $\vartheta_{\rm avg}$. Their absolute values remain smaller than about $3\%$ for all values of $\sqrt{s}$. The similarities in slopes can be understood considering that only the IFI corrections depend asymmetrically on the scattering angle. On the other hand, ISR corrections are symmetric for $\vartheta_{\rm avg} \to -\vartheta_{\rm avg}$ and FSR contributions do not depend on $\vartheta_{\rm avg}$,  because of the different spin of the involved emitting particles. By inspection of the numerical values of the full NLO corrections and the FSR+IFI ones, we can conclude that the NLO corrections are numerically driven by the ISR corrections. This is expected because of the enhancement due to the collinear logarithm involving the electron mass. The higher-order radiative corrections are of the order of some $0.1\%$, with increasing size as the angle $\vartheta_{\rm avg}$ increases, for $\sqrt{s} = 0.4~$GeV and $\sqrt{s} = 0.77~$GeV. At $\sqrt{s} = 1.1~$GeV, they weigh $1\%$ and are almost independent of $\vartheta_{\rm avg}$. The slightly increased size of these corrections compared to smaller values of $\sqrt{s}$ is due to the presence of the $\rho$ peak at a lower energy w.r.t. $\sqrt{s}$.

On the r.h.s. plots of Fig.~\ref{fig:FsQED_thav}, we show the predictions obtained with different approaches for the pion form factor, namely factorised (yellow), GVMD (orange) and FsQED (red), at NLOPS accuracy. In the bottom panels, the relative differences between the GVMD (orange) and FsQED (red) approaches and the factorised one are shown, as defined in the following equation
\begin{equation}
\label{eq:Kformfactors}
    \Tilde{K}_{\rm FF} = \left(\dv{\sigma_{\rm FF}}{\vartheta_{\rm avg}}\right)\left(\dv{\sigma_{\rm Factorised}}{\vartheta_{\rm avg}}\right)^{-1}-1\,,\\[5pt]
\end{equation}
where FF $=$ GVMD, FsQED.
{At $\sqrt{s} = 0.4$~GeV the differences between the FsQED- and GVMD-based predictions are almost constant, around $2\%$. This shift can be ascribed to the difference btween the GS and BW form factors, which are used in the factorised/FsQED and GVMD approaches,  respectively. In fact $|F^\text{BW}_\pi|^2/|F^\text{GS}_
\pi|^2\simeq 1.02$ for $\sqrt{s}=0.4$~GeV. The differences between the FsQED and the factorised approach are small, spanning the range between about $-0.3\%$ and $+0.3\%$, increasing monotonically with $\vartheta_{\rm avg}$. At $\sqrt{s} = 0.77$~GeV, the difference between the FsQED (and GVMD) and the factorised approaches increases by an order of magnitude w.r.t. $\sqrt{s} = 0.4$~GeV, spanning the range between about $-2.5\%$ and $+2.5\%$, while the shift between the GVMD and FsQED approaches is $0.01\%$. A similar situation occurs at $\sqrt{s} = 1.1$~GeV, where the FsQED- and GVMD-based calculations are shifted of about $0.3\%$. The differences of the FsQED approach from the factorised one range from about $-1.2\%$ at small $\vartheta_{\rm avg}$ to $+1.2\%$ at large $\vartheta_{\rm avg}$. We remark that the effects of the three approaches on angular distributions are not negligible. The slight shift of the $\Tilde{K}$ ratios from being centred on $\Tilde{K}=0$ determines a non-negligible difference in the charge asymmetry, defined in Eq.~\eqref{eq:FBasym}, whose shape as a function of $\sqrt{s}$ is heavily affected by the presence of $F_\pi$ in the loop, as will be shown later in Sec.~\ref{res_asymmetry}. }
\begin{figure}[t]
    \centering
    \includegraphics[width=\linewidth]{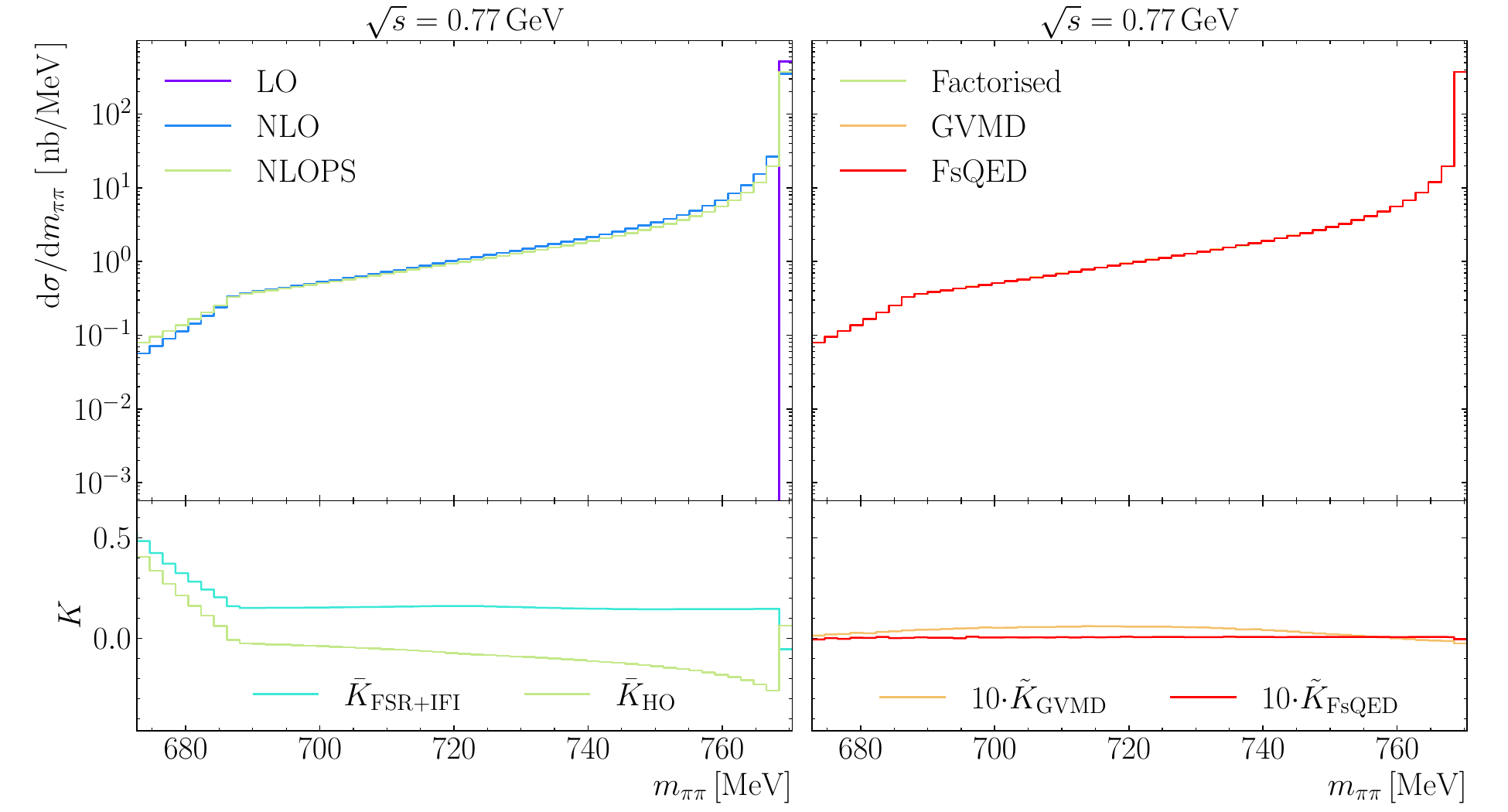}
    \caption{The same as in Fig.~\ref{fig:FsQED_thav}, but plotted against the $\pi^+\pi^-$ invariant mass $m_{\pi\pi}$ at $\sqrt{s}~=~0.77$~GeV.}
    \label{fig:FsQED_mpp}
\end{figure}

In Fig.~\ref{fig:FsQED_mpp}, the differential cross section as a function of the $\pi^+\pi^-$ invariant mass is shown. This is the example of a fully exclusive differential cross section w.r.t. photon radiation. In fact, the LO cross section is contained in a single bin, at $\sqrt{s}$ and the rest of the plot is due to the LO real radiation matrix element for $e^+ e^- \to \pi^+ \pi^- \gamma$, and multi-photon corrections given by the Parton Shower. No photon recombination criteria are adopted. The upper plot on the left panel of Fig.~\ref{fig:FsQED_mpp} shows the predictions obtained with \textsc{BabaYaga@NLO} in mode LO (purple bin at the nominal $\sqrt{s}$ value), NLO (blue line) and NLOPS (light green). The blue line displays the typical tail towards $\pi^+ \pi^-$-invariant masses lower than $\sqrt{s}$ given by the emission of radiation. The knee at around $m_{\pi \pi} = 0.685$~GeV is induced by the cuts of the event selection. In the lower panel, the fraction of ${\cal O(\alpha)}$ real FSR+IFI w.r.t. the complete radiative matrix element is shown (cyan line), where, for this observable, the IFI contribution is zero after integration of the pions over a symmetrical angular range. As can be seen, the FSR contribution is positive and almost flat, of the order of 15\%, reaching the value of 50\% at the lower limit of the $m_{\pi \pi}$ range. The leading logarithmic multi-photon contribution is quantified by the light-green line, being negative, of the order of $-30\%$ in the next-to-largest $m_{\pi \pi}$ bin, and decreasing in size at lower $m_{\pi \pi}$ values. After the knee the higher order contribution becomes large and positive, up to the order of $40\%$. 

On the right panels of Fig.~\ref{fig:FsQED_mpp}, the same differential distribution $\dd\sigma / \dd m_{\pi \pi}$ is shown for different form factor approaches and with NLOPS running mode of the generator. In the lower panel, the relative differences of the predictions with FsQED (red line) and GVMD (orange line) approaches w.r.t. the factorised one are presented. Since for $m_{\pi \pi} < \sqrt{s}$ only the real radiation matrix element contributes, by construction the FsQED-based predictions are equal to the ones obtained with the factorised approach. The differences between the GVMD approach and the other ones reflect the differences between $F_{\pi}^{\rm BW}$ and $F_{\pi}^{\rm GS}$ reported in the lower-right panel of Fig.~\ref{fig:ffpi-inputs}, ranging from almost zero close to the $\rho$ peak to about $+0.6\%$ at lower $m_{\pi \pi}$ values. 

We close this section showing in Fig.~\ref{fig:ddiff_fact} the double differential cross section of $e^+e^-\to f^+f^-$, with $f =\{e,\,\mu,\,\pi\}$, at $\sqrt{s}=0.5$~GeV (left) and $\sqrt{s}=0.77$~GeV (right), w.r.t. the final-state three-momenta, at NLOPS accuracy. For the $\pi^+ \pi^-$ final state the factorised approach to the form factor is used. The plots show the separation of the $e,\,\mu$ and $\pi$ signatures in a typical $e^+e^-$ annihilation experiment, allowing for the extraction of a correct identification of the produced particle. The peak in the top right corner corresponds to an $e$ signature, whereas the second peak more on the bottom-left indicates the production of a $\mu$ final state. The peak at smaller momenta corresponds to the production of a $\pi$ final state. For a centre-of-mass energy of $\sqrt{s}=0.5$ GeV the three peaks are well-separated, since the energy is sufficiently small to discriminate the particle masses. However, for higher energies, the three final states cannot be correctly separated. This reflects what was already observed in~\cite{CMD-3:2023alj}.

\begin{figure}[t]
    \centering
    \includegraphics[width=\textwidth]{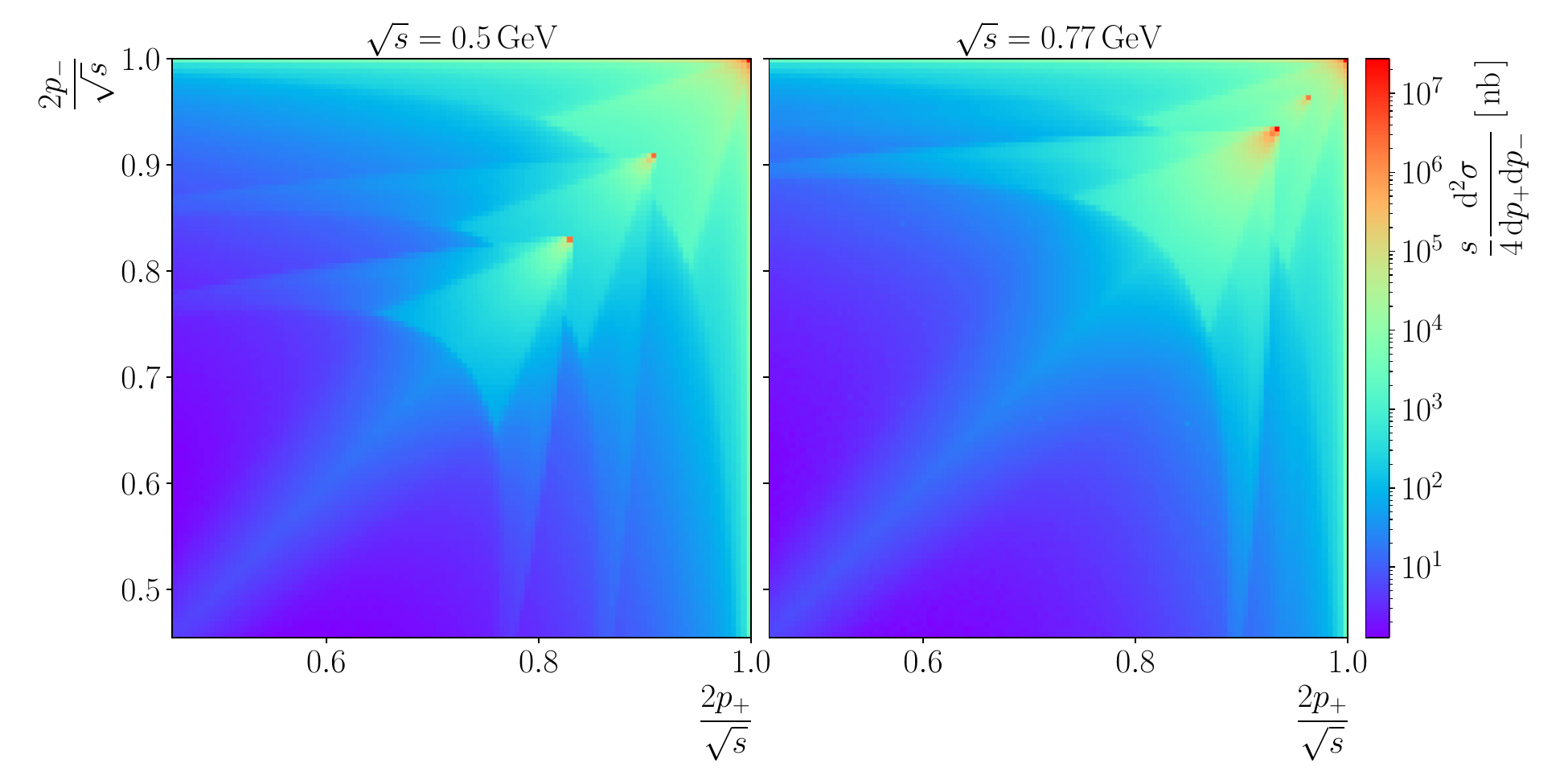}
    \caption{Double differential cross section  w.r.t. the final-state three-momenta $p_+$ and $p_-$ for the process $e^+e^-\to f^+f^-$ with $f = \{e,\,\mu,\,\pi\}$  at $\sqrt{s}=0.5$ GeV (left) and $\sqrt{s}=0.77$ GeV (right).}
    \label{fig:ddiff_fact}
\end{figure}

\subsection{Results for the charge  asymmetry}\label{res_asymmetry}

The pion charge asymmetry $A_{\rm FB}$, defined in Eq.~\eqref{eq:FBasym} is a crucial observable for this process as it is useful to understand to which level the simulation tools reproduce the data. This is demonstrated by the angular asymmetry of the differential distributions $\dv{\sigma}{\vartheta_{\rm avg}}$ illustrated in Fig.~\ref{fig:FsQED_thav} and discussed in Sect.~\ref{res_scalar_differential}. For this reason, we show in Fig.~\ref{fig:AFB_CMD3} the predictions for the charge asymmetry as a function of $\sqrt{s}$ obtained with \textsc{BabaYaga@NLO} using the pion form factor parameterisation measured by the CMD-3 experiment. In this plot, the effects on $A_{\rm FB}$ at NLO and NLOPS accuracy and using the three different approaches to the pion form factor are shown. In particular, the purple and blue lines contain the predictions with the factorised form factor and radiative corrections at NLO and NLOPS accuracy, respectively; green and light-green lines refer to the GVMD approach, with NLO and NLOPS radiative corrections, respectively; the predictions based on the FsQED approach are shown at NLO in orange and at NLOPS in red. For the sake of illustration, the CMD-3 data points are superimposed. 

\begin{figure}[t]
    \centering    
\includegraphics[width=\linewidth]{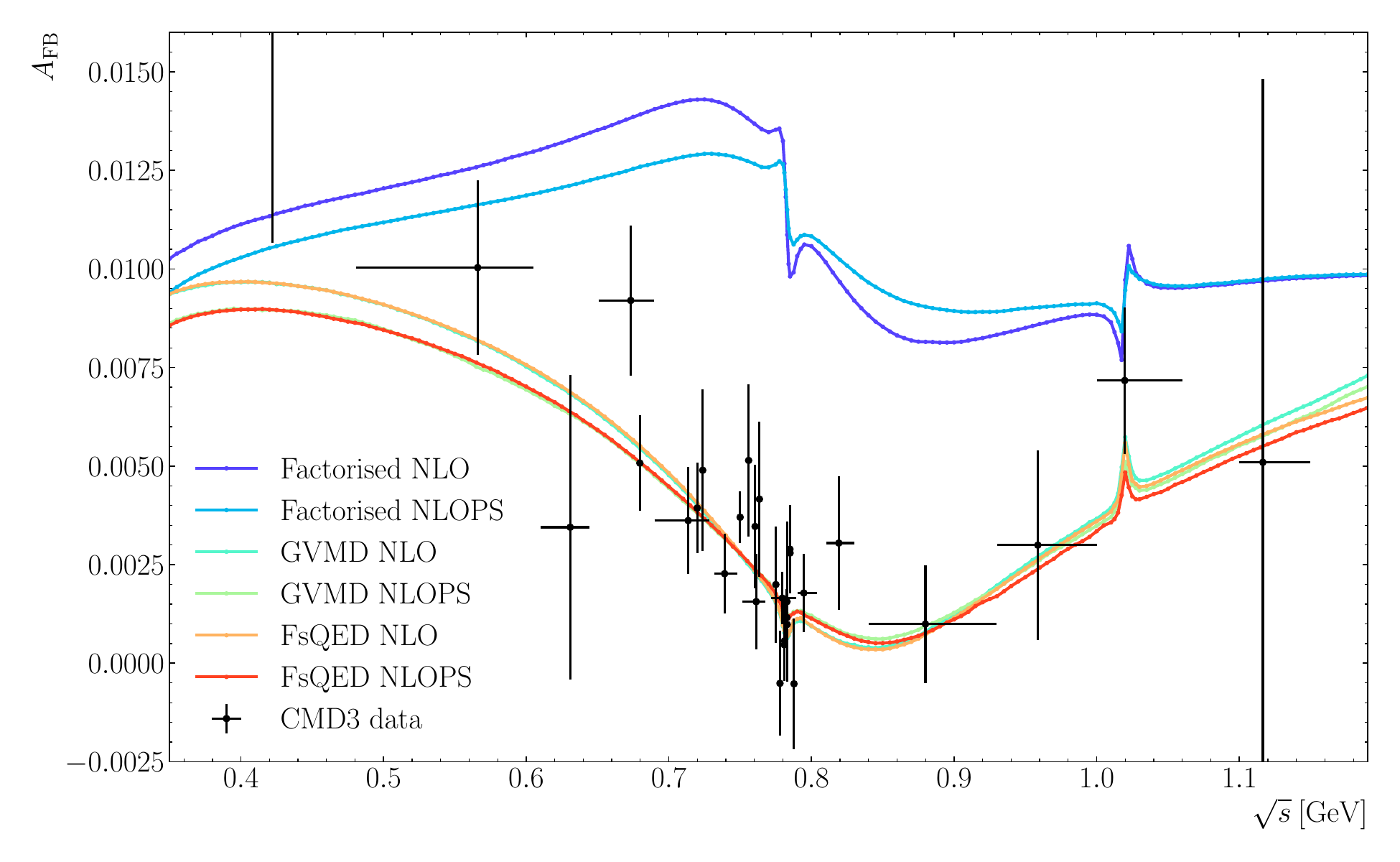}
    \caption{Charge asymmetry, as defined in Eq.~\eqref{eq:FBasym}, using the pion form factor measured at the CMD-3 experiment. We show in similar colours the result for each of the approaches discussed before at NLO and NLOPS accuracy. The black crosses show the CMD3 experimental data with their uncertainties.}
    \label{fig:AFB_CMD3}
\end{figure}

The pion charge asymmetry is very small, since it is induced by radiative corrections. This makes it particularly sensitive to all the contributions which change the shape of the angular distributions, while effects due to the overall normalisation tend to cancel out. In this respect, by looking at the right-column plots of Fig.~\ref{fig:FsQED_thav}, we can expect a small effect of the different approaches to the pion form factor at low centre-of-mass energies. Larger effects can be expected both around the $\rho$ peak and above it. This is indeed the case in Fig.~\ref{fig:AFB_CMD3}, where the three NLO predictions in the first centre-of-mass energy bins are of the order of  0.1\%. As $\sqrt{s}$ increases, the predictions using the factorised form factor increase up to the $\rho$ peak, whereas those using the GVMD and FsQED approaches decrease until eventually reaching about $A_{\rm FB} = 0$. In this region, we can appreciate the relevance of the insertion of the pion form factor in the virtual corrections in modelling the data, even with large error bars. On the other hand, the predictions based on the factorised approach have a different shape with respect to the data. 
The effect of the $\rho - \omega$ interference is visible for the three approaches at $\sqrt{s}\simeq m_\omega$ together with the $\rho-\phi$ interference at $\sqrt{s}\simeq m_\phi$. 
Remarkably, we find a very good agreement between the GVMD and FsQED approaches\footnote{In Fig.~23 of \cite{CMD-3:2023alj}, the GVMD and FsQED approaches exhibit a discrepancy in the $\rho$ peak region. We verified that this is mainly due to the non-inclusion of the term $- \Im F_\pi(s)/s\left[\lim_{s'\to s^-}\Im\bar{\delta}_V^\text{IFI}(\lambda^2,s')(s-s')\right]$ in the real part of the virtual IFI correction in the FsQED approach.}, except for small differences due to the approximation of the pion form factor as a sum of BW functions in the former case.

While the NLO corrections, in particular the IFI contribution, are responsible for the non-trivial shape of $A_{\rm FB}$, the higher-order radiative corrections beyond NLO tend to give a small perturbation on top of the NLO calculation. The higher-order contributions are in general negative and in agreement with the slopes of the $K_{\rm HO}$ factors in Fig.~\ref{fig:FsQED_thav}. They weigh at most about 0.1\% for all the form factor approaches, except for the factorised one. In this case, the corrections are of the order of 0.2\% for centre-of-mass energies that are close to the $\rho $ peak, whereas they are positive for $m_\rho < \sqrt{s} < m_\phi$. For the GVMD and FsQED approaches, the higher-order contributions are larger in the low-energy region. They become very small in the $\rho$ peak region and above.

\section{Summary and outlook}
\label{sec:conc}
Given the importance of the $e^+ e^- \to \pi^+ \pi^- (\gamma)$ process in energy scan measurements of the pion form factor, we have performed
a novel calculation of the radiative corrections to pion pair production in $e^+ e^-$ collisions. In our computation, the full set of NLO corrections is matched to a PS to account for the exclusive emission of multiple photons. In the PS implementation, all the contributions due to ISR, IFI and FSR are resummed to all orders. 
To this end, we have constructed an original PS algorithm by modelling ISR according to QED and by treating photon emission from the final-state pions as in sQED.

The inclusion of the pion form factor in the calculation allows to keep into account the internal structure of the pion. The pion form factor is introduced in the virtual NLO calculation according to three different schemes. The first one is a factorised approach, where the sQED virtual corrections multiply the form factor evaluated at scale $s$. Alternatively, the other two schemes, the GVMD model and the FsQED approach, allow to include the form factor in the loop momentum integration. This represents the first fully exclusive implementation of form factor approaches beyond the factorised approximation in a MC event generator. 

Our efforts led to the development of an updated version of the \textsc{BabaYaga@NLO} event generator that is available for precision measurements of the pion form factor in energy scan experiments. Each of the three pion form factor approaches can be optionally switched on for event generation. Moreover, several form factor parameterisations are available in the code and any other one can be easily added by the user. This can be useful for sound estimates of the theoretical uncertainties in the theoretical predictions induced by hadronic effects. 

We have shown that the radiative corrections to the pion pair production are largely dominated, not surprisingly, by ISR. However, also FSR and IFI play a role, especially at the level of the differential cross sections. Most importantly, the inclusion of the composite structure of the pion in loops is crucial for a reliable modelling of the charge asymmetry, as emphasised in previous studies. The contribution of PS exponentiation is relevant for sub-percent measurements of the pion form factor.
The theoretical accuracy of 
the present NLOPS approach to two pion production as in \textsc{BabaYaga@NLO} will be addressed in a separate publication.
Preliminary comparisons between our MC results and the predictions of other codes can be found in~\cite{strong2020}.
Clearly, the approach adopted in this article can be extended to other hadronic channels, such as ${e^+ e^- \to K^+ K^- (\gamma)}$.

Concerning the prospects of our work, we plan to implement the radiative process $e^+ e^- \to \pi^+ \pi^- \gamma$ in \textsc{BabaYaga@NLO} with a NLOPS accuracy. Because of the importance of this process in the measurement of the pion form factor via the radiative return method, this effort would 
lead to the development of an event generator 
independent of the MC code PHOKHARA~\cite{Rodrigo:2001kf,Czyz:2002np,Czyz:2003ue,Czyz:2004rj,Campanario:2019mjh}, which is the standard tool for the simulation of radiative processes at flavour factories.

For radiative processes, i.e. the 
hadronic channel $e^+ e^- \to \pi^+ \pi^- \gamma$ 
and the leptonic processes 
$e^+ e^- \to \mu^+ \mu^- \gamma \, / \, e^+ e^- \gamma$,
the matching between NLO corrections and PS algorithm can be done in \textsc{BabaYaga@NLO} by following the same procedure applied to  $e^+ e^- \to \gamma \gamma$~\cite{Balossini:2008xr,CarloniCalame:2019dom}. This would allow to study the impact of additional radiation on the measurements of the pion form factor via the radiative return method, as recently scrutinised in~\cite{BaBar:2023xiy}. 

Over the long term, we are interested to improve the 
matching of fixed-order corrections to the PS in \textsc{BabaYaga@NLO}, in order to provide results at NNLOPS 
accuracy. As a proof of concept, we plan to combine PS resummation with the 
gauge-invariant subsets of exact NNLO corrections associated with dipole radiation, 
such as the dominant ISR contributions in s-channel $2\to2$
processes and the radiation from electron/positron legs in t-channel Bhabha scattering.

\acknowledgments
We are grateful to Gilberto Colangelo, Martin Hoferichter, Stefano Laporta, Barbara Pasquini, Simone Rodini, Peter Stoffer and Yannick Ulrich for useful discussions and exchange of information. We acknowledge Fedor Ignatov for carefully reading the manuscript and providing valuable feedback and information.
We also wish to thank Achim Denig, Andrey S. Kupich and Graziano Venanzoni for interest in our work and the colleagues of the Working Group ``Radiative corrections and Monte Carlo tools for low-energy hadronic cross section in $e^+ e^-$ collisions'' for fruitful collaboration. We are indebted to the Mainz Institute for  Theoretical Physics (MITP) of the Cluster of Excellence PRISMA$+$ (Project ID 390831469) for its hospitality and support during the workshop ``The Evaluation of the Leading Hadronic Contribution to the Muon $g\!-\!2$: Consolidation of the MUonE Experiment and Recent Developments in Low Energy $e^+e^-$ Data''. We acknowledge financial support by the Italian Ministero dell'Universit\`a e Ricerca (MUR) and European Union - Next Generation EU through the research grant number 20225X52RA ``MUS4GM2: Muon Scattering for $g\!-\!2$'' under the program Prin~2022. The work of E.B. is supported by the Prin-2022 project ``MUS4GM2''. F.P.U. is grateful to 
the Wilhelm and Else Heraeus Foundation for financial support to attend the ``Simon Eidelman School on Muon Dipole Moments and Hadronic Effects'', 2-6 September 2024, Nagoya University.

\appendix
\newcommand{\pu}{p_1}
\newcommand{\pdd}{p_2}
\newcommand{\pddv}{\mathbf p_2}
\newcommand{\kd}{k_2}
\newcommand{\kdv}{\mathbf k_2}
\newcommand{\qv}{\mathbf q}
\newcommand{\uv}{\mathbf u}
\newcommand{\Pv}{\mathbf P}
\newcommand{\uq}{\mathbf u_q}
\section{Dispersive integrals of \texorpdfstring{\protect\boldmath$\Im\hspace{-1pt}\text{D}_0$}{ImD0}}
\label{app:dimreg}

As discussed in Sec.~\ref{sec:disp}, the FsQED approach requires the evaluation of two different dispersive integrals of the imaginary part of the box diagrams, both defined in Eq.~\eqref{EQ:disp_ints_sec3}. 
In this appendix, we detail the calculation of such integrals. 
The convention used for the momenta is shown in Fig.~\ref{FIG:Cut}. 

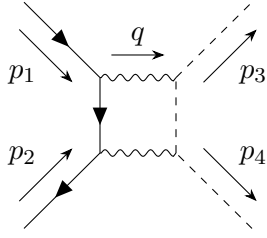
\begin{figure}[t]
\begin{equation*}
\begin{tikzpicture}
\begin{feynman}
    \vertex (a);
    \vertex[right=1cm of a] (b);
    \vertex[below=1cm of a] (i);
    \vertex[below=1cm of b] (j);
    \vertex[above left=1cm and 1cm of a] (c);
    \vertex[below left=1cm and 1cm of i] (d);
    \vertex[above right=1cm and 1cm of b] (e);
    \vertex[below right=1cm and 1cm of j] (f);
    \diagram* {
      (a) -- [photon, momentum = $q$] (b),
      (c) -- [fermion, momentum'=$p_1$] (a) -- [fermion] (i) -- [fermion, reversed momentum'=$p_2$] (d),
      (f) -- [scalar, reversed momentum'=$p_4$] (j) --[scalar] (b) -- [scalar,momentum'=$p_3$] (e),
      (i) -- [photon] (j),
    };
\end{feynman}
\end{tikzpicture}
\end{equation*}
\vspace{-5mm}
\caption{Convention on momenta for the box diagram.}
\label{FIG:Cut}
\end{figure}

\subsection{Dispersive integration}

The two integrals we have to compute are
\begin{align}\label{Eq:int1ovs}
\begin{split}
        \mathcal{I}_{1/s'}(t) = \Im \int_{4m_\pi^2}^\infty  \frac{\dd s'}{s'} \int \frac{\dd^4 q}{i \pi^2}\,&\frac{1}{(q^2-\lambda^2)[(q-p_3-p_4)^2-s'-\lambda^2]}\,\times \\
   &\frac{1}{
   (q^2-2 p_3\cdot q)(q^2-2\pu \cdot q)} \,,
\end{split}
\end{align}
\begin{align}\label{Eq:intsovs}
\begin{split}
\mathcal{I}_{s'/s'}(t)= \Im \int_{4m_\pi^2}^\infty  \dd s' \int \frac{\dd^4 q}{i \pi^2}\,&\frac{1}{(q^2-\lambda^2) [(q-p_3-p_4)^2-s'-\lambda^2]} \, \times\\
&\frac{1}{ (q^2-2 p_3\cdot q)(q^2-2\pu \cdot q)} \,,
\end{split}
\end{align}
where the poles are dealt with the standard $i \epsilon$ prescription.  
We focus on the direct box contribution, as the result can be easily extended to the crossed box via the substitution $p_3 \to p_4$, which gives $\mathcal{I}(u)$. 

To compute the imaginary part of the above integrals we can resort to the Cutkosky rules~\cite{Cutkosky:1960sp}, \textit{i.e.} the discontinuity of the amplitude can be obtained by cutting the propagators with the prescription
\begin{equation}
\frac{1}{q^2-\mu^2+i\epsilon} \;\longrightarrow\; -2\pi i \, \delta(q^2-\mu^2)\hspace{1pt}\theta(q^0) \,,
\end{equation}
with the energy flowing in the same direction of time. 
This is a crucial point of our calculation, as it allows us to easily exchange the integrals in $q$ and $s'$. Therefore, we obtain
\begin{equation}
  \mathcal{I}_{1/s'}(t) =
    2\int_{4m_\pi^2}^\infty  \frac{\dd s'}{s'}\int \dd^4 q \
    \frac{\delta(q^2-\lambda^2) \theta(q^0) \delta(q^2-2 q_0\sqrt{s} +s -s'-\lambda^2)
      \theta(\sqrt{s
  }-q^0)}{(q^2-2 \pu\cdot q)(q^2-2 p_3\cdot q)} \,,
  \label{eq:iml1}
\end{equation}
where the evaluation is carried in the centre-of-mass frame, \textit{i.e.} $\pu+\pdd\equiv(\sqrt s, \mathbf{0})$. 
By using $\delta(q^2-\lambda^2)$ to integrate over $q_0$ we obtain
\begin{equation}
  \mathcal{I}_{1/s'}(t)
    =
   2\int_{4m_\pi^2}^\infty  \frac{\dd s'}{s'}
   \int \frac{\dd^3 \mathbf{q}}{2 q_0}
   \frac{
   \delta(s-2 q_0\sqrt{s} -s')
     \theta(\sqrt{s
       }-q^0) }
     {(\lambda^2-2 \pu \cdot q)(\lambda^2-2 p_3\cdot q)} \,,
       \label{eq:iml2}
\end{equation}
that sets $q_0 = \sqrt{\mathbf{q}^2 + \lambda^2}$.
By using the delta constraint $\delta(s-2 \sqrt{s} q_0 -s')$ and the condition $q_0 \geq \lambda$, we obtain the relations
\begin{align}
    q_0  =  \frac{s-s'}{2 \sqrt s}\,,  \qquad  
  4 m_\pi^2 \le   s'   \le s - 2 \sqrt s \lambda \,,
\end{align}
that set the upper bound of the dispersive integration. Notice that the above relations hold  irrespective of the variable which is chosen to deal with the $\delta$ function constraint.
    
We can introduce the change of variable 
\begin{align}
    w  =  \frac{s-s'}{2 \sqrt s} \,,  \qquad
  \lambda\le   w   \le \frac{s-4m_\pi^2}{2\sqrt{s}} \,,
\end{align}
so that the integral reads
  \begin{equation}\label{eq:iml3}
    \mathcal{I}_{1/s'}(t)
    =  \int_{\lambda}^{\frac{s-4m_\pi^2}{2\sqrt{s}}}\dd w \int\frac {\dd^3\mathbf{q}}{q_0} \frac{\delta(q_0-w)}{(s - 2\sqrt s w)}
\frac{1}{(\lambda^2 - 2p_1 \cdot q )(\lambda^2- 2 p_3 \cdot q)} \,,
\end{equation}
which is strongly reminiscent of the eikonal integral. In the $\lambda\to 0$ limit, we can drop the $\lambda^2$ terms of the two propagators, as the integral differs by a $\order{\lambda}$ term, as will be shown in the following section. 
Moreover, the integral can be split in two contributions
\begin{equation}
\begin{aligned}
    \mathcal{I}_{1/s'}(t)
    & =  \frac{1}{4}\int_{\lambda}^{\frac{s-4m_\pi^2}{2\sqrt{s}}}\dd w \int\frac {\dd^3\mathbf{q}}{q_0} \frac{\delta(q_0-w)}{(s - 2\sqrt s w)}
\frac{1}{(p_1 \cdot q )( p_3 \cdot q)}
  \\[2pt]
     & =  
     \frac{1}{4 s}
    \int\dd w \int\frac {\dd^3\mathbf{q}}{q_0}
\frac{\delta(q_0-w)}  
     {(p_1 \cdot q )( p_3 \cdot q)} \\[2pt]
     &+ \frac{1}{2\sqrt s}
   \int\dd w \int\frac {\dd^3\mathbf{q}}{q_0} \frac{w}{(  s - 2\sqrt s w)}
\frac{\delta(q_0-w)}  
     {(p_1 \cdot q )( p_3 \cdot q)} 
     \\[5pt]
     & = \mathcal{I}_1(t) + \mathcal{I}_2(t) \,,
  \label{eq:iml4}
\end{aligned}
\end{equation}
where the integral in $q$ can be evaluated taking the $\lambda\to 0$ limit inside the integral, as will be justified in the following section. 

The integral in $q$ cannot be done first in $\lambda$ mass regularisation. This would require the analytic integration of the $\text{D}_0$ retaining the full $\lambda^2,s'$ dependency, which is not feasible analytically. For this reason, we integrate $\mathcal{I}_1$ over $w$ by means of the delta constraint, so that the loop energy lies in the range $\lambda \leq q_0 \leq (s-4m_\pi^2)/2\sqrt{s}$. Therefore, the integral over the photon three-momentum is proportional to the eikonal integral
\begin{equation}
    {\cal I}_1 (t)=      \frac{1}{ 4 s}
    \int_{|\mathbf{q}|\leq \omega}\frac {\dd^3\mathbf{q}}{q_0}
    \frac{1}{ (p_1 \cdot q )( p_3 \cdot q)}= \frac{1}{4 s} \mathcal{L}(t) \,,
\end{equation}
where $\omega =\frac{\sqrt{s}}{2}\beta_\pi^2 + \order{\lambda^2}$. The analytic expression for ${\cal L}(t)$ is given in~\cite{tHooft:1978jhc} where the $t$ contribution is obtained for $i=1,j=3$. The explicit expression of the eikonal integral in our notation is
\begin{equation}
\begin{aligned}
     \mathcal{I}_1=\frac{\pi}{s f}    \Bigg[& 2\log\frac{m_e^2+m_\pi^2-t + f}{2 m_e m_\pi}\log\frac{s \beta_\pi^4}{\lambda^2}\\[2pt]
     &+\frac{1}{4}\log^2 \frac{1-\beta_e}{1+\beta_e}-\frac{1}{4}\log^2 \frac{1-\beta_\pi}{1+\beta_\pi}\\[2pt]
     &+\text{Li}_2\left(1-\frac{\sqrt{s}}{2}\frac{(1+\beta_e)(m_e^4-(f + m_\pi^2-t)^2)}{(f+(m_e^2-m_\pi^2)^2-t)(f+(m_e^2+m_\pi^2)^2-t}\right)\\[2pt]
     &+\text{Li}_2\left(1-\frac{\sqrt{s}}{2}\frac{(1-\beta_e)(m_e^4-(f + m_\pi^2-t)^2)}{(f+(m_e^2-m_\pi^2)^2-t)(f+(m_e^2+m_\pi^2)^2-t}\right)\\[2pt]
     &-\text{Li}_2\left(1-\frac{(1+\beta_\pi)(f + m_\pi^2-m_e^2-t)}{(f+m_e^2+m_\pi^2-t)^2-4m_e^2m_\pi^2}\right)\\[2pt]
&-\text{Li}_2\left(1-\frac{(1-\beta_\pi)(f + m_\pi^2-m_e^2-t)}{(f+m_e^2+m_\pi^2-t)^2-4m_e^2m_\pi^2}\right)
\Bigg] \,,
    \end{aligned}
    \label{eq:I1lambda}
\end{equation}
where $f\equiv f(t) = \sqrt{t^2+(m_e^2-m_\pi^2)^2-2 (m_e^2+m_\pi^2)t}$. For the second integral, the limit of vanishing photon mass translates into the equality $q_0 = |\mathbf{q}|$. Hence, the integration in spherical coordinates is
\begin{equation}
\begin{aligned}
    {\cal I}_2(t) & =  \frac{1}{ 2 \sqrt s}\int_0^{(s-4m_\pi^2)/(2\sqrt{s})}  \dd |\mathbf{q}| 
    \frac{1}{(  s - 2\sqrt s |\mathbf{q}|)}  \int \dd \Omega
\frac{|\mathbf{q}|^2}{(p_1 \cdot q)(p_3 \cdot q)}
     \\
     &= \frac{\pi}{2 s}\log \frac{s}{4m_\pi^2}
\int_0^\pi \dd \cos\theta \int_0^1 \dd x\frac{1}{(P\cdot q)^2} \,,  
\end{aligned}
\label{eq:imfinite}
\end{equation}
where we introduced the linear combination of momenta $P=x p_1 +(1-x)p_3$ via a Feynman parameterisation. The result reads
\begin{equation}
    \mathcal{I}_2(t) =   \frac{\pi}{s}\log \frac{s}{4m_\pi^2} \frac{1}{ {f(t)}}\log\frac{m_\pi^2+m_e^2 - t + {f(t)}}{m_\pi^2+m_e^2 - t - {f(t)}} \,,
\end{equation}
with
\begin{equation}
      f(t) = \sqrt{t^2+(m_e^2-m_\pi^2)^2-2 (m_e^2+m_\pi^2)t} \,.
\end{equation}
The second integral needed, Eq.~\eqref{Eq:intsovs}, can be obtained in the same exact way, leading to a single contribution 
\begin{equation}
      \mathcal{I}_{s'/s'}(t) =\frac{1}{4}\mathcal{L}(t)\,.
\end{equation}
  
The above discussion can be translated in dimensional regularisation, obtaining the same result. After the step~\eqref{eq:iml1}, the integrals become
\begin{align}
        \mathcal{I}_{1/s'}(t)& =
\mathcal{C}_D\int_{4m_\pi^2}^\infty  \frac{\dd s'}{s'}\int \dd^{D-1} q \
    \frac{\delta(q^2) \theta(q^0) \delta(q^2-2 q_0\sqrt{s} +s -s')
      \theta(\sqrt{s
  }-q^0)}{2( \pu\cdot q)( p_3\cdot q)} \,,
\end{align}
\begin{align}
    \mathcal{I}_{s'/s'}(t) &=
\mathcal{C}_D\int_{4m_\pi^2}^\infty  \dd s'\int \dd^{D-1} q \
    \frac{\delta(q^2) \theta(q^0) \delta(q^2-2 q_0\sqrt{s} +s -s')
      \theta(\sqrt{s
  }-q^0)}{2( \pu\cdot q)( p_3\cdot q)} \,,
\end{align}
where $D=4-2\epsilon_\text{IR}$ and $\mathcal{C}_D=(2\pi\mu)^{4-D} $. In particular, Eq.~\eqref{eq:iml2} and  Eq.~\eqref{eq:iml3} are the same with $\lambda=0$ and $\dd^3 \mathbf{q} \to \dd^{D-1} \mathbf{q}  $. The step of Eq.~\eqref{eq:iml4} is the same with $\lambda=0$, $\dd^3 \mathbf{q} \to \dd^{D-1} \mathbf{q}  $ and ${\cal I}_2$ is finite in the $D\to 4$ limit. Therefore ${\cal I}_2$ is still given by Eq.~\eqref{eq:imfinite} and ${\cal I}_1$ is the eikonal integral in dimensional regularisation, which becomes Eq.~\eqref{eq:I1lambda} after the replacement
\begin{equation}\label{eq:translation_rule}
    \frac{1}{\epsilon_\text{IR}} \to \log \frac{\lambda^2}{\mu^2} + \gamma_E - \log 4\pi\, .
\end{equation}

We also carried out the calculation in dimensional regularisation integrating explicitly in $q$ first and then in $s'$, retaining the full electron and pion mass dependence. This procedure is far more intricate, as the analytic expression of the integrals in $q$ are non-trivial and we cannot use the eikonal integral. However, the result we obtained is the same. This robust check ensures that our results for the dispersive integrals of the imaginary part of the $\text{D}_0$ are truly independent of the regularisation scheme and the order of loop and dispersive integrations. In the following, we give further mathematical details on the treatment of the $\lambda\to 0$ expansion in the mass regularisation scheme.

\subsection{Integral convergence}

In this section we justify more rigorously our treatment of the vanishing small terms. We carry out the discussion only for $\mathcal{I}_{1/s'}$ as it is valid for both integrals. Starting from Eq.~\eqref{eq:iml3} and integrating over $\dd w$ we obtain
  \begin{equation}
    \mathcal{I}_{1/s'}(t)
    = \int_{|\mathbf{q}|\le\omega_\lambda}
    \frac {\dd^3\mathbf{q}}{q_0} \frac{1}{(s - 2\sqrt s q_0)}
\frac{1}{(\lambda^2 - 2p_1 \cdot q )(\lambda^2- 2 p_3 \cdot q)} \,,
\end{equation}
where
\begin{equation}
\omega_\lambda =\sqrt{\frac{(s-4 m_\pi^2)^2}{4 s}-\lambda^2} \,.
  \end{equation}
By setting $ \mathbf{q}= \lambda \mathbf{u}$ and $u_0=\sqrt{|\mathbf{u}|^2+1}$, we obtain
\begin{equation}
 \mathcal{I}_{1/s'}(t)
    = \int_{|\mathbf{u}|\le\omega_\lambda/\lambda}
    \frac {\dd^3\mathbf{u}}{u_0} \frac{1}{(s - 2\sqrt s \lambda u_0)}
\frac{1}{(\lambda - 2p_1 \cdot u )(\lambda- 2 p_3 \cdot u)} \,,
\end{equation}
It is now straightforward to check that the integral of the difference of the above integrand and the same integrand with the replacement
\begin{equation}
\frac{1}{(\lambda - 2p_1 \cdot u )(\lambda- 2 p_3 \cdot u)}\to
\frac{1}{ 4(p_1 \cdot u )( p_3 \cdot u)}
\end{equation}
leads to a convergent integral times a coefficient of order $\lambda$. In more detail
\begin{equation}
\mathcal{I}_{1/s'}(t)
    = \int_{|\mathbf{u}|\le \omega_\lambda/\lambda}
    \frac {\dd^3\mathbf{u}}{u_0} \frac{1}{(s - 2\sqrt s \lambda u_0)}
\frac{1}{(p_1 \cdot u )( p_3 \cdot u)} + {\cal R} \,,
\end{equation}
where
\begin{align}
\begin{split}
{\cal R} & \le  \int_{|\mathbf{u}|\le \omega_\lambda/\lambda}
    \frac {\dd^3\mathbf{u}}{u_0} \frac{1}{ (s - 2\sqrt s \lambda u_0)}
    \frac{\lambda |2 (p_1 \cdot u )+2( p_3 \cdot u)-\lambda |}{4 (\lambda - 2p_1 \cdot u )(\lambda- 2 p_3 \cdot u)(p_1 \cdot u )( p_3 \cdot u) } \\[2pt]
    & \le  \int_{|\mathbf{u}|<\omega_\lambda/\lambda}
    \frac {\dd^3\mathbf{u}}{u_0} \frac{1}{ 4 m_\pi^2}
    \frac{\lambda (|2 (p_1 \cdot u )+2 ( p_3 \cdot u)|+\lambda) }{4 (\mu - 2p_1 \cdot u )(\mu- 2 p_3 \cdot u)(p_1 \cdot u )( p_3 \cdot u) } \\[5pt]
& =  \lambda^2 c_1(\mu,\lambda)+ \lambda c_2(\mu,\lambda) \,,
\end{split}
\end{align}
where $\mu$ is an arbitrary mass $\mu>\lambda$ such that no pole appears in the integration domain. We have
\begin{align}
\lim_{\lambda \to 0} c_j(\mu,\lambda)  = c_j(\mu,0) \in \mathbb{R}
\quad \Rightarrow \quad \lim_{\lambda \to 0} {\cal R} =0 \,.
\end{align}
Next, we show that we can set $\lambda \to 0$ {\em inside} the ${\cal I}_2$ integral, \textit{i.e.}
\begin{equation}
  \int_{|\mathbf{q}|\le \omega_\lambda}
  \!\!\dd^3\mathbf{q} \frac{1}{(  s - 2\sqrt s q_0)}
\frac{1}  
     {(p_1 \cdot q )( p_3 \cdot q)}
     =  \int_{|\mathbf{q}|\le \omega_\lambda}
  \!\!\dd|\mathbf{q}| \dd \Omega \frac{|\mathbf{q}|^2}{(  s - 2\sqrt s q_0)}
\frac{1}  
     {(p_1 \cdot q )( p_3 \cdot q)} \,.
\end{equation}
For $\lambda \to 0$, the integrand converges pointwise to
\begin{align}
\frac{1}{(  s - 2\sqrt s|\mathbf{q}|)}
\frac{1}  
     {(p_1 \cdot u )( p_3 \cdot u)}, \quad u \equiv (1,\mathbf{q}/|\mathbf{q}|)
\end{align}
and we have
\begin{align}
\left |
\frac{|\mathbf{q}|^2}{(  s - 2\sqrt s q_0)}
\frac{1}  
     {(p_1 \cdot q )( p_3 \cdot q)}
     \right | \le
     \frac{1}{(  s - 2\sqrt s \sqrt{|\mathbf{q}|^2+\mu^2})}
\frac{1}  
     {(p_1 \cdot u )( p_3 \cdot u)}, \quad \lambda < \mu
\end{align}
where $u \equiv (1,\mathbf{q}/|\mathbf{q}|)$ and $\mu$ is an arbitrary {\em constant} mass such that no pole occurs inside the integration domain.
     
The two properties above ensure that we can use Lebesgue's dominated convergence theorem~\cite{rudin1976principles} and take the limit inside the integral.

\bibliographystyle{JHEP}
\bibliography{twopions}

\end{document}